\newif\ifdraft
\newif\iffull
\newif\ifcomment
\newif\iflatexdiff
\newif\ifbibtex
\newif\ifpreprint
\latexdifftrue   
\bibtextrue      
\preprinttrue    
\fulltrue        
\newif\ifplbpaper
\newif\ifapspaper
\apspapertrue
\def\snntitle{$\snn$}
\ifpreprint
\def\snntitle{$\snnbf$}
\fi
\def\dvers{v5.0}       
\def\dtitle{Centrality dependence of particle production \\in p--Pb collisions at \snntitle = 5.02 TeV}
\def\stitle{Particle production and centrality in p--Pb}
\def\bibstname{utphys}
\ifpreprint
\documentclass[ALICE,manyauthors,12pt]{cernphprep}
\usepackage[comma,square,numbers,sort&compress]{natbib}
\else 
\ifplbpaper
\documentclass[final,3p,12pt]{elsarticle}
\biboptions{comma,square,numbers,sort&compress}
\def\bibstname{utphys}
\fi
\ifapspaper
\documentclass[10pt,aps,superscriptaddress,altaffilletter,nobibnotes,nofootinbib]{revtex4-1}
\newenvironment{frontmatter}{}{\maketitle}
\def\bibstname{apsrev4-1}
\fi
\fi
\usepackage{graphicx}  
\usepackage{dcolumn}   
\usepackage{bm}        
\usepackage{amssymb}   
\usepackage{amsfonts}
\usepackage{graphics}
\usepackage{grffile}   
\usepackage{epsfig}
\usepackage{units}
\usepackage{hyperref}
\usepackage[usenames]{color}
\usepackage[normalem]{ulem} 
\usepackage{color}
\usepackage[utf8x]{inputenc}
\usepackage[T1]{fontenc}
\usepackage{pbox}
\iflatexdiff
\RequirePackage{color}\definecolor{RED}{rgb}{1,0,0}\definecolor{BLUE}{rgb}{0,0,1}

\fi

\newcommand{\ZDC}          {\rm{ZDC}}
\newcommand{\ZNA}          {\rm{ZNA}}
\newcommand{\ZNC}          {\rm{ZNC}}
\newcommand{\ZPA}          {\rm{ZPA}}
\newcommand{\ZPC}          {\rm{ZPC}}
\newcommand{\SPD}          {\rm{SPD}}

\newcommand{\VZERO}        {\rm{VZERO}}
\newcommand{\VZEROA}       {\rm{VZERO-A}}
\newcommand{\VZEROC}       {\rm{VZERO-C}}

\newcommand{\VZAND}        {\rm{VZERO-AND}}

\newcommand{\pp}           {pp}

\newcommand{\PbPb}         {\mbox{Pb--Pb}}
\newcommand{\pPb}          {\mbox{p--Pb}}
\newcommand{\Pbp}          {\mbox{Pb--p}}

\newcommand{\pA}           {\mbox{p--A}}

\newcommand{\Nch}          {\ensuremath{N_\mathrm{ch}}}
\newcommand{\dNdeta}       {\ensuremath{\mathrm{d}N_\mathrm{ch}/\mathrm{d}\eta}}

\newcommand{\s}            {\ensuremath{\sqrt{s}}}
\newcommand{\pt}           {\ensuremath{p_{\rm T}}}

\newcommand{\snn}          {\ensuremath{\sqrt{s_{\rm NN}}}}
\newcommand{\snnbf}        {\ensuremath{\mathbf{{\sqrt{s_{\mathbf NN}}}}}}
\newcommand{\sonly}        {\ensuremath{\sqrt{s}}}

\newcommand{\TAB}          {\ensuremath{T_\mathrm{pPb}}}

\newcommand{\Npart}        {\ensuremath{N_\mathrm{part}}}
\newcommand{\Nparttar}     {\ensuremath{N_\mathrm{part}^\mathrm{target}}}
\newcommand{\avNpart}      {\ensuremath{\langle N_\mathrm{part} \rangle}}

\newcommand{\Ncoll}        {\ensuremath{N_\mathrm{coll}}}
\newcommand{\avNcoll}      {\ensuremath{\langle N_\mathrm{coll} \rangle}}

\newcommand{\avTAB}        {\ensuremath{\langle T_\mathrm{pPb} \rangle}}

\newcommand{\etac}         {\ensuremath{\eta_{\rm cms}}}
\newcommand{\dNdetac}      {\ensuremath{d}N_\mathrm{ch}/\ensuremath{d}\etac}

\newcommand{\signn}        {\ensuremath{\sigma^{\rm inel}_{\rm NN}}}

\newcommand{\qpa}          {\ensuremath{Q_{\rm pPb}}}
\newcommand{\rpa}          {\ensuremath{R_{\rm pPb}}}

\iflatexdiff

\renewcommand{\xout}[1]    {\textcolor{red}{\sout{#1}}}
 
\newcommand{\old}[1]       {{\textcolor{red}{\sout{#1}}}}

\else

\renewcommand{\xout}[1]    {}
\newcommand{\old}[1]       {\relax}

\fi

\graphicspath{{./img/}}
\makeatletter
\def\input@path{{./src/}}
\makeatother
\ifdraft
\usepackage{lineno}
\linenumbers
\modulolinenumbers[5]
\usepackage{fancyhdr}
\fi
\begin{document}
\ifpreprint
\begin{titlepage}
\PHyear{2014}
\PHnumber{281}        
\PHdate{17 November}  
\title{\dtitle}
\ShortTitle{\stitle}
\iffull
\Collaboration{ALICE Collaboration~\thanks{See Appendix~\ref{app:collab} for the list of collaboration members}}
\else
\Collaboration{ALICE Collaboration}
\fi
\ShortAuthor{ALICE Collaboration} 
\ifdraft
\begin{center}
\today\\ \color{red}DRAFT \dvers\ \hspace{0.3cm} \$Revision: 831 $\color{white}:$\$\color{black}\vspace{0.3cm}
\end{center}
\fi
\else
\begin{frontmatter}
\title{\dtitle}
\iffull
\input{authors-paper.tex}            
\else
\ifdraft
\author{ALICE Collaboration \\ \vspace{0.3cm} 
\today\\ \color{red}DRAFT \dvers\ \hspace{0.3cm} \$Revision: 831 $\color{white}:$\$\color{black}}
\else
\author{ALICE Collaboration}
\fi
\fi
\fi
\begin{abstract}
We report measurements of the primary charged particle pseudorapidity
density and transverse momentum distributions in \pPb\ collisions at
$\snn = 5.02$ TeV, and investigate their correlation with experimental
observables sensitive to the centrality of the collision.  Centrality
classes are defined using different event activity estimators,
i.e. charged particle multiplicities measured in three different
pseudorapidity regions as well as the energy measured at beam rapidity
(zero-degree).  The procedures to determine the centrality, quantified
by the number of participants (\Npart), or the number of
nucleon-nucleon binary collisions (\Ncoll), are described.  We show
that, in contrast to \PbPb\ collisions, in \pPb\ collisions large
multiplicity fluctuations together with the small range of
participants available, generate a dynamical bias in centrality
classes based on particle multiplicity.  We propose to use the
zero-degree energy, which we expect not to introduce a dynamical bias,
as an alternative event-centrality estimator.  Based on zero-degree
energy centrality classes, the \Npart\ dependence of particle
production is studied.  Under the assumption that the multiplicity
measured in the Pb-going rapidity region scales with the number of
Pb-participants, an approximate independence of the multiplicity per
participating nucleon measured at mid-rapitity of the number of
participating nucleons is observed. Furthermore, at high-\pt\
the \pPb\ spectra are found to be consistent with the pp spectra
scaled by \Ncoll\ for all centrality classes.  Our results represent
valuable input for the study of the event activity dependence of hard
probes in \pPb\ collision and, hence, help to establish baselines for
the interpretation of the \PbPb\ data.

\ifdraft 
\ifpreprint
\end{abstract}
\end{titlepage}
\else
\end{abstract}
\end{frontmatter}
\newpage
\fi
\fi
\ifdraft
\thispagestyle{fancyplain}
\else
\end{abstract}
\ifpreprint
\end{titlepage}
\else
\end{frontmatter}
\fi
\fi
\setcounter{page}{2}

\section{Introduction}
\label{sec:introduction}
Proton--lead collisions are an essential component of the heavy ion
programme at the Large Hadron Collider (LHC)~\cite{Salgado:2011wc}.
Measurements of benchmark processes in \pPb\ collisions serve as an
important baseline for the understanding and the interpretation of the
nucleus--nucleus data.  These measurements allow one to
disentangle hot nuclear matter effects which are characteristic of
the formation of the quark-gluon plasma (QGP) from cold nuclear matter
effects. The latter are the effects due to the presence of the nuclei
themselves and not the QGP, for example $k_{\rm T}$ broadening,
nuclear modification of parton densities, and partonic energy loss in
cold nuclear matter.

Of particular interest are studies of nuclear effects on parton
scatterings at large momentum transfer (hard processes).  To this end,
the nuclear modification factor defined as the ratio of particle or
jet transverse momentum (\pt) spectra in minimum bias (MB) p--Pb to those
in pp collisions scaled by the average number of binary p--nucleon
(p--N) collisions \avNcoll\ is measured~\cite{alice_RpA}. The latter
is given by the ratio of p--N and p--Pb inelastic cross-sections times
the mass number $A$.  In the absence of nuclear effects, the nuclear
modification factor is expected to be unity.  In heavy ion collisions,
binary scaling is found to hold in measurements of prompt
photons~\cite{Chatrchyan:2012vq} and electroweak
probes~\cite{Aad:2012ew,Chatrchyan:2011ua}, which do not strongly
interact with the medium.  The observation of binary scaling in p--Pb
demonstrates that the strong suppression of hadrons~\cite{RAA},
jets~\cite{Abelev:2013kqa} and heavy flavour hadrons~\cite{RAAD,
Abelev:2013yxa} seen in Pb--Pb collisions is due to strong final state
effects.  Centrality dependent measurements of the nuclear
modification factor $\rpa (\pt, {\rm cent})$, defined as
\begin{equation} \label{eq:Rpa}
\rpa (\pt , {\rm cent}) =
\frac{{\rm d}N^{\rm pPb}_{\rm cent}/{\rm d}\pt} { \langle N_{\rm coll}^{\rm cent} \rangle  {\rm d}N^{\rm pp}/ {\rm d}\pt}\, ,
\end{equation}
require the determination of the average $\Ncoll ^ {\rm cent}$ for each centrality
class.

Moreover, it has been recognised that the study of p--Pb collisions is
also interesting in its own right.  Several
measurements~\cite{Abelev:2012ola,Abelev:2013bla,Abelev:2013haa,ABELEV:2013wsa}
of particle production in the low and intermediate transverse momentum
region clearly show that \pPb\ collisions cannot
be explained by an incoherent superposition of \pp\ collisions.
Instead the data are compatible with the presence of
coherent~\cite{Dusling:2013oia} and collective ~\cite{Bozek:2011if}
effects.  Their strength increases with multiplicity indicating a
strong collision geometry dependence.  In order to corroborate this
hypothesis a more detailed characterisation of the collision geometry is
needed.

The Glauber model~\cite{Miller:2007ri} is generally used to calculate
geometrical quantities of nuclear collisions (A--A or p--A). In this
model, the impact parameter $b$ controls the average number of
participating nucleons (hereafter  referred as ``participants'' or
also ``wounded nucleons'' ~\cite{Bialas:1976,Bialas:2007eg}),
\Npart\ and the corresponding number of collisions \Ncoll.  It is expected that
variations of the amount of matter overlapping in the collision region will
change the number of produced particles, and parameters such
as \Npart\ and \Ncoll\ have traditionally been used to describe those
changes quantitatively, and to relate them to \pp\ collisions.

Using the Glauber model one can calculate the probability distributions ${\cal \pi}_{\nu} (\nu)$, where $\nu$ stands for \Npart\ or \Ncoll .
Since $\nu$ cannot be measured directly it has to be related via a model to an observable $M$, generally called   centrality estimator, via the conditional probability ${\cal P}(M | \nu ) $ to observe  $M$ for a given $\nu$.  
For each collision system and centre of mass energy,
the model has to be  experimentally validated by comparing the measured probability distribution
${\cal P}_{\rm meas}(M)$ to the one calculated from the convolution  ${\cal P}_{\rm calc}(M) = \sum_{\nu} {\cal P}(M | \nu )  {\cal \pi}_{\nu} (\nu)$. 
Once the model has been validated, for each event class defined by an $M$-interval the average $\nu$ is calculated. 
In order to unambigously determine $\nu$, one chooses observables whose mean values depend monotonically on  $\nu$.
Note that in p--A collisions
the impact parameter is only loosely correlated to $\nu$. 
Hence, although one uses traditionally the term centrality to refer to these measurements, the relevant parameters are \Npart\  and \Ncoll .

The procedure described above can be easily extended to several estimators.
Of particular interest are estimators from kinematic regions that are causally disconnected after the collision.
The measurement of a finite correlation between them unambiguously establishes their connection to the common collision geometry.
Typically these studies are performed with observables from well separated pseudorapidity ($\eta$) intervals, e.g. at zero-degree (spectators, slow-nucleons, deuteron break-up probability) and multiplicity in the rapidity plateau.

The use of centrality estimators in p--A collisions based on multiplicity or summed
energy in certain pseudo-rapidity intervals is motivated by the
observation that they show a linear dependence with \Npart \ or \Ncoll
.  This is also in agreement with models for the centrality dependence
of particle production (e.g. the Wounded Nucleon Model
~\cite{Bialas:1976,Bialas:2007eg}),
or also string models like
FRITIOF~\cite{Pi:1992ug}).  The total rapidity integrated multiplicity
of charged particles measured in hadron--nucleus collisions ($N_{\rm
ch}^{\rm h-A}$) at centre of mass energies ranging from 10 to 200~GeV
(E178~\cite{E178}, PHOBOS~\cite{Back:2004mr}) is consistent with a linear
dependence on \Npart : $N_{\rm ch}^{\rm h-A} = N_{\rm ch}^{\rm
pp} \cdot \Npart / 2$.  The ratio of particle pseudorapidity ($\eta$)
densities in d--Au and pp collisions exhibits a dependence on $\eta$,
which implies that the scaling behaviour has a strong rapidity
dependence with an approximate \Npart-scaling at $\eta = 0$ and an
approximate scaling with the number of target participants ($\Nparttar
= \Npart -1$) in the Au-going direction~\cite{Back:2004mr}.  In d--Au
collisions at RHIC ($\snn = 200$~GeV), the PHENIX and STAR
collaborations~\cite{Adare:2013nff, Adams:2003im} have used the
multiplicity measured in an $ \eta$-interval of width 0.9 centered at
$\eta \approx -3.5$ (Au-going direction) as a centrality
estimator. The multiplicity distribution has been successfully
described by the Glauber model assuming \Nparttar - scaling.  Finally,
in centrality averaged p--Pb collisions at the LHC ($\snn = 5.02$~TeV)
the primary charged particle pseudorapidity density at $\eta = 0$
scaled by the mean number of participants is found to be consistent
with the corresponding value in pp collisions interpolated to the same
$\snn$~\cite{alice_pA_dndeta}.

At RHIC, the deuteron dissociation probability can be accurately
modelled by a Glauber calculation and measured using the zero degree
calorimeters in the d-direction ~\cite{Adams20058,Adare:2013nff}.  The
mean number of participants has been determined for centrality classes
obtained with the multiplicity estimator described above and used to
calculate the deuteron break-up probability.  Inferred and measured
probabilities are consistent, demonstrating the correlation between
collision geometry and multiplicity, and providing a stringent test for
the \Npart\ determination.

Since for example hard scatterings can significantly contribute to the
overall particle multiplicity, correlations between high-\pt\ particle
production and bulk multiplicity can also be induced after the
collisions and, hence, they are not only related to the collision
geometry.  Therefore, the use of \Ncoll\ from the Glauber model to
scale cross-sections of hard processes from pp to p--A has to undergo
the same scrutiny as the correlation of the centrality estimator to
the collision geometry. This is necessary also due to the enhanced
role of
multiplicity fluctuations in p--A.  While the average of centrality
estimators vary monotonically with $\nu$, for a full description of
the conditional probability ${\cal P}(M | \nu)$ fluctuations of $M$
for a fixed $\nu$ have to be taken into account.  In \PbPb\
collisions, these multiplicity fluctuations have little influence on
the centrality determination.  The range of $\nu$ is large and ${\cal
P}(M | \nu)$ converges with increasing $\nu$ rapidly to a Gaussian
with small width relative to the the range of $\nu$.  However,
in \pPb\ collisions, the range of multiplicities used to select a
centrality class is of similar magnitude as the fluctuations, with the
consequence that a centrality selection based on multiplicity may
select a biased sample of nucleon--nucleon collisions 
(for a discussion of this effect in d+Au see~\cite{Adare:2013nff}).

%
%
In essence, by selecting high (low) multiplicity one chooses not only
large (small) average \Npart, but also positive (negative)
multiplicity fluctuations leading to deviations from the binary
scaling of hard processes.  These fluctuations are\ partly related to
qualitatively different types of collisions.  High multiplicity
nucleon-nucleon collisions show a significantly higher particle mean
transverse momentum. They can be understood as ``harder'' collisions,
i.e. with higher 4-momentum transfer squared $Q^2$ or as 
nucleon-nucleon collisions where multiple
parton-parton interactions (MPI) take place.

In contrast, a centrality selection that is not expected to
induce a bias on the binary scaling of hard processes is provided by
the energy measurement with the Zero Degree Calorimeters (ZDC) due to
their large $\eta$-separation from the central barrel detectors.  They
detect the so-called ``slow'' nucleons produced in the interaction by
nuclear de-excitation processes, or knocked out by wounded
nucleons~\cite{Sikler:2003ef,oppedisano:2006}.  The relationship of
the energy deposited in the \ZDC\ to the number of collisions requires
a detailed model to describe the slow nucleon production.  A heuristic
approach, based on a parameterization of data from low energy
experiments, is discussed in the present paper.

We will show that centrality estimators using forward neutron energy
and those using central multiplicity give consistent results
for \Npart\ and \Ncoll, demonstrating their connection to the
collision geometry.  Based on the considerations outlined above we
study two different procedures for centrality estimation.  The first
procedure is to determine the centrality with charged particle
multiplicity.  The collision geometry is determined by fitting the
measured multiplicity distribution with the \Ncoll\ distribution
obtained from the Glauber model~\cite{Miller:2007ri}, convolved with a
negative binomial distribution (NBD).  Due to the possible dynamical
bias introduced by the multiplicity selection \Ncoll\ should 
in this case not be
used to scale hard cross-sections.  Additional effort is needed to
understand the bias or to extend the Glauber model to
include additional dynamical fluctuations. Several possible
directions have been discussed, for example Glauber-Gribov
fluctuations of the proton size~\cite{gribov1969} as well as
fluctuations of the number of hard-scatterings per collisions due to
the impact parameter dependence and purely statistical (Poissonian)
fluctuations~\cite{hijing}.

The second procedure requires a centrality selection with minimal bias
and, therefore, uses the ZDC signal.  To relate the ZDC signal to the
collision geometry we have developed a heuristic model for slow
nucleon emission, based on a parameterization of data from low energy
experiments.  This heuristic approach however can provide only a
model-dependent \Ncoll\ determination.  However, one can study the
correlation of two or more observables out of which at least one is
expected to scale linearly with \Ncoll.  Examples are i) the
target-going multiplicity proportional to the number of wounded target
nucleons ($\Nparttar=\Npart-1=\Ncoll$), ii) the multiplicity at
mid-rapidity proportional to the number of participants ($\Npart
= \Ncoll + 1$), iii) the yield of hard probes, like high-\pt\
particles at mid-rapidity proportional to \Ncoll. These scalings can
be used as an ansatz when calculating \Ncoll\ based on an event
selection using the ZDC.

Both alternatives are discussed in the present paper.  The paper is
organised as follows.  Section \ref{sec:multiplicity} describes the
experimental conditions, the event selection, and the event
characterisation using the multiplicity distributions of charged
particles measured in various $\eta$ ranges, or the energy collected
in the ZDC.  Section \ref{sec:GlauberFit} describes the centrality
determination based on charged particle distributions using an
NBD-Glauber fit to extract the average geometrical quantities for
typical centrality classes.  Section \ref{sec:geometry} presents a
phenomenological model describing the relation of the energy deposited
in the \ZDC\ calorimeter and \Ncoll. Section \ref{sec:bias} discusses
the various effects leading to a bias in the centrality measurements
based on particle multiplicity.  Section \ref{sec:hybrid} introduces a
hybrid method, where we use the ZDC to characterize the event
activity, and base the determination of \Ncoll\ on the assumption
that \Npart-scaling holds for the central pseudorapidity multiplicity
density or $\Npart^{\rm target} $-scaling for particle production in
the target region.  Section \ref{sec:results} discusses the
implications of the different choices of a centrality estimator on the
physics results, such as the nuclear modification factors, or the
pseudorapidity density of charged particles at mid-rapidity.
Section \ref{sec:conclusions} summarizes and concludes the paper.

\section{Experimental Conditions}
\label{sec:multiplicity}
The data were recorded during a dedicated LHC run of 4 weeks in
January and February 2013. Data have been taken with two beam
configurations, by inverting the direction of the two particle
species, referred to as p--Pb and Pb--p, respectively, for the
situations where the proton beam is moving towards positive
rapidities, or vice versa.  The two-in-one-magnet design of the LHC
imposes the same magnetic rigidity of the beams in the two rings
,implying that the ratio of beam energies is fixed to be exactly equal
to the ratio of the charge/mass ratios of each beam.  Protons at 4 TeV
energy collided onto fully stripped $^{208}_{82}$Pb ions at 1.58 TeV
per nucleon energy resulting in collisions at $\snn=5.02$~TeV in the
nucleon--nucleon centre-of-mass system (cms), which moves with a
rapidity of $\Delta y_{\rm NN}=0.465$ in the direction of the proton
beam.  In the following, we will use the convention that $y$ stands
for $y_{\rm cms}$, defined such that the proton moves towards positive
$\eta_{\rm cms}$, while $\eta$ stands for $\eta_{\rm lab}$.

The number of colliding bunches varied from 8 to 288. The proton and
Pb bunch intensities were ranging from 0.2 $\times 10^{12}$ to 6.5
$\times 10^{12}$ and from 0.1 $\times 10^{12}$ to 4.4 $\times
10^{12}$, respectively. The luminosity at the ALICE interaction point
was up to 5 $\times 10^{27}$cm$^{−2}$s$^{−1}$ resulting in a 10 kHz
hadronic interaction rate.  The RMS width of the interaction region is
6.3 cm along the beam direction and of about 60 $\mu$m in the
direction transverse to the beam.

The ALICE apparatus and its performance in the LHC Run 1 are described
in~\cite{aliceapp} and~\cite{Abelev:2014ffa}, respectively.  The main
detector components used for the centrality determination are: the
Silicon Pixel Detector (\SPD), two cylindrical layers of hybrid
silicon pixel assemblies covering $|\eta|<2.0$ for the inner layer and
$|\eta|<1.4$ for the outer layer for vertices at the nominal
interaction point, with {93.5}\% active channels; the Time Projection
Chamber (TPC), a large cylindrical drift detector covering
$|\eta|<0.9$; the \VZERO\ scintillator counters, covering the full
azimuth within $2.8<\eta<5.1$~(\VZEROA) and
$-3.7<\eta<-1.7$~(\VZEROC); and the Zero Degree Calorimeters~(\ZDC),
two sets of neutron (\ZNA\ and \ZNC) and proton (\ZPA\ and \ZPC)
calorimeters positioned at $\pm$112.5~m from the interaction point,
with an energy resolution of about 20\% for the neutron and 24\% for
the proton calorimeters.

The p--Pb trigger, configured to have high efficiency for hadronic
events, requires a signal in both the \VZEROA\ and
\VZEROC~(\VZAND\ requirement). 
Beam--gas and other machine-induced background collisions with
deposited energy above the thresholds in the \VZERO\ or
\ZDC\ detectors are suppressed by requiring the signal arrival time to be
compatible with a nominal \pPb\ interaction.  The fraction of
remaining beam-related background after all requirements is estimated
from control triggers on non-colliding or empty bunches, and found to
be negligible.

The resulting event sample corresponds to a so-called \emph{visible
cross-section} of $2.09 \pm 0.07 \, {\rm barn}$ measured in a van der
Meer scan~\cite{MGagliardi2013}.  From Monte Carlo simulations we
expect that the sample consists mainly of non-single diffractive (NSD)
collisions and a negligible contribution of single-diffractive (SD)
and electromagnetic interactions.  The \VZAND\ trigger is not fully
efficient for NSD events.  Previous Monte Carlo studies (for details
see~\cite{alice_pA_dndeta}) have shown that the inefficiency is
observed mostly for events without reconstructed vertex, i.e. with no
particles produced at central rapidities.  Given the fraction of such
events in data (1.5\%), the corresponding inefficiency was found to be
2.2\% with a large systematic uncertainty of 3.1\%. Correcting for this
inefficiency would mainly concern the most peripheral class (80-100\%)
where the correction amounts up to 11\% $\pm$ 15.5\%.  For the results
reported in this paper, centrality classes have been defined as
percentiles of the visible cross-section and the measurements are not
corrected for trigger inefficiency.

The centrality determination is performed by exploiting the rapidity
coverage of the various detectors.  The raw multiplicity distributions
measured in the Central Barrel are modelled by assuming particle
production sources are distributed according to a NBD.  The
zero-degree energy of the slow nucleons emitted in the nucleon
fragmentation requires more detailed models.

In this context, the main estimators used for centrality in the
following are: 
\begin{itemize}
\item CL1: the number of clusters in the outer layer of the silicon pixel detector, $|\eta|<1.4$; 
\item V0A: the amplitude measured by the VZERO hodoscopes on the A-side
(the Pb-going side in the \pPb\ event sample), $2.8<\eta<5.1$;
\item V0C: the amplitude measured by the VZERO hodoscopes on the C-side
(the p-going side in the \pPb\ event sample), $-3.7<\eta<-1.7$;
\item V0M: the sum of the amplitudes in the VZERO hodoscopes on the A- and C-side (V0A+V0C);
\item ZNA: the energy deposited in the neutron calorimeter on the A-side (the Pb-going side in the \pPb\ event sample).
\end{itemize}

\section{Centrality from Charged Particle Distributions} 
\label{sec:GlauberFit}
\subsection{NBD-Glauber Fit} \label{subsec:GlauberFit}

\begin{figure}[tbh!f]
 \centering \includegraphics[width=0.75\textwidth]{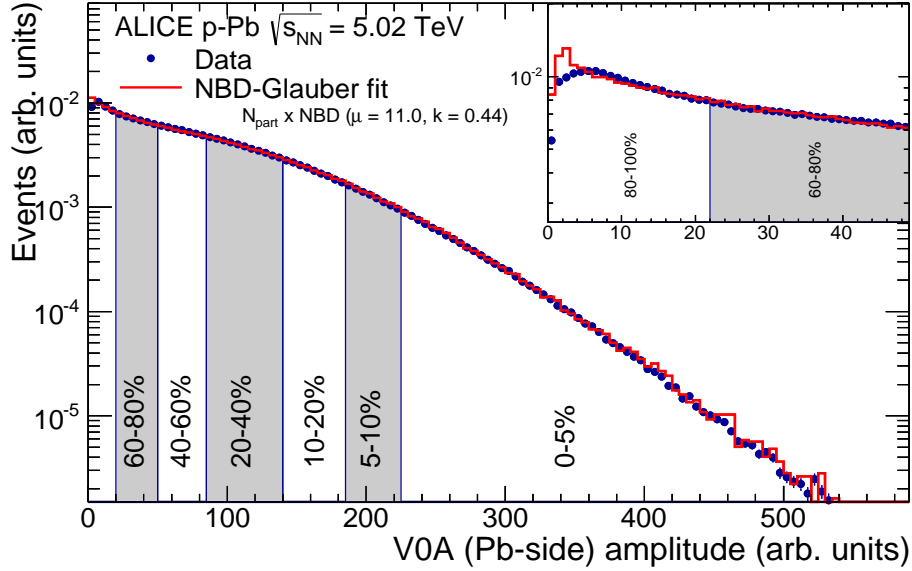} 
 \caption{ (color online) Distribution of the sum of amplitudes in the
   V0A hodoscopes (Pb-going), as well as the NBD-Glauber
   fit (explained in the text). Centrality classes are indicated by
   vertical lines. The inset shows a zoom-in on the most peripheral events.
 \label{fig:V0AGlau}}
\end{figure}

To determine the relationship between charged particle multiplicity and
the collision properties, such as the number of participating nucleons
\Npart, binary pN collisions \Ncoll, or nuclear overlap \TAB ($=
\Ncoll / \signn$), it is customary to use the Glauber Monte Carlo
model combined with a simple model for particle
production~\cite{glauber1959,glauber1955,glauber2006,glauberMC,glauberMC2}. The
method was used in \PbPb\ collisions and is described in detail in
\cite{Alice:centrality}.  In the Glauber calculation, the nuclear
density for $^{208}_{82}$Pb is modelled by a Woods-Saxon distribution for a
spherical nucleus
\begin{equation}
\rho(r) = \rho_{0} \frac{1}{1 + \exp \left(\frac{r-R}{a}\right) }
\end{equation}
with $\rho_{0}$ the nucleon density, which provides the overall
normalization, a radius of ($R=6.62 \pm 0.06$)~fm and a skin depth of
$a=(0.546 \pm 0.010)$~fm, based on data from low energy
electron-nucleus scattering experiments~\cite{DeJager:1987qc}.
Nuclear collisions are modelled by randomly displacing the projectile
proton and the target Pb nucleus in the transverse plane.  A
hard-sphere exclusion distance of $0.4$~fm between nucleons is
employed.  The proton is assumed to collide with the nucleons of the
Pb-nucleus if the transverse distance between them is less than the
distance corresponding to the inelastic nucleon-nucleon cross-section
of $70\pm 5$~mb at $\sqrt{s} = 5.02$~TeV, estimated from interpolating
data at different centre of mass energies~\cite{Nakamura:2010zzi}
including measurements at 2.76 and 7 TeV~\cite{abelev:2012sj}. The
VZERO-AND cross-section measured in a van der Meer
scan~\cite{MGagliardi2013} was found to be compatible, assuming
negligible efficiency and EM contamination corrections, with the
Glauber-derived p-nucleus inelastic cross-section (2.1 $\pm$ 0.1 b).
%
The Glauber-MC determines on an event-by-event basis the properties of
the collision geometry, such as \Npart, \Ncoll\ and \TAB, which must
be mapped to an experimental observable.  

Assuming that the average V0A multiplicity is proportional to the number of
participants in an individual \pA\ collision, the probability
distribution $P(n)$ of the contributions $n$ to the amplitude from
each p-nucleon collisions can be described by the NBD, defined as:

\begin{equation}
  P(n; \mu, k) =  \frac{\Gamma(n+k)}{\Gamma(n+1)\Gamma(k)} \cdot
  \frac{(\mu/k)^n}{(\mu/k+1)^{n+k}}\,,
\end{equation}
where $\Gamma$ is the gamma function, $\mu$ the mean amplitude per participant and the dispersion
parameter $k$ is related to the relative width given by $\sigma/\mu =
\sqrt{1/\mu + 1/k}$.  From the closure of the NBD under convolution, it
follows that the conditional probability ${\cal P}(n | \Npart )$,
i.e. $\Npart$ repeated convolutions, is equal to $P(n; \Npart \mu , \Npart
k)$.

\begin{table}[thb!f]
 \centering \footnotesize  
\begin{tabular}{c|cc|cc}
 System & \multicolumn{2}{c}{\pp}  & \multicolumn{2}{c}{\pPb} \\
 Distribution & $\mu$ & $k$ & $\mu$ & $k$  \\ 
 \hline 
 V0A & \phantom{0}9.6  & 0.56 & 11.0 & 0.44 \\
 V0M & 25.2 & 0.82 & 23.6 & 1.08 \\
 CL1 & \phantom{0}9.8  & 0.64 & 8.74 & 0.76 \\
\end{tabular}
 \caption{Fit parameters of the \Npart $\times$ NBD for \pp\ collisions at 7 TeV and
   \pPb\ multiplicity distributions.
\label{tab:NBDpar}} 
\end{table}
\begin{table}[thb!f]
 \centering \footnotesize 
\hspace{-1cm}
\begin{tabular}{c|cccccccc}
 Centrality (\%)& $\langle b \rangle (\rm fm)$ & $\sigma$ (\rm fm)& $\langle\TAB\rangle$
 (mb$^{-1}$)& $\sigma$ (mb$^{-1}$)& $\langle\Npart\rangle$ & $\sigma$ & $\langle\Ncoll\rangle$ &
 $\sigma$  \\ 
 \hline 
\phantom{0}0 - \phantom{00}5    & 3.12 & 1.39 & 0.211\phantom{0}  & 0.0548 & 15.7\phantom{0} & 3.84 & 14.7\phantom{0} & 3.84  \\ 
\phantom{0}5 - \phantom{0}10   & 3.50 & 1.48 & 0.186\phantom{0}  & 0.0539 & 14.0\phantom{0} & 3.78 & 13.0\phantom{0} & 3.78  \\ 
10 - \phantom{0}20  & 3.85 & 1.57 & 0.167\phantom{0}  & 0.0549 & 12.7\phantom{0} & 3.85 & 11.7\phantom{0} & 3.85  \\ 
20 - \phantom{0}40  & 4.54 & 1.69 & 0.134\phantom{0}  & 0.0561 & 10.4\phantom{0} & 3.93 & \phantom{0}9.36 & 3.93  \\ 
40 - \phantom{0}60  & 5.57 & 1.69 & 0.0918 & 0.0516 & \phantom{0}7.42 & 3.61 & \phantom{0}6.42 & 3.61  \\ 
60 - \phantom{0}80  & 6.63 & 1.45 & 0.0544 & 0.0385 & \phantom{0}4.81 & 2.69 & \phantom{0}3.81 & 2.69  \\ 
80 - 100 & 7.51 & 1.11 & 0.0277 & 0.0203 & \phantom{0}2.94 & 1.42 & \phantom{0}1.94 & 1.42  \\  
\phantom{0}0 - 100  & 5.56 & 2.07 & 0.0983 & 0.0728 & \phantom{0}7.87 & 5.10 & \phantom{0}6.87 & 5.10 \\ 
\end{tabular}
 \caption{Geometric properties ($b$, \TAB, \Npart, \Ncoll) of
   \pPb\ collisions for centrality classes defined by cuts in
   V0A. The mean values and the $\sigma$ values are obtained with a Glauber
   Monte Carlo calculation, coupled to an NBD to fit the V0A
   distribution.
\label{tab:Ncoll}} 
\end{table}
\begin{table}[thb!f] 
 \centering \footnotesize  
\hspace{-1cm}
\begin{tabular}{c|c|ccc|ccc|c|c}
 Centrality (\%)& $\langle\Ncoll^{b}\rangle$  & $\langle\Ncoll^{\rm CL1}\rangle$ & $\langle\Ncoll^{\rm V0M}\rangle$ & $\langle\Ncoll^{\rm V0A}\rangle$ & \pbox{2.5cm}{Sys. \\Glauber} & \pbox{2.5cm}{Sys. \\MC-closure}  & \pbox{2.5cm}{Sys. \\Tot}  & $\langle\Ncoll^{\rm ZNA}\rangle$ & \pbox{2.5cm}{Sys. \\SNM} \\
 \hline 
 \phantom{0}0 -\phantom{00}5  & 14.4\phantom{0} & 15.6\phantom{0} & 15.7\phantom{0} & 14.8\phantom{0} &  10\%  (3.7\%) & \phantom{0}3\% & 10\% & 15.7\phantom{0} & 7\%\\ 
 \phantom{0}5 -\phantom{0}10  & 13.8\phantom{0} & 13.6\phantom{0} & 13.7\phantom{0} & 13.0\phantom{0} &  10\%  (3.5\%) & \phantom{0}1\% & 10\% & 13.9\phantom{0} & 5\%\\ 
 10 -\phantom{0}20            & 12.7\phantom{0} & 12.0\phantom{0} & 12.1\phantom{0} & 11.7\phantom{0} &  10\%  (3.2\%) & \phantom{0}2\% & 10\% & 12.4\phantom{0} & 2\%\\ 
 20 -\phantom{0}40            & 10.2\phantom{0} & \phantom{0}9.49 & \phantom{0}9.55 & \phantom{0}9.36 &  8.8\% (3.1\%) & \phantom{0}2\% & 9\%  & \ 9.99\phantom{0} & 2\%\\ 
 40 -\phantom{0}60            & \phantom{0}6.30 & \phantom{0}6.18 & \phantom{0}6.26 & \phantom{0}6.42 &  6.6\% (4.3\%) & \phantom{0}3\% & 7.2\% & \phantom{0}6.53 & 4\%\\ 
 60 -\phantom{0}80            & \phantom{0}3.10 & \phantom{0}3.40 & \phantom{0}3.40 & \phantom{0}3.81 &  4.3\% (6.7\%) & 20\%           & 20\%  & \phantom{0}3.04 & 4\%\\ 
 80 -100                      & \phantom{0}1.44 &\phantom{0}1.76  & \phantom{0}1.72 & \phantom{0}1.94 &  2.0\% (9.3\%) & 23\%           & 23\%  & \phantom{0}1.24 & 8\%\\
 \phantom{0}0 -100            & \phantom{0}6.88 & \phantom{0}6.83 & \phantom{0}6.87 & \phantom{0}6.87 &  8\%  (3.4\%) & - & 8\% &\phantom{0}6.88 & - \\
\end{tabular}
 \caption{Comparison of $\langle \Ncoll \rangle$ values. In the first column results are listed for
   centrality classes obtained by ordering the events according to the
   impact parameter distribution ($b$). In the next three columns
   \avNcoll\ values are given for the various centrality estimators
   CL1, V0A, V0M.  The systematic uncertainty on \avNcoll\ (in
   parenthesis on \TAB) is obtained by changing all Glauber parameters by
   1$\sigma$; the second column is obtained from the MC-closure test; those two are added in quadrature to
   obtain the total systematic uncertainty on \avNcoll.  The last column gives
   the \avNcoll\ values obtained for the ZNA (see
   Sec.~\ref{sec:geometry}) and the uncertainty on the Slow Nucleon Model (SNM, see
   Sec.~\ref{sec:geometry}).
\label{tab:NcollCompare}} 
\end{table}

To obtain the NBD parameters $\mu$ and $k$, the calculated V0A
distribution, obtained by convolving the Glauber \Npart\ distribution
with ${\cal P}(n | \Npart )$, is fitted to the measured V0A
distribution.  
The fit is performed excluding the low V0A
amplitude region, V0A$<$~10. We note however that fitting with the
full range gives consistent results.  The measured V0A distribution
together with the NBD-Glauber distribution for the best fit are shown in
Fig.~\ref{fig:V0AGlau}. Similar fits have been performed to V0M and
CL1 and the corresponding fit parameters are listed in Table
\ref{tab:NBDpar}.  The values of the parameters $\mu$ and $k$ are
similar to those obtained by fitting the corresponding multiplicity
distributions in \pp\ collisions at 7 TeV. Since the raw distribution
is sensitive to experimental parameters such as noise and gain, one
cannot expect identical values even in the case of perfect
\Npart-scaling and therefore the comparison is only qualitative.

\begin{figure}[tbh!f]
 \centering
 \includegraphics[width=0.75\textwidth]{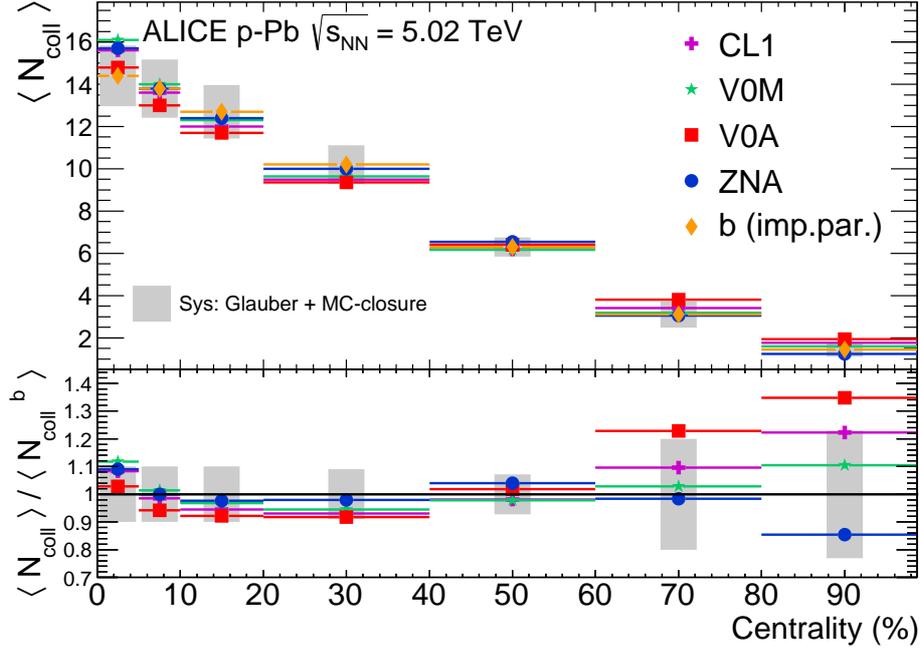}
 \caption{(color online) Values of $\langle \Ncoll \rangle$ extracted from CL1, V0M, V0A, ZNA and by
   ordering the events according to the
   impact parameter distribution ($b$). The systematic uncertainty, given by the quadrature
   sum of the uncertainty from the Glauber parameters and the
   MC-closure test, are drawn around the values obtained with $b$.
  \label{fig:Ncoll}}
\end{figure}

For a given centrality class, defined by selections in the measured
distribution, the information from the Glauber MC in the corresponding
generated distribution is used to calculate the mean number of
participants \avNpart, the mean number of collisions \avNcoll, and the
average nuclear overlap function $\langle T_{\rm pPb} \rangle$.  These
are given in Table~\ref{tab:Ncoll}, with the corresponding $\sigma$ values.
Since the event selection dominantly selects NSD events, it is
important to note that the number of participants in the Glauber
calculation would increase by only 2.5\% for NSD events. This was
estimated with a modified Glauber calculation to exclude SD
collisions~\cite{alice_pA_dndeta}.

The systematic uncertainties are evaluated by varying the Glauber
parameters (radius, skin depth and hard-sphere exclusion distance)
within their known uncertainty. The uncertainties on \avNcoll\ are
listed in Table~\ref{tab:NcollCompare}, by adding all the deviations
from the central result in quadrature. The uncertainties range from
about 4-5\% in peripheral collisions to about 10\% in central
collisions.  Note that, as $\TAB = \Ncoll / \signn$, the uncertainties
on \signn\ and \avNcoll\ largely cancel in the calculation of
\TAB. However, edge effects in the nuclear overlap are large for
\TAB\ in peripheral collisions.

The procedure was tested with a MC-closure test using HIJING
\pPb\ simulations~\cite{hijing} with nuclear modifications of the
parton density (shadowing) and elastic scattering switched off.  In
the MC-closure test the V0A distribution obtained from a detailed
detector simulation coupled to HIJING was taken as the input for the
fit with the NBD-Glauber method.  The difference between the
\avNcoll\ values calculated from the fit, and those from the MC truth
used in the HIJING simulation range from 3\% in central to 23\% in
peripheral events (see Table~\ref{tab:NcollCompare}). The large uncertainty in the peripheral events arises from the small absolute values of \Ncoll \ itself. In this case a small absolute uncertainty results in a
large relative deviation.  The total
uncertainty on \avNcoll\ for each centrality class with the CL1, V0M
or V0A estimators is obtained by adding the uncertainty from the
variation of the Glauber parameters with those from the respective
MC-closure test in quadrature.

The NBD-Glauber fit is repeated for the multiplicity distribution of
the SPD clusters (CL1) and for the sum of 
V0A and V0C, V0M, in the same centrality classes as for
V0A.  The \avNcoll\ values as a function of centrality are given in
Table~\ref{tab:NcollCompare} and shown in Fig.~\ref{fig:Ncoll} for the
various estimators. In addition, the events from the MC-Glauber calculation were
  ordered according to their impact parameter, and the values of
  \avNcoll\ were extracted for the same centrality classes.  The
variation of \Ncoll\ between different centrality estimators is small and of similar magnitude as the
systematic uncertainty obtained by adding in quadrature the
uncertainty from the Glauber model and from the MC-closure test. This
implies that the \avNcoll\ determination with the NBD-Glauber fit is
robust and independent of the centrality estimator used. 

\subsection{Glauber-Gribov corrections}\label{subsec:gribov}
Event-by-event fluctuations in the configuration of the incoming
proton can change its scattering cross-section~\cite{gribov1969}.  In
the Glauber MC this phenomenon is implemented by an effective
scattering cross-section~\cite{heiselberg1991,guzey2006,alvioli2013}.
At high energies, the configuration of the proton is taken to be
frozen over the time scale of the p--A collision. Analogously to the
studies in~\cite{AtlasCent,Loizides:2014vua}, the effect of these
frozen fluctuations of the projectile proton is evaluated with a
modified version of the Glauber MC, referred to as
“Glauber-Gribov”. This version includes event-by-event variations of
the nucleon-nucleon cross-section. Here we have used the same values
of the parameter $\Omega$, which controls the width of the probability
distribution of \signn, as used in~\cite{AtlasCent}, namely $\Omega=
0.55$ and $1.01$, where $\Omega= 0.0$ corresponds to the standard
Glauber.

\begin{figure}[t!]
 \centering
 \includegraphics[width=0.495\textwidth]{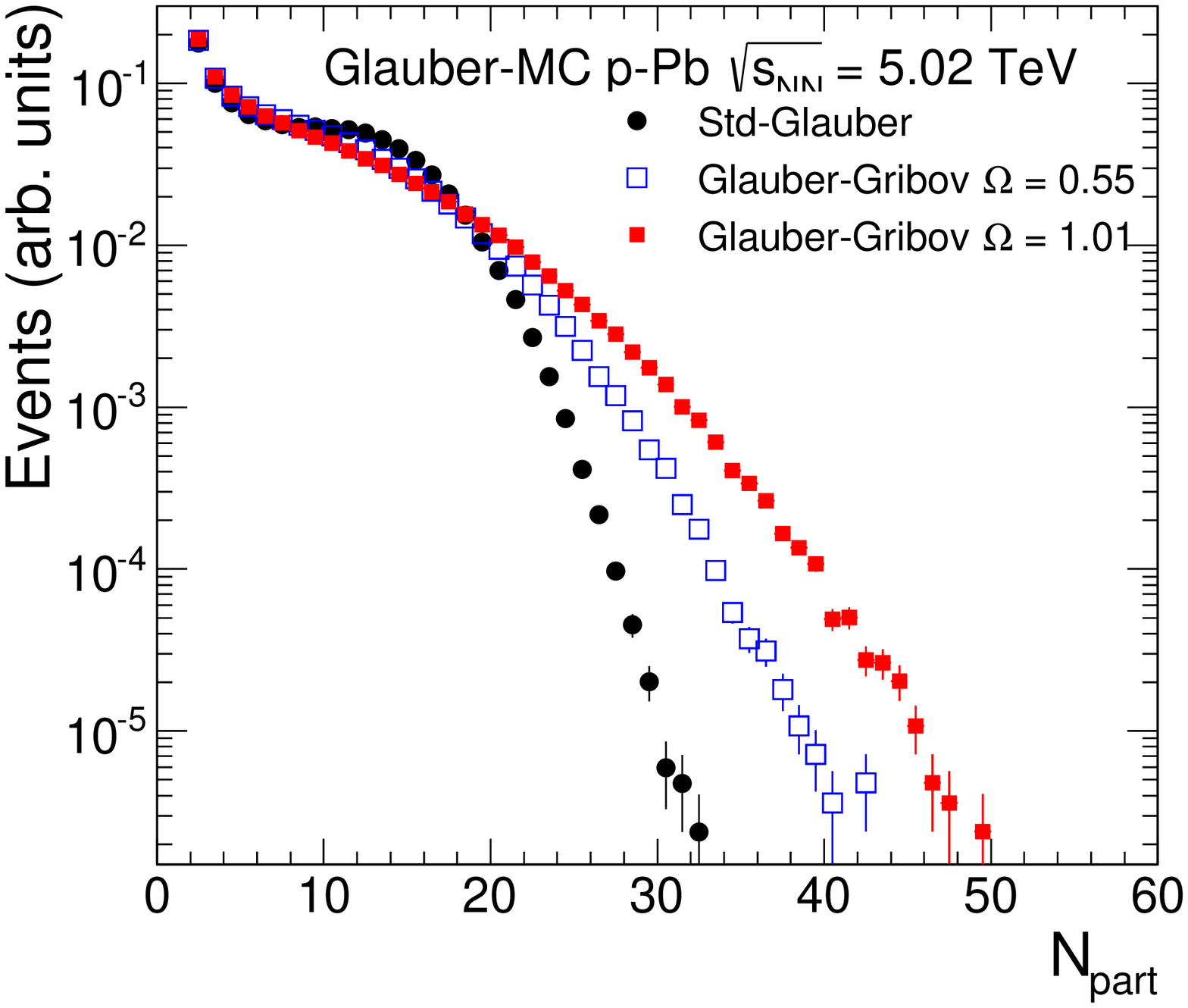}
 \includegraphics[width=0.49\textwidth]{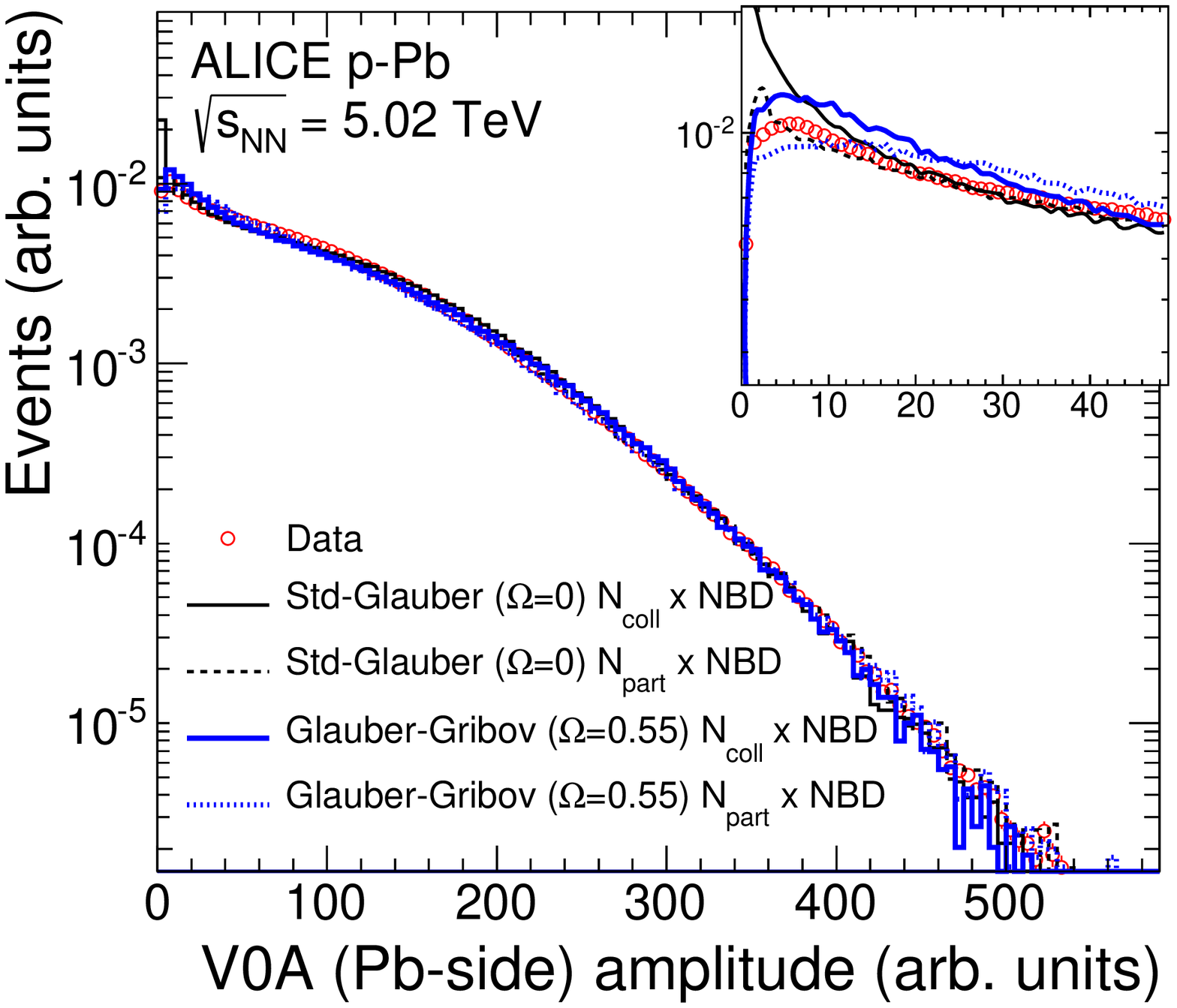}
 \caption{(color online) Left: Glauber and Glauber-Gribov Monte Carlo $N_{part}$
   distributions for 5.02 TeV \pPb\ collisions. Right: Measured V0A
   distribution compared to Glauber and Glauber-Gribov fits assuming
   \Npart\ or \Ncoll\ scaling. The inset shows a zoom-in on the most peripheral events.
  \label{fig:Gribov}}
\end{figure}

The distribution of the number of participants, \Npart, obtained from
the two Glauber-Gribov parameter variations are shown in the left panel of
Fig.~\ref{fig:Gribov} together with a standard \Npart\ distribution
obtained using a fixed inelastic cross-section, $\signn = 70$~mb. The
Glauber-Gribov \Npart\ distributions are much broader than the Glauber
distribution due to the cross-section fluctuations. We note that by
construction the total inelastic \pPb\ cross-section is unaltered by
the proton fluctuations.

\begin{table}[thb!f] 
 \centering \footnotesize  
\begin{tabular}{c|cc}
 Distribution & $\mu$ & $k$  \\ 
 \hline 
 Std-Glauber and \Ncoll $\otimes$ NBD    & 12.2 & 0.58 \\
 Std-Glauber and \Npart $\otimes$ NBD    & 11.0 & 0.44 \\
 Glauber-Gribov and \Ncoll $\otimes$ NBD & 12.6 & 1.35 \\
 Glauber-Gribov and \Npart $\otimes$ NBD & 11.0 & 0.60 \\
\end{tabular}
 \caption{Fit parameters of the V0A
   distributions using standard Glauber and Glauber-Gribov ($\Omega = 0.55$)
   distributions of \Npart\ and \Ncoll coupled to a NBD.
\label{tab:NBDparGribov}} 
\end{table}
\begin{table}[thb!f] 
 \centering \footnotesize  
\hspace{-1cm}
\begin{tabular}{c|cc|cc}
  & \multicolumn{2}{c}{Std-Glauber}  &  \multicolumn{2}{c}{Glauber-Gribov}   \\
 Centrality (\%)& \Npart\ x NBD & \Ncoll x NBD &  \Npart\ x NBD & \Ncoll x NBD \\
 \hline 
 \phantom{0}0 -\phantom{0}\phantom{0}5   & 14.8\phantom{0} & 15.3\phantom{0} & 17.8\phantom{0} & 19.2\phantom{0} \\ 
 \phantom{0}5 -\phantom{0}10  & 13.0\phantom{0} & 13.4\phantom{0} & 14.4\phantom{0} & 15.2\phantom{0}\\ 
 10 -\phantom{0}20 & 11.7\phantom{0} & 12.0\phantom{0} & 12.0\phantom{0} & 12.5\phantom{0} \\ 
 20 -\phantom{0}40 & \phantom{0}9.36 & \phantom{0}9.62 & \phantom{0}8.82 & \phantom{0}9.04 \\ 
 40 -\phantom{0}60 & \phantom{0}6.42 & \phantom{0}6.40 & \phantom{0}5.68 & \phantom{0}5.56 \\ 
 60 -\phantom{0}80 & \phantom{0}3.81 & \phantom{0}3.42 & \phantom{0}3.33 & \phantom{0}2.89 \\ 
 80 -100& \phantom{0}1.94 &\phantom{0}1.85 & \phantom{0}1.72 & \phantom{0}1.43 \\
 \phantom{0}0 -100 & \phantom{0}6.87 & \phantom{0}6.87 & \phantom{0}6.73 & \phantom{0}6.75 \\
\end{tabular}
 \caption{\Ncoll\ values obtained for various fits of the V0A, using
   Std-Glauber ($\Omega=0.0$) and Glauber-Gribov ($\Omega=0.55$)
   distributions for \Npart\ or \Ncoll, coupled to a NBD.
\label{tab:NcollGribov}} 
\end{table}

The Glauber-Gribov distributions of \Npart\ and \Ncoll, coupled to a
NBD, were fitted to the measured distribution of V0A. The right panel
of Fig.~\ref{fig:Gribov} shows the V0A distribution together with
various fits performed with the standard Glauber or the Glauber-Gribov
distribution, using $\Omega = 0.55$, and assuming that the signal
increases proportionally either to \Npart\ or to \Ncoll. As before, no
attempt is made to describe the most peripheral region (below
$\sim$90\%), where trigger efficiency is not 100\%. The extracted
parameters are given in Table~\ref{tab:NBDparGribov}.

The standard NBD-Glauber fits yield satisfactory results using either
the \Npart\ or the \Ncoll-scaling, which result in similar average
number of collisions \avNcoll\, evaluated for each of the centrality
intervals as shown in Table \ref{tab:NcollGribov}.  The Glauber-Gribov
fits with $\Omega=0.55$ provide an equally good description of the
measured V0A distribution as the standard Glauber, indicating that the
fits cannot discriminate between the models. The broader
\Npart\ distributions in the Glauber-Gribov models require smaller
intrinsic fluctuations in the NBD at fixed \Npart. No satisfactory fit
is obtained with $\Omega=1.01$. As expected, the corresponding values
of \avNcoll\, also shown in Table \ref{tab:NcollGribov}, are larger
(smaller) for central (peripheral) than those obtained from the
standard Glauber, as a consequence of the different shapes of the
\Npart\ distributions in these models (see Fig.~ \ref{fig:Gribov}
(left)). Both assumptions that the multiplicity distribution is
proportional to \Npart\ or \Ncoll\ are found to give an equally good
description of the experimental data (see Fig.~\ref{fig:Gribov}, and
parameters reported in Table~\ref{tab:NBDparGribov}). The
difference in the extracted geometric quantities is within 10\% for 0-60\% and sligthly increase for the most peripheral, which is of similar order as the uncertainty derived from the Glauber parameters (see
the last two columns of Table~\ref{tab:NcollGribov}).

\section{Centrality from the Zero-Degree Energy} 
\label{sec:geometry}

The energy measured in the zero degree calorimeters (\ZDC) can be used
to determine the centrality of the collision. The \ZDC\ detects the so
called ``slow'' nucleons produced in the interaction: protons in the
proton \ZDC\ (ZP) and neutrons in the neutron \ZDC\ (ZN).  The
multiplicity of slow nucleons is expected to be monotonically related
to \Ncoll~\cite{Sikler:2003ef} and can therefore be used as a
centrality estimator.

Emitted nucleons are classified as ``black'' or ``grey''. This
terminology originates from emulsion experiments where it was related
to the track grain density.  Black particles, typically defined to
have velocity $\beta \lesssim 0.25$ in the nucleus rest frame, are
produced by nuclear evaporation processes, while grey particles,
$0.25 \lesssim \beta \lesssim 0.7$, are mainly nucleons knocked out
from the nucleus.
Experimental results at lower energies show that the features of the
emitted nucleons, such as angular, momentum and multiplicity
distributions, are weakly dependent on the projectile energy in a wide
range from 1~GeV up to 1~TeV~(\cite{Sikler:2003ef} and references
therein).  These observations suggest that the emission of slow
particles is mainly dictated by nuclear geometry.

To relate the energy deposited in the \ZDC\ to the number of binary
collisions quantitatively requires a model to describe the production
of slow nucleons. Since there are no models available that are able to
describe the slow nucleon emission at LHC energies, we relied on the
weak dependence on collision energy and followed a heuristic approach.
For this purpose we developed a model for the slow nucleon emission
(SNM) based on the parameterization of experimental results at lower
energies.

\begin{figure}[t!f]
 \centering 
 \includegraphics[width=0.39\textwidth]{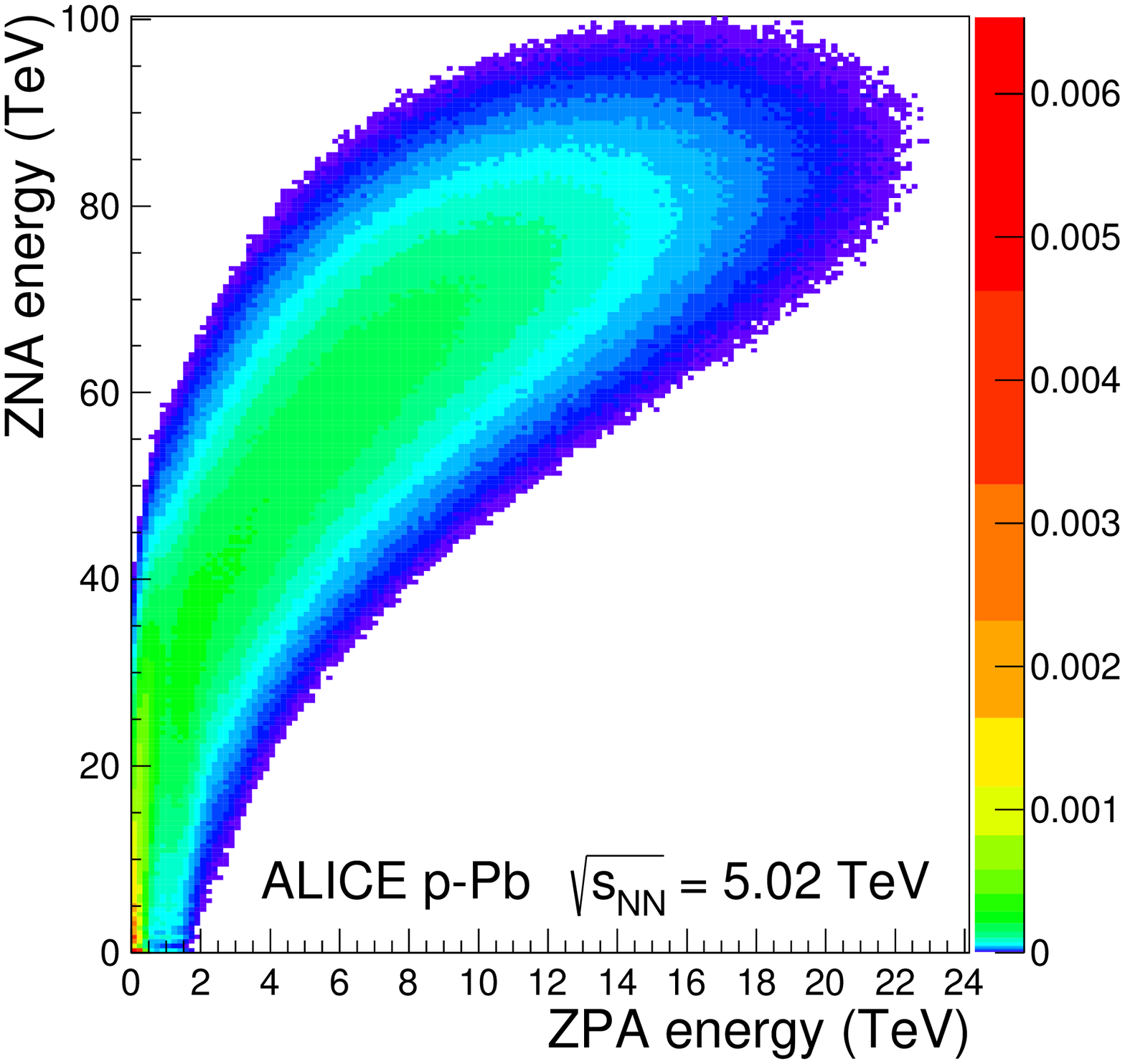} 
 \includegraphics[width=0.46\textwidth]{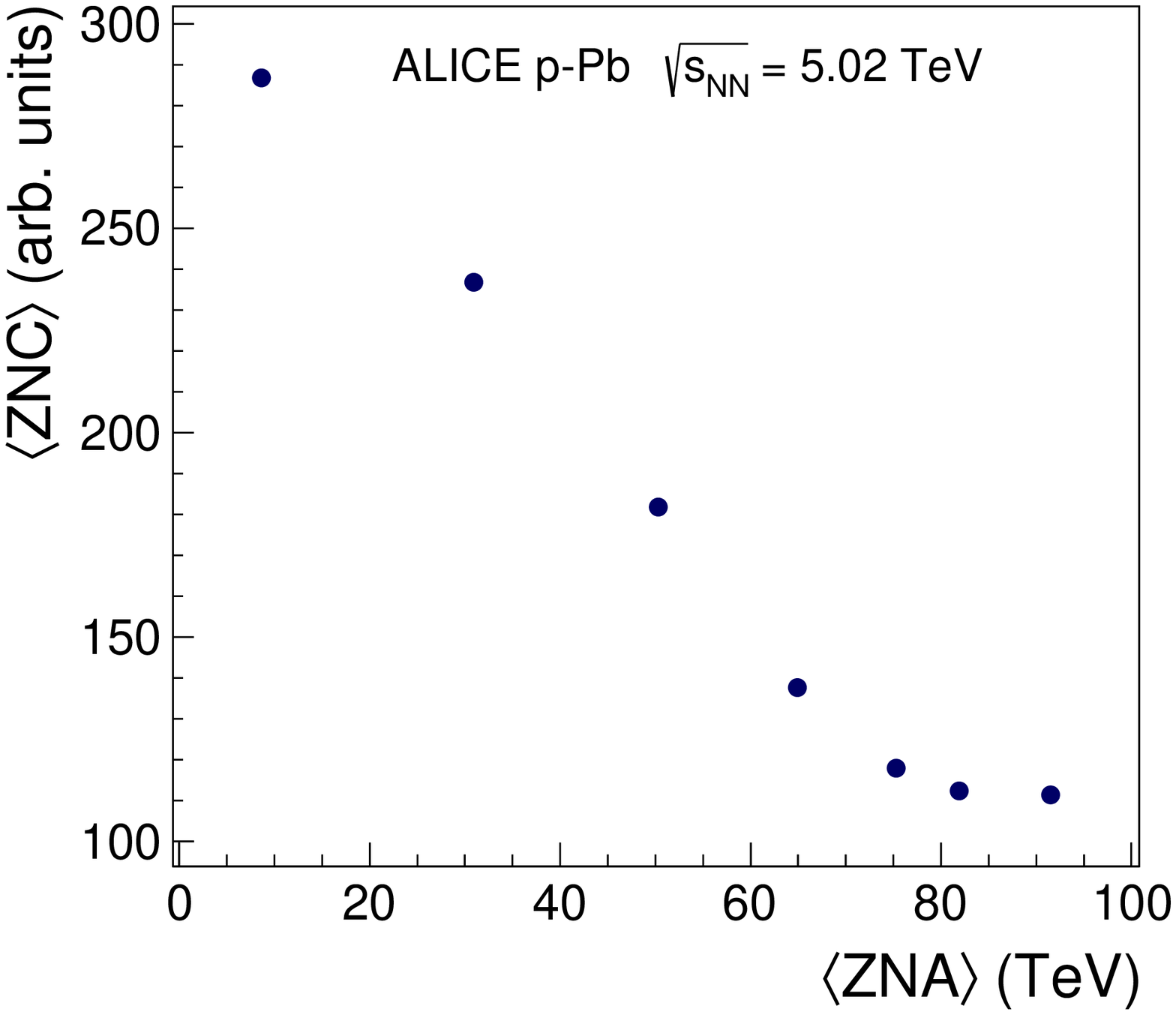} 
\caption{(color online) Left: Correlation between Pb remnant neutron (ZNA) and proton (ZPA) calorimeter energies. 
Right: Average signal on the p remnant side (ZNC) versus average
energy detected by ZNA in centrality bins selected using ZNA.}
\label{fig:zdccorr}
\end{figure}

In the left panel of Fig.~\ref{fig:zdccorr} it is shown that the
energy detected by the neutron calorimeter on the Pb-remnant side
(ZNA) is correlated with the energy detected in the proton ZDC (ZPA),
up to the onset of a saturation in the emission of neutrons. This
saturation effect is commonly attributed to the black
component~(\cite{Sikler:2003ef} and references therein).  The energy
detected by ZP is lower. This is due both to the lower number of
protons in the Pb nucleus and to the lower acceptance for emitted
protons that are affected by LHC magnetic fields. Furthermore,
contrary to ZN, ZP response and energy resolution strongly depend on
the proton impact point.  In the following we focus on the ZN spectrum
for these reasons.

The energy released in the ZNA is anti-correlated with the signal in
the neutron calorimeter placed on the p-remnant side (ZNC) (see
Fig.~\ref{fig:zdccorr}, right). The p-remnant side ZN signal cannot be
easily calibrated in energy units due to the lack of peaks in the
spectrum.  Events characterized by low \Ncoll \ values, corresponding
to low energy deposit in ZNA, have the largest contribution in ZNC.
This implies that the participant contribution cannot be neglected for
very peripheral events, where the sample is also partially
contaminated by electromagnetic processes. Therefore supposing that no
nucleons are emitted in the limit that there is no collision, the
model is not expected to provide a complete and reliable description
for very peripheral data.

In the following, we briefly summarize the main ingredients of the developed
heuristic model for slow nucleon emission.  The average
number of emitted grey protons is calculated as a function of \Ncoll \
using a second order polynomial function:
\begin{equation}
\langle N_{\rm grey \, p}\rangle = c_{0}+c_{1} {N_{\rm coll}}+c_{2}{N_{\rm coll}}^{2}.  
\end{equation}
This relationship was found to be in good agreement with grey proton
data measured by E910 in p--Au collisions with 18~GeV/c proton
beam~\cite{Chemakin:1999jd}.  The coefficient values taken from the
E910 fit are rescaled to Pb nuclei using the ratio
$(\rm{Z_{Pb}/Z_{Au}})$: $c_{0}=-0.24$, $c_{1}=0.55$,
$c_{2}=0.0007$. The linear term is the dominant contribution while the
quadratic term is negligible.  Neglecting in this context a possible
saturation effect for black protons, we approximate the average number
of black protons using the ratio between ``evaporated'' and ``direct''
proton production measured by the COSY experiment in p--Au
interactions at~2.5~GeV~\cite{Letourneau:2002}: ${\langle N_{\rm
black\, p}\rangle = 0.65 \langle N_{\rm grey \, p}\rangle}$.

The average number of slow neutrons is obtained using the following formula:
\begin{equation}\label{eq:nslown}
\langle N_{\rm slow \ n}\rangle = \alpha N_{LCF}+ \left( a-\frac{b}{c+N_{LCF}}\right) 
\end{equation}
where $N_{\rm LCF}$ is the number of Light Charged Fragments, namely
the number of fragments with Z$<8$.  Since we cannot directly measure
the number of light charged fragments in ALICE, we assumed that
$N_{\rm LCF}$ is proportional to the number of slow protons as
measured by COSY~\cite{Letourneau:2002}:
$N_{\rm LCF} = \gamma \cdot \langle N_{\rm slow \, p} \rangle$
where the proportionality factor $\gamma= 1.71$ is obtained through a minimization procedure.
The first term in eq.(\ref{eq:nslown}) describes the grey neutron
production that linearly increases with \Ncoll \ and hence with
$N_{\rm LCF}$.  The second term reproduces the saturation in the
number of black nucleons, and is based on a parameterization of
results from the COSY experiment where the neutron yield is related to
$N_{\rm LCF}$~\cite{Letourneau:2002}.  The values of the parameters
$\alpha$, $a$, $b$ and $c$ are obtained through a minimization
procedure and are: $\alpha=0.48$, $a=50$, $b = 230$, $c = 4.2$.

The relative fraction of black and grey neutrons is evaluated assuming
that $90 \%$ of the emitted neutrons are black, as measured in proton
induced spallation reactions in the energy range between 0.1 and
10~GeV~\cite{Kowalczyk:2008cw}.  The number of nucleons emitted from
$_{82}^{208}$Pb is finally calculated event by event as a function
of \Ncoll, assuming binomial distributions with probabilities $p
=\langle N_{\rm slow \, p} \rangle /82$ for protons and $p= \langle
{N_{\rm slow \, n}\rangle/126}$ for neutrons.

The kinematical distributions of the black and the grey components are
described by independent statistical emission from a moving frame:
black nucleons are emitted from a stationary source, while grey
nucleons from a frame slowly moving along the beam direction with
$\beta_{\rm grey}=0.05$.
The angular distribution for grey tracks is forward peaked in the
polar angle $\theta$, while black nucleons are assumed to be uniformly
distributed, in agreement with the experimental
observations~\cite{Chemakin:1999jd, Dabrowska:1993cw}.

\begin{figure}[t!f]
 \centering 
 \includegraphics[width=0.68\textwidth]{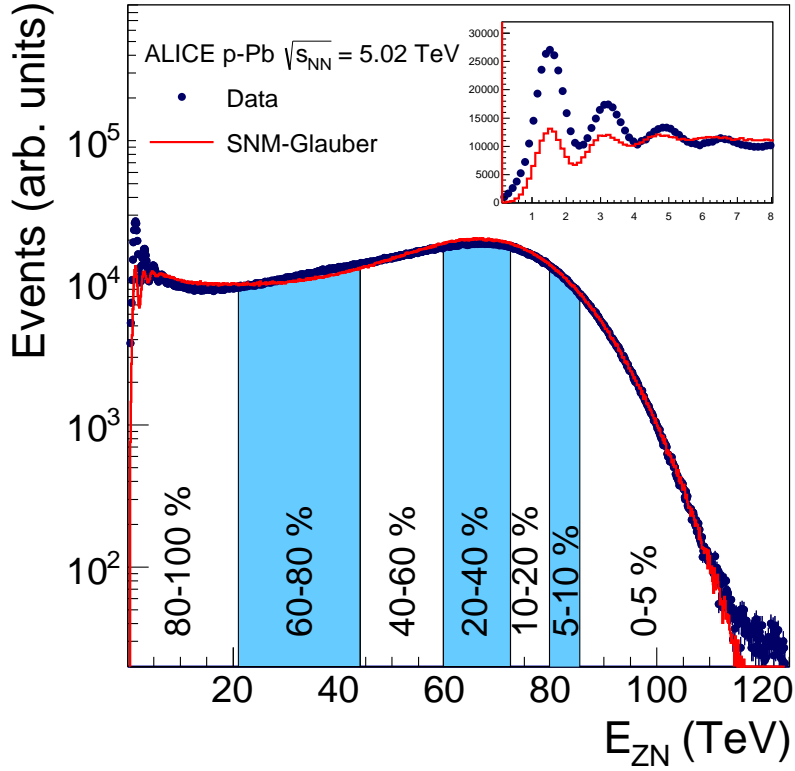} 
\caption{(color online) Distribution of the neutron energy spectrum measured in the Pb-remnant side ZN calorimeter. 
 The distribution is compared with the corresponding
 distribution from the SNM-Glauber model (explained in the text) shown
 as a line. Centrality classes are indicated in the figure. The inset
 shows a zoom-in on the most peripheral events.
\label{fig:zdcspec}}
\end{figure}

The neutron calorimeter has full geometric acceptance for neutrons 
emitted from the Pb nucleus, as estimated through Monte Carlo simulations. 
Experimentally, a fraction of triggered events (4.4\%) does not produce
a signal in ZN, these are very peripheral events with no neutron emission.
The convolution of ZN acceptance and efficiency has been calculated
coupling an event generator based on the SNM to HIJING~\cite{hijing}
and using a full GEANT~3~\cite{geant3ref2} description of the ALICE
experimental apparatus. Taking into account the experimental
conditions (beam crossing angle and detector configuration), we obtain
that 94\% of the events have a signal in the neutron calorimeter, in
good agreement with the experimental acceptance (95.6\%).  Since the
events without ZNA signal have the same CL1, V0A and V0M distributions
as the those in the 80-100\% centrality bin they are attributed to
this bin.

The SNM, coupled to the probability distribution for \Ncoll \
calculated from the Glauber MC as in Sec.~\ref{sec:GlauberFit}, is
fitted to the experimental distribution of the \ZDC\ energy in
Fig.~\ref{fig:zdcspec}.  The detector acceptance and resolution are
fixed to the experimental values. The parameters that are obtained by
fitting the data are: $\gamma$, $a$, $b$, $c$ and $\alpha$.  The main
features of the measured energy distribution in the neutron
calorimeter on the Pb-side are reasonably well described by the SNM.
The \avNcoll, reported in Table~\ref{tab:NcollCompare} and in
Fig.~\ref{fig:Ncoll}, is then calculated for centrality classes
defined by dividing the energy spectrum in percentiles of the hadronic
cross-section.  The systematic uncertainty on the \Ncoll\ values
reported in Table~\ref{tab:NcollCompare}, has been evaluated by
varying the model parameters within reasonable ranges: i) using for
the relative fraction of black over grey protons ${\langle N_{\rm
black\, p}\rangle = 0.43 \langle N_{\rm grey \, p}\rangle}$ from
spallation reaction results~\cite{Kowalczyk:2008cw}, ii) including a
saturation effect for black protons, iii) decreasing the ratio of
black over grey neutrons to $0.5$ as obtained from
DPMJET~\cite{Roesler:2000he}, iv) neglecting the linear term in
Eq.~\ref{eq:nslown} and assuming complete saturation for the neutrons,
v) varying $\gamma$ by $\pm 10\%$ and vi) assuming different
parametrization for the fluctuations in the number of slow nucleons
for a fixed \Ncoll\ value.  We note that this uncertainty
correspond to the variation of the SNM parameters, therefore it is
meant as the uncertainty \emph{within} our SNM and does not reflect
any possible other model that could describe nucleus fragmentation.
When using the \Ncoll\ values for the ZNA centrality estimator, the
total systematic uncertainty on \Ncoll\ is obtained by adding the
uncertainties from the Glauber and SNM parameters in quadrature.

\begin{figure}[t!f]
 \centering 
 \includegraphics[width=1.0\textwidth]{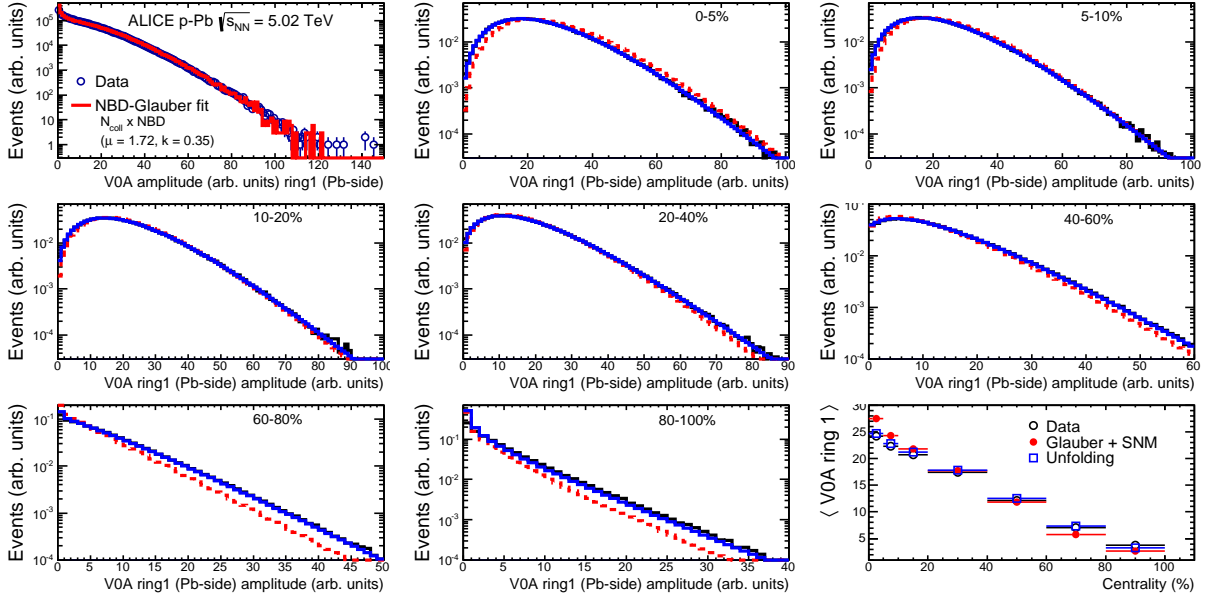} 
\caption{(color online) V0A ring1 signal distributions. 
The top left panel shows the distribution for MB events together with
a NBD-Glauber fit. The remaining panels show the distributions and
mean values for centrality classes selected with ZNA.  These are
compared to those obtained by the convolution of the ${\cal P}(\Ncoll
| cent_{\rm ZNA})$ distributions from the SNM with the NBD from the
NBD-Glauber fit to V0A ring1.  Data are also compared to the
distributions obtained with an unfolding procedure, where the \Ncoll\
distributions have been fitted to the data using the parameters from
the NBD-Glauber fit. The bottom right panel compares the mean values
of these distributions as a function of the centrality.
\label{fig:convolution}}
\end{figure}

Within the Glauber-model, the consistency between measurements
of \Ncoll\ in largely separated rapidity regions establishes their
relation to centrality.  To this end, we correlate the ZNA
measurements to the amplitudes measured in the innermost ring of
the \VZEROA\ detector (V0A ring1), since this ring covers the most
forward rapidity in the Pb-going direction.
 The \Ncoll\ distributions (${\cal P}(\Ncoll| cent_{\rm ZNA})$) for
centrality classes selected with ZNA ($cent_{\rm ZNA}$) are obtained
from the SNM-Glauber fit.  These are convolved with the NBD obtained
from the NBD-Glauber fit to the MB V0A ring1 distribution.
Fig.~\ref{fig:convolution} compares the distributions of V0A ring1
obtained from these convolutions to the ones measured in the same ZNA
centrality classes.  As expected, the distributions in the most
peripheral events, where the SNM does not provide a reliable
description of the data, are not well reproduced by the Glauber-MC
convolution. In all other classes, the experimental distributions are
well reproduced. The deviations are consistent with the ones between
$\Ncoll^{\rm ZNA}$ (see Table~\ref{tab:NcollCompare}) and $\Ncoll^{\rm
Pb-side}$ (see Table~\ref{tab:NcollCompareH}) assuming that the
target-going charged-particle multiplicity measured in V0A ring1 is
proportional to the number of wounded target nucleons.

In addition, Fig.~ \ref{fig:convolution} shows the results of an
unfolding procedure.  For each V0A ring1 distribution selected by a
ZNA centrality class, we find the \Ncoll\ distribution that, convolved
with the $\rm NBD_{\rm MB}$, fits the data, i.e. the parameters of the
fit are the relative contributions of each \Ncoll\ bin.  The unfolded
distributions (shown in blue) agree well with the data for all
centrality bins, apart from a small discrepancy in the 80-100\%
distribution at low amplitude, which is affected by trigger and event
selection efficiency, however.  The $\Ncoll^{\rm MB}$ distribution
which results from the sum of the unfolded distributions of all
centrality bins agree well with the one from Glauber-MC.  The
existence of \Ncoll\ distributions that folded with NBD agree with
measured signal distributions is a necessary condition for ZNA to
behave as an unbiased centrality selection.  In contrast, it is worth
noting that a centrality selection based on central multiplicity, as
CL1, has no such solution; i.e. no such good agreement can be found
when the V0A ring1 distributions are selected by ordering the events
according to CL1.  The biases related to centrality selection will be
discussed in the next section.  The assumption that the ZNA selection
is bias free will be used in Sec.~\ref{sec:hybrid} as an ansatz for
the hybrid method.

\section{Discussion of potential biases on centrality}
\label{sec:bias}
\subsection{Multiplicity Bias}
\label{section:multbias}
\begin{figure}[t!f]
 \centering
 \includegraphics[width=1.\textwidth]{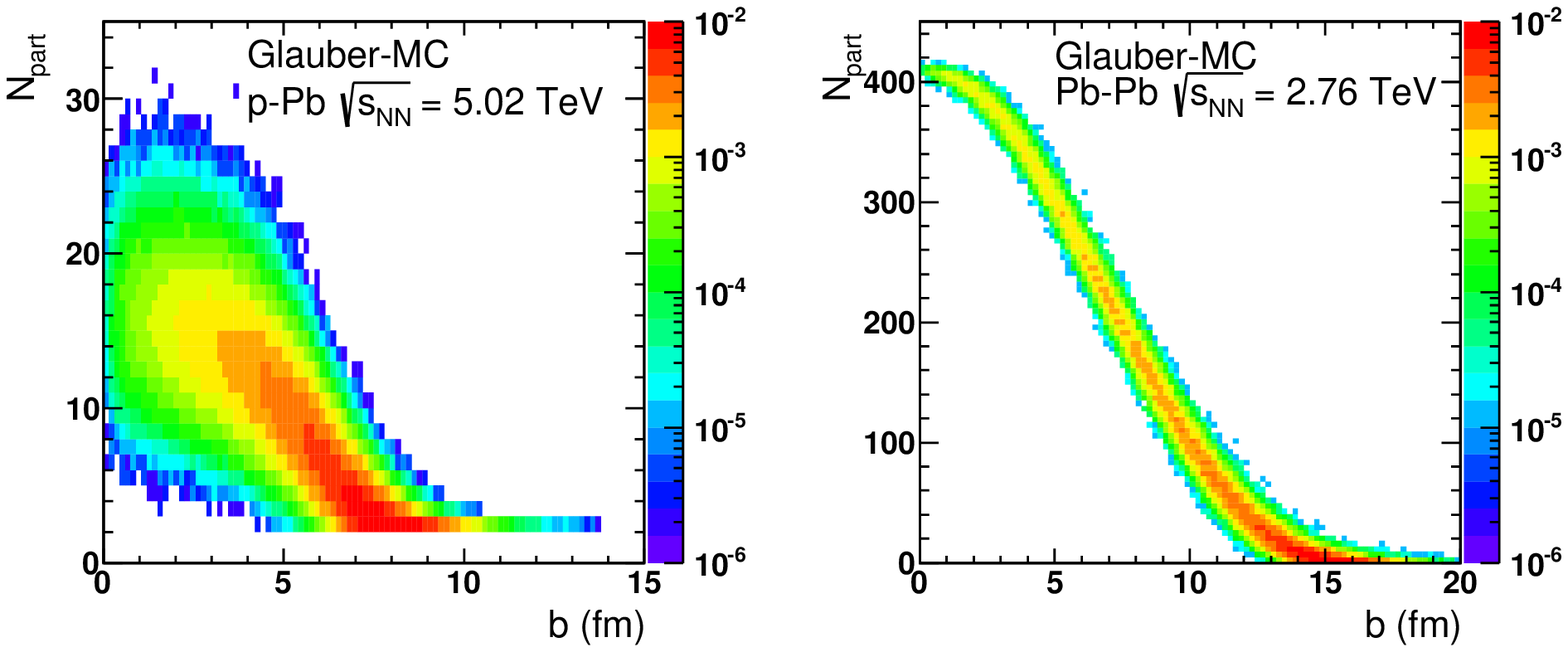}
 \includegraphics[width=1.\textwidth]{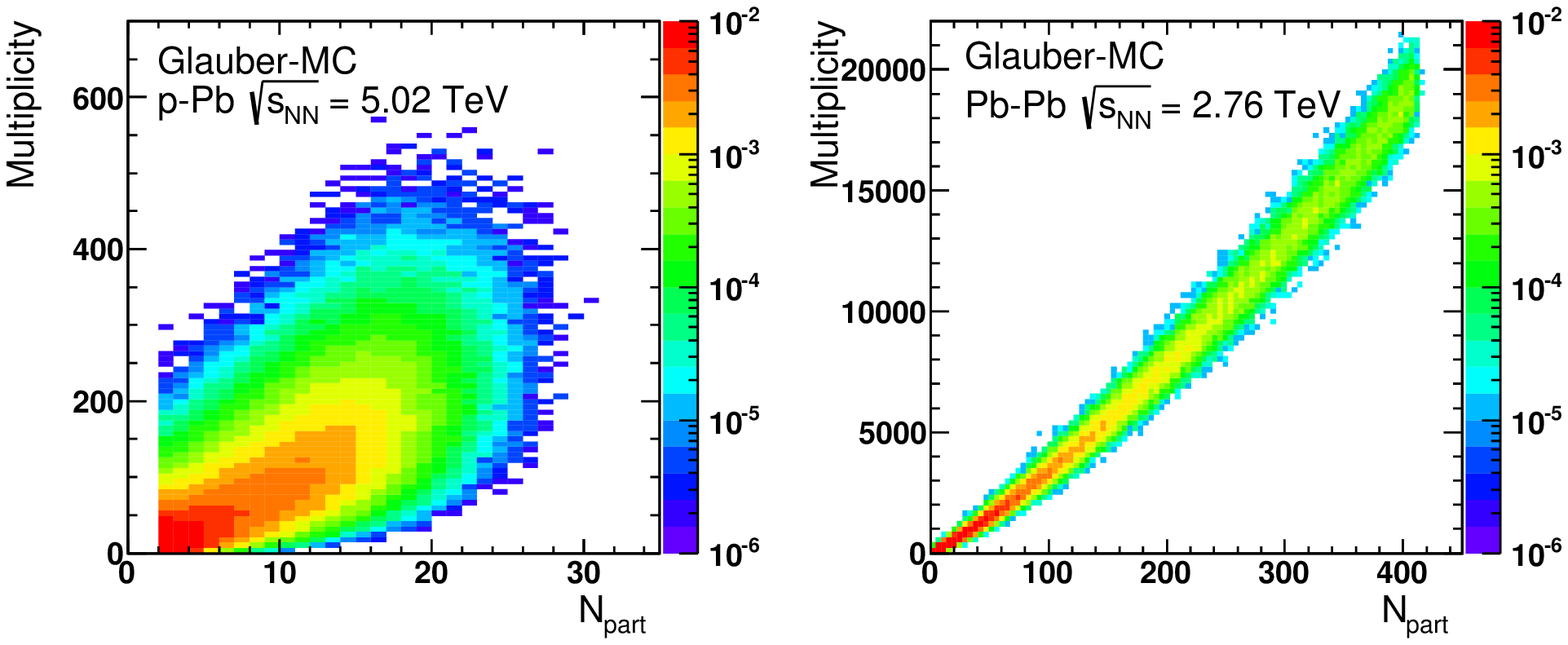}
 \caption{(color online) Top: Scatter plot of number of
   participating nucleons versus impact parameter; Bottom: Scatter plot of multiplicity versus 
   the number of  participating nucleons  from the
   Glauber fit for V0A. The quantities are calculated with a Glauber
   Monte Carlo of \pPb\ (left) and \PbPb\ (right) collisions.
  \label{fig:corrGlau}}
\end{figure}

Section \ref{subsec:GlauberFit} describes the NBD-Glauber fitting
procedure used to determine the collision geometry in terms of
\Ncoll\ and \Npart, for each centrality class.  The NBD is used to
account for multiplicity fluctuations at fixed \Npart.  In contrast to
\PbPb\ collisions, for \pPb\ collisions these multiplicity
fluctuations are sizeable compared to the width of the
\Npart\ distribution, as illustrated in Fig.~\ref{fig:corrGlau}.  For
large fluctuations, a centrality classification of the events based on
multiplicity may select a sample of nucleon-nucleon collisions which
is biased compared to a sample defined by cuts on the impact parameter
$b$.

\begin{figure}[t!f]
 \centering
 \includegraphics[width=0.49\textwidth]{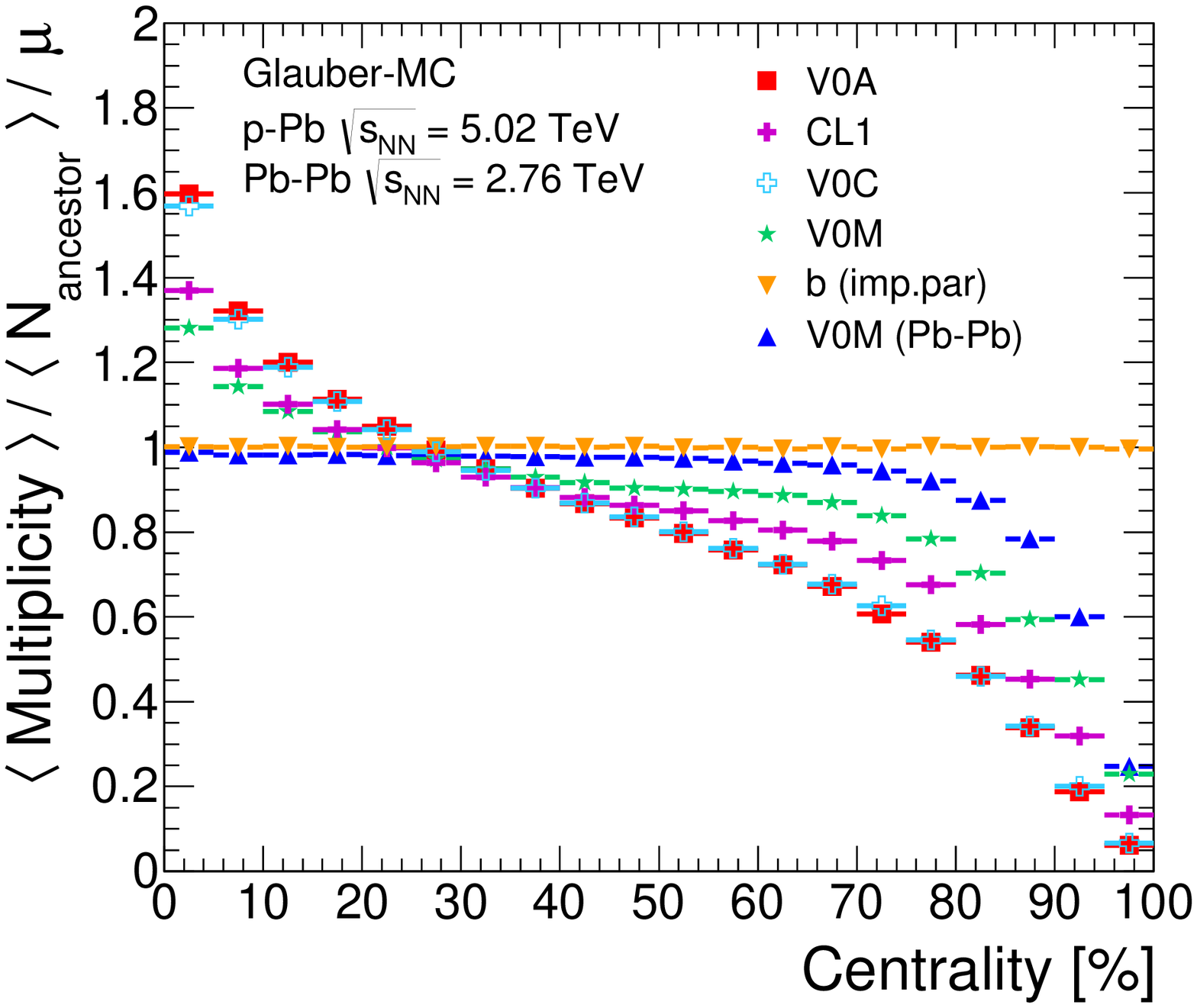}
 \includegraphics[width=0.49\textwidth]{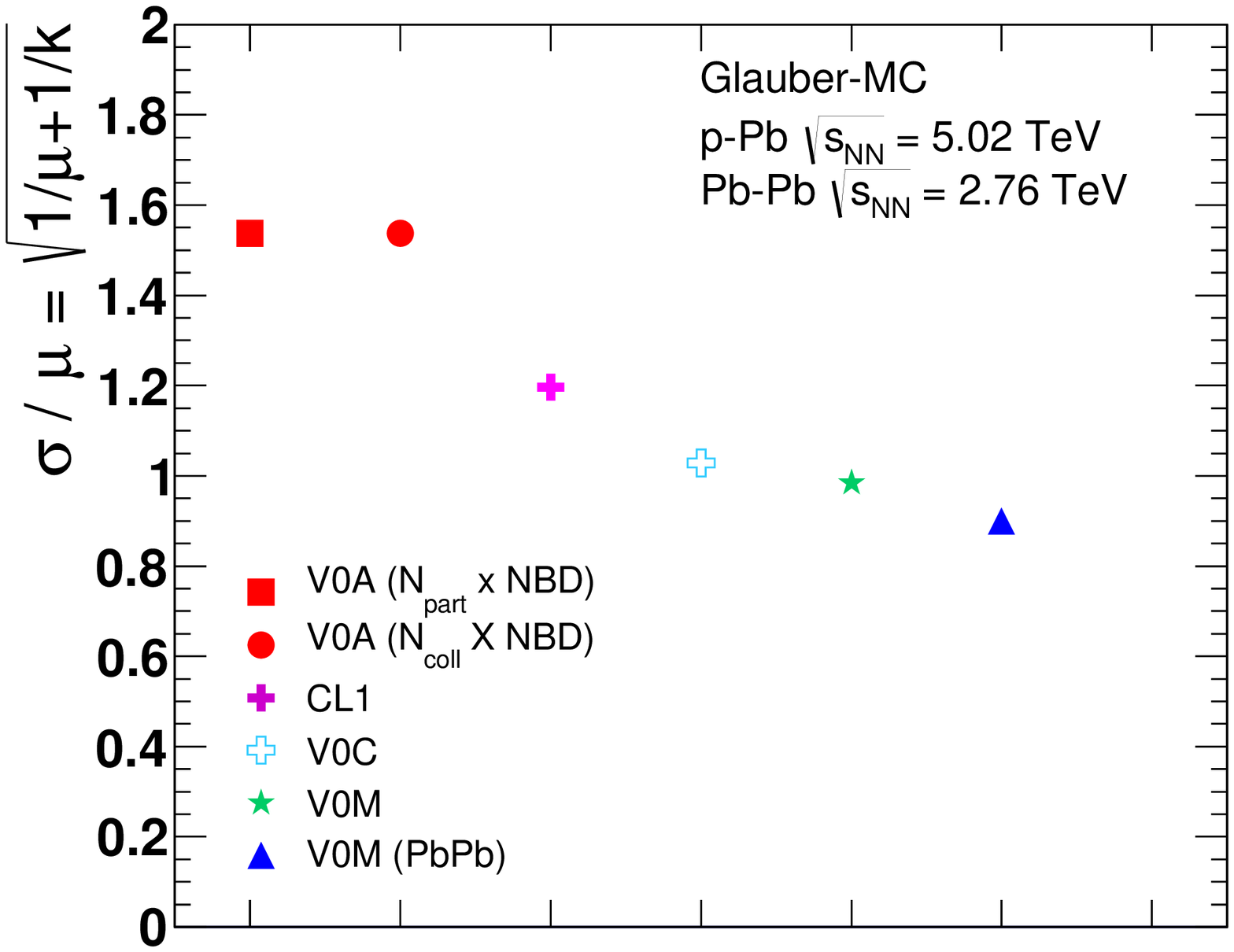}
 \caption{(color online) Left: Multiplicity fluctuation bias quantified as the mean
   multiplicity per $\avNpart / \mu$ from the NBD-Glauber MC in \pPb\
   and \PbPb\ calculations. Right: Relative width of the NBD obtained
   from the NBD-Glauber fit of various multiplicity distributions in
   \pPb\ and \PbPb\ calculations.
  \label{fig:multdist}}
\end{figure}

This selection bias, which occurs for any system with large relative statistical
fluctuations in particle multiplicity per nucleon-nucleon collision,
can be quantified using the Glauber fit itself. The left panel of
Fig.~\ref{fig:multdist} shows the ratio between the average
multiplicity per average participant and the average multiplicity of the NBD
as a function of centrality.  In \PbPb\ collisions, where the width of
the plateau of the \Npart\ distribution is large with respect to
multiplicity fluctuations, the ratio deviates from unity only for the
most peripheral collisions. As expected, in \pPb\ collisions the ratio
differs from unity for all centralities with large deviations for the
most central and most peripheral collisions; the most central
(peripheral) collisions have on average much higher (lower)
multiplicity per participant. When selecting event classes using impact parameter $b$ intervals, there is no deviation from unity, as expected.  The right panel
of Fig.~\ref{fig:multdist} shows for each centrality estimator the
relative width of the NBD distribution ($\sigma / \mu)$.  As expected,
the estimators with the largest bias on the multiplicity per
participant correspond to those with the largest relative width.

It is instructive for the further discussion  to consider the clan
model \cite{clan}, which is the standard physical explanation of the
NBD distribution in the context of particle production in pp
collisions.  In this model particle sources, called ancestors, are
produced independently according to a Poisson distribution with mean
value, $\langle N \rangle = k \cdot \ln{(1+\mu/k)}$.  Each ancestor
can produce on average $\mu/\langle N \rangle$ particles, e.g. by
decay and fragmentation, and a clan contains all particles that stem
from the same ancestor.  Hence, the bias observed above also
corresponds to a biased number of clans, which are sources of particle
production.  Analogously, in all recent Monte Carlo generators a large
part of the multiplicity fluctuations is indeed due to the
fluctuations of the number of particle sources, i.e. multiple
semi-hard ($Q^2 \gg \Lambda_{\rm QCD}$) parton-parton scatterings
(MPI).  

As an example, the HIJING generator accounts for fluctuations
of the number of MPI per NN interaction via an NN overlap function
$T_{\rm NN} (b_{\rm NN})$, where $b_{\rm NN}$ is the NN impact
parameter, i.e. the impact parameter between the proton and each
wounded nucleon of the Pb-nucleus.  The probability for inelastic NN
collisions is given as one minus the probability to have no
interaction:
\begin{equation}
{\rm d} \sigma_{inel} = \pi {\rm d}b_{\rm NN}^2 [1 - e^{-(\sigma_{\rm soft} +
    \sigma_{\rm hard}) T_{\rm NN} (b_{\rm NN}))}] \, ,
\label{eq:hijing1}
\end{equation}
where $\sigma_{\rm soft}$ is the geometrical soft cross-section of
57 mb \cite{hijing} related to the proton size and $\sigma_{\rm hard}$ the energy
dependent pQCD cross-section for $2 \rightarrow 2$ parton scatterings.
%
%
Further, as in the clan model, there is a Poissonian probability 
\begin{equation}
P(n_{\rm hard}) = \frac{{\langle n_{\rm hard} \rangle } ^
  {n_{\rm hard}}} {n_{\rm hard}!} e^{- \langle n_{\rm hard} \rangle }
\end{equation}
for
multiple hard collisions with an average number determined by $b_{\rm
  NN}$:
\begin{equation}
 \langle n_{\rm hard} \rangle = \sigma_{\rm hard} T_{\rm NN} (b_{\rm
  NN}) \, .
\end{equation}

Hence, the biases on the multiplicity discussed above correspond to a
bias on the number of hard scatterings ($n_{\rm hard}$) and $\langle
b_{\rm NN} \rangle $ in the event.  The latter correlates fluctuations
over large rapidity ranges (long range correlations).  As a
consequence, for peripheral (central) collisions we expect a lower
(higher) than average number of hard scatterings per binary collision,
corresponding to a nuclear modification factor less than one (greater than one).
\begin{figure}[ht]
\begin{minipage}[t]{0.49\linewidth}
\centering
 \includegraphics[trim = 0cm 0cm 0cm 0cm, clip  = true, width=\textwidth]{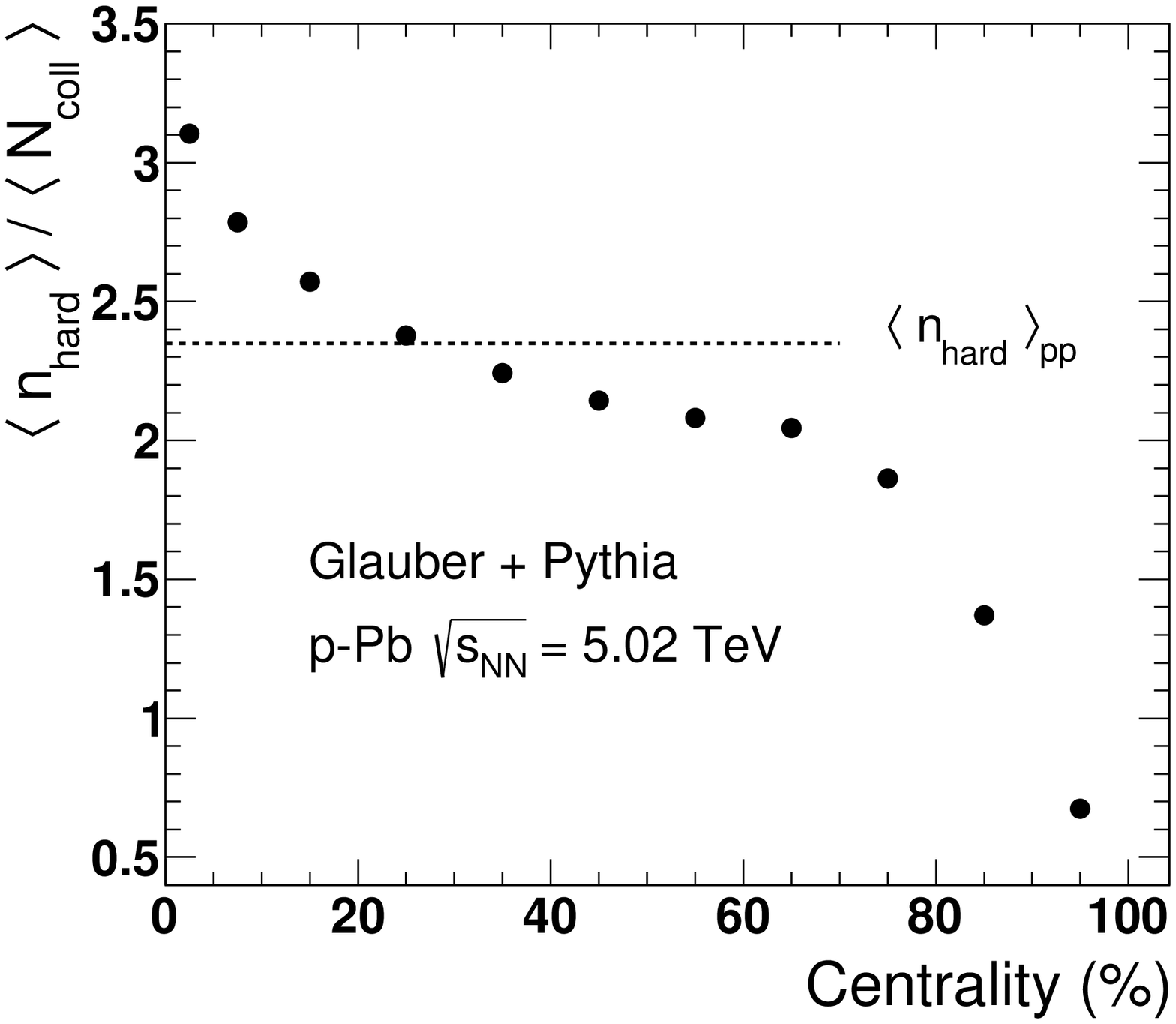}
\caption{ Number of hard scatterings (MSTI(31) in PYTHIA6) per \Ncoll\  as a function
    of the centrality calculated with a toy MC that couples a pp
    PYTHIA6 calculation to a \pPb\ Glauber MC (described in the text).}
\label{fig:Nhard}
\end{minipage}
\hspace{0.1cm}
\begin{minipage}[t]{0.49\linewidth}
\centering
\includegraphics[width=\textwidth]{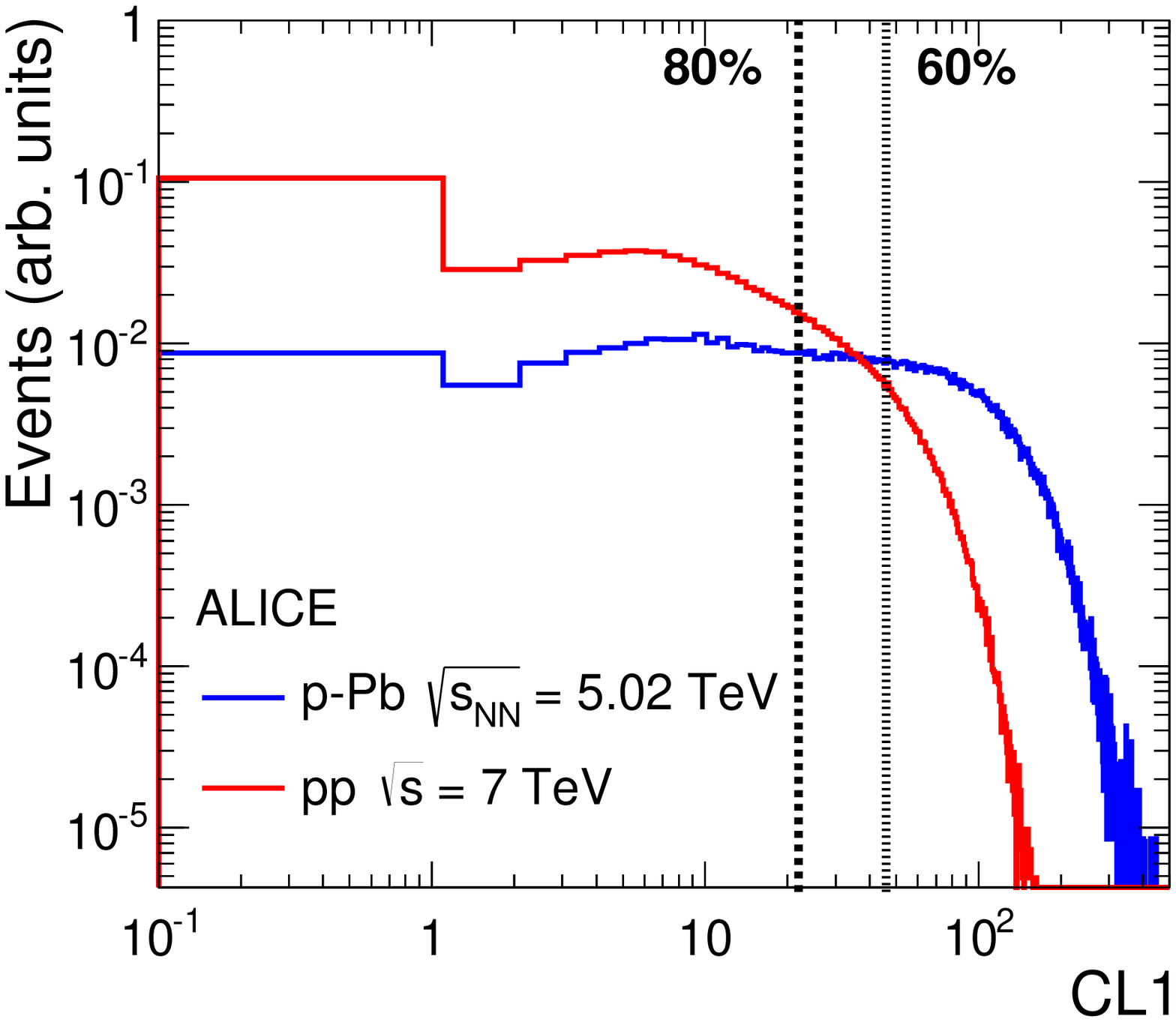}
\caption{(color online) Multiplicity distribution used as centrality estimators in
    \pPb\ collisions, compared to the distribution in \pp\ collisions
    at \sonly~=~7~TeV. The dashed lines mark the 80\%
    and the 60\% percentile of the \pPb\ cross-section respectively.}
\label{fig:jetveto}
\end{minipage}
\end{figure}

In general, the number of binary pN collisions, \avNcoll, is used
to scale the reference \pp\ yields and obtain the nuclear modification
factor, used to quantify nuclear matter effects. However, for
centrality classes based on multiplicity, owing to the bias induced by
such selection, hard processes do not simply scale with \Ncoll\ but
rather with an effective number of collisions, obtained by scaling the
${\ensuremath{\langle N_\mathrm{coll}^\mathrm{Glauber} \rangle}}$ by
the number of hard scatterings per pN collision: ${\ensuremath{\langle
    N_\mathrm{coll}^\mathrm{Glauber} \rangle}} {\ensuremath{\langle
    n_\mathrm{hard} \rangle}}_{pN} / {\ensuremath{\langle
    n_\mathrm{hard} \rangle}}_{pp}$.  As discussed in the HIJING
example above, the number of hard scatterings per pN collision is
simulated in Monte Carlo models. In this specific MC, even without
bias, the total number of hard scatterings deviates from simple
\Ncoll-scaling due to energy conservation at high \Ncoll .  Instead,
with the objective to study a baseline corresponding to an incoherent
and unconstrained superposition of nucleon-nucleon collisions, the
PYTHIA ~\cite{Sjostrand:2006za} event generator has been coupled to
the \pPb\ MC Glauber calculation. For each MC Glauber event PYTHIA is
used to generate \Ncoll\ independent \pp\ collisions.  In
the following we refer to this model as G-PYTHIA.  In this model, the
number of hard scatterings per pN collision shows a strong deviation
from \Ncoll\ scaling which is illustrated in Fig.~\ref{fig:Nhard} and
resembles the bias observed in Fig.~\ref{fig:multdist}.

\begin{figure}[t!f]
 \centering
 \includegraphics[trim=0cm  -0.6cm 0cm 0cm, clip=true,width=0.75\textwidth]{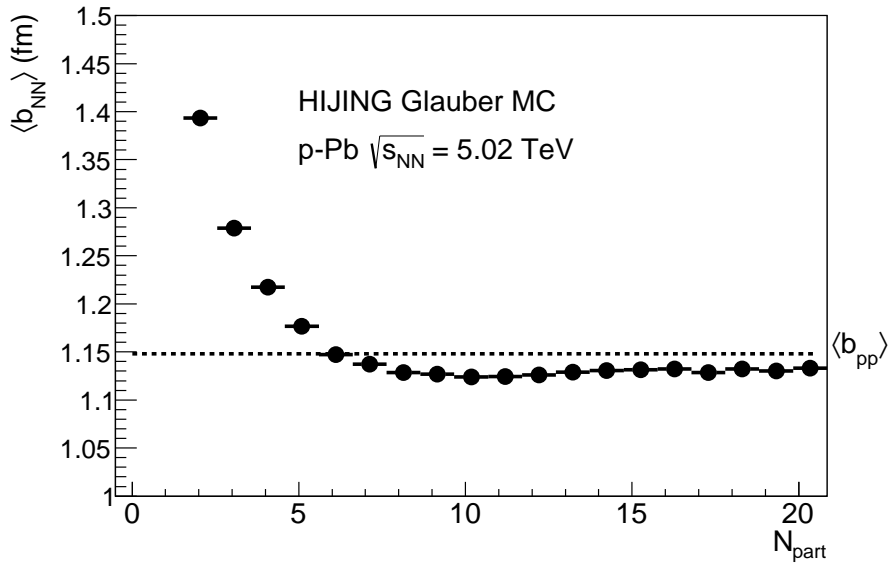} 
 \caption{Average nucleon-nucleon impact parameter as a function of
   the number of participants for \pPb\ at \snn = 5.02~TeV from a
   Glauber MC calculation as implemented in HIJING (no shadowing, no
   elastic scattering).  The result depends on the modelling of the
   spatial parton density in the nucleon.  In HIJING it is
   approximated by the Fourier transform of a dipole form factor.}
  \label{fig:bNN}
\end{figure}

\subsection{Jet-veto bias}
Additional kinematic biases exist for events containing high-\pt\ particles.
These particles arise from the fragmentation of partons produced in
parton-parton scattering with large momentum transfer.  Their contribution to the overall multiplicity rises with parton
energy  and, thus, can introduce a trivial correlation between the
centrality estimator and the presence of a high-\pt\ particle in the
event.  In particular, for very peripheral collisions, the
multiplicity range that governs the centrality for the bulk of soft
collisions can represent an effective veto on hard processes, leading
to a $\rpa < 1$.  
This bias is illustrated in 
Fig.~\ref{fig:jetveto}. It shows a multiplicity distribution
which is used as centrality estimators in \pPb\ collisions, compared
to the same distribution in \pp\ collisions at \sonly~=~7~TeV. The
dashed lines mark the 80\% and the 60\% percentile of the
\pPb\ cross-section.  The fraction of the \pp\ cross-section selected with
the 80-100\% (60-100\%) \pPb\ multiplicity cut is 0.8 (0.97).  The
80\% cut in \pPb\ is smaller than the multiplicity range covered in
pp, therefore resulting in an effective veto on the large multiplicity
events produced by hard processes.

\subsection{Geometric bias} \label{subsec:geobias}
The $b_{\rm NN}$ dependence of particle production postulated in 
section \ref{section:multbias} leads to a purely geometrical, centrality estimator
independent bias for peripheral \pPb\ collisions \cite{Jia:2009mq}.  
As illustrated in
Fig.~\ref{fig:bNN}, the mean impact parameter between two nucleons
($b_{\rm NN}$), calculated from a Monte Carlo Glauber simulation, is
almost constant for central collisions, but rises significantly for
\Npart $<6$. This reduces the average number of MPIs for most
peripheral events, enhancing the effect of the bias leading to a
nuclear modification factor less than (greater than) one for
peripheral (central) collisions.

In summary, based on simplified models we have identified three
different possible biases that are expected to lead to deviations from
unity at high \pt of the nuclear modification factors in peripheral
and central collisions.  As will be discussed and studied in the
following sections, the effect decreases with increasing rapidity
separation between the $\rpa$ measurement and the centrality
estimator.

For the estimators we used, the main biases are:
\begin{enumerate}
\item CL1: strong bias due to the full
  overlap with tracking region. Additional bias 
  from the “jet veto effect”, as jets contribute to the multiplicity and
  shift events to higher centralities (\pt-dependent) ;
\item V0M: reduced bias since the \VZERO\ hodoscopes are
  outside the tracking region;
\item V0A: reduced bias because of the enhanced
  contribution from the Pb fragmentation region;
\item ZNA: no  bias expected.
\end{enumerate}
In addition, independent of the centrality estimator, there is a
geometrical bias for peripheral collisions (see
Sec.\ref{subsec:geobias}).

\section{The Hybrid Method} 
\label{sec:hybrid}
\subsection{Basis and assumptions of the method} 
The hybrid method presented in the following section aims to provide
an unbiased centrality estimator and relies on two main
assumptions. The first is to assume that an event selection based on
ZN does not introduce any bias on the bulk at mid-rapidity and on
high-\pt\ particle production.  This assumptions is based on the
results from the unfolding procedure presented in
Sec.\ref{sec:geometry} and the full acceptance of ZN for neutrons
emitted from the Pb nucleus, also discussd in
Sec.\ref{sec:geometry}. In addition consistent results where obtained
with proton calorimeter ZP, in the region of its full acceptance.
Therefore we do not expect a significant bias from the ZN selection
and herein this is taken as ansatz.  This selection was also used in
the method proposed in Sec.\ref{sec:geometry}, however the
\Ncoll\ determination provded by the SNM-Glauber model is
model-dependent. In contrast, in the hybrid method, the
\Ncoll\ determination is based -as an ansatz- on a particular scaling
for particle multiplicity (the second assumption), e.g. we assume that
the charged-particle multiplicity measured at mid-rapidity scales with
the number of participants.

To obtain more insight into the particle production mechanisms, we
study the correlation of various pairs of observables that, in
ZN-centrality classes, are expected to scale linearly with \Npart \ or
\Ncoll.  One of these observables is the charged-particle density
\dNdeta\ in $|\eta|<2.0$, measured with the SPD. The charged particle
pseudorapidity density is obtained from the measured distribution of
tracklets, formed using the position of the primary vertex and two
hits, one on each SPD layer~\cite{alice_pA_dndeta}. At larger
pseudorapidities, where a direct multiplicity measurement is not
available, we study the raw signals of the four rings of \VZEROA\ and
\VZEROC\ detectors separately.  We exploit both beam configurations,
\pPb\ and \Pbp\, in order to cover the widest possible rapidity range.
To take into account the impact of secondary particles, the
pseudorapidity coverage of the \VZERO\ detector rings with respect to
the primary charged particles was calculated with a full detector
simulation based on DPMJET~\cite{Roesler:2000he,geant3ref2} and it is
given in Table~\ref{tab:rapidity} in the centre-of-mass system
(cms), which moves with a rapidity of $\Delta y_{\rm NN}=0.465$ in the
direction of the proton beam (see Sec.~\ref{sec:multiplicity}).

\begin{table}[thb!f]
 \centering \footnotesize  
\begin{tabular}{c|c|c}
 Ring &  $\langle\eta_{\rm cms}\rangle$ (\pPb) & $\langle\eta_{\rm cms}\rangle$ (\Pbp) \\ 
 \hline 
 VZERO-A ring 1 &  -5.39 &  4.45 \\
 VZERO-A ring 2 &  -4.80 &  3.87 \\
 VZERO-A ring 3 &  -4.28 &  3.35 \\
 VZERO-A ring 4 &  -3.82 &  2.89 \\
 VZERO-C ring 1 &   3.34 & -4.26 \\
 VZERO-C ring 2 &   2.82 & -3.74 \\
 VZERO-C ring 3 &   2.33 & -3.25 \\
 VZERO-C ring 4 &   1.86 & -2.78 \\
\end{tabular}
 \caption{Average pseudorapidity covered by \VZERO\ detector rings in
   \pPb\ and \Pbp\ collisions.
\label{tab:rapidity}} 
\end{table}

The information about charged particle multiplicity, dominated by soft
particles, is complemented with observables from hard
processes which are expected to scale with the number of binary
collisions, such as the yield of high-\pt\ (10~<\pt\ <~20~GeV/$c$) particles
measured at mid-rapidity ($|\eta|<0.3$).

In order to compare these observables on the same scale and also, at
first order, to neglect detector efficiency and acceptance effects, we
use so called normalized signals $\langle S \rangle_{\rm i}/\langle S
\rangle _{MB}$. These are obtained dividing $\langle S \rangle _{\rm i}$,
i.e. the mean value of \dNdeta, number of raw SPD tracklets or raw
\VZERO\ signal in a given ZN-centrality class $i$, by the corresponding
mean values in minimum bias collisions.

\begin{figure}[t!f]
 \centering
\includegraphics[width=0.45\textwidth]{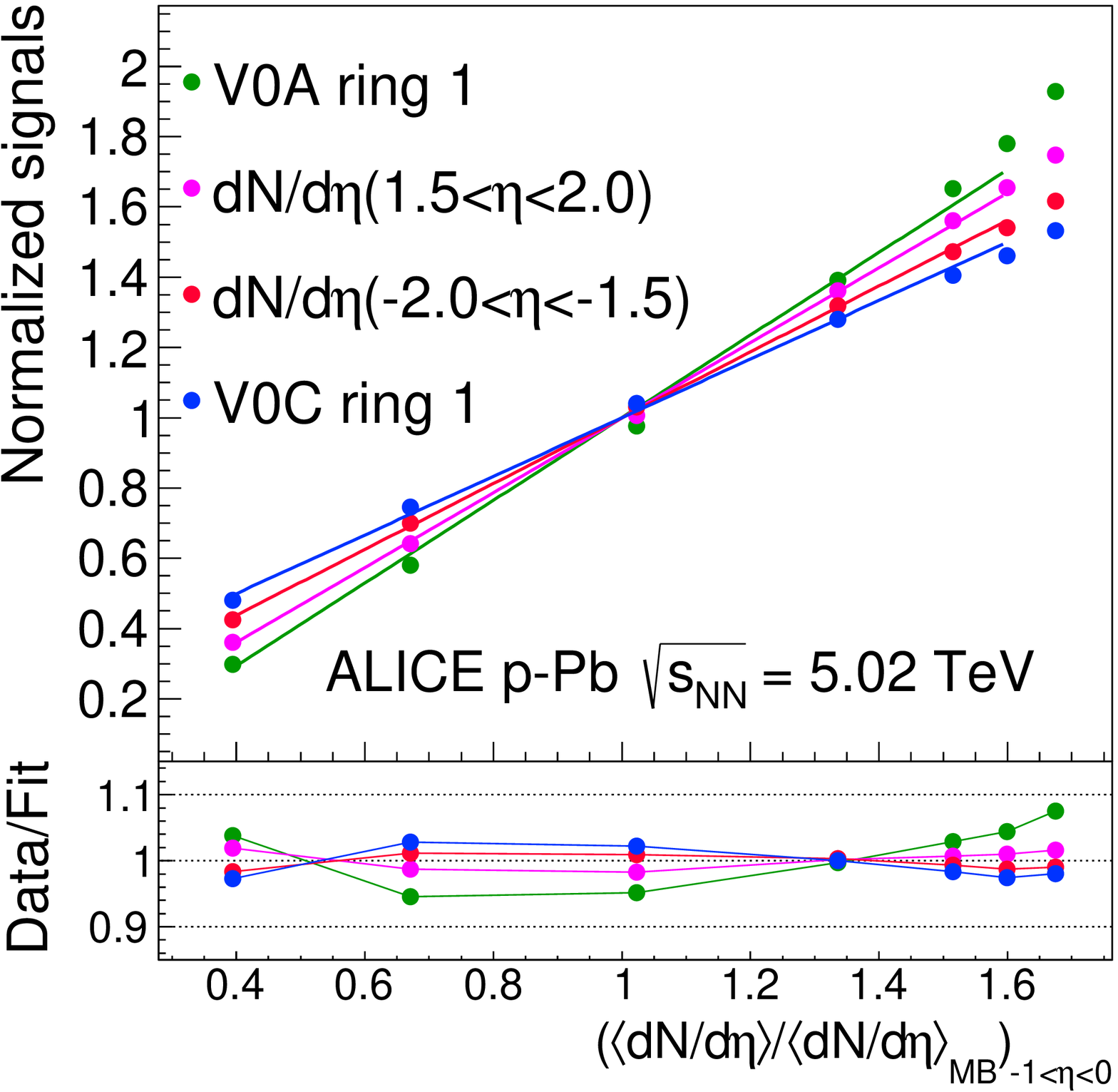} 
\hspace{0.5cm}
\includegraphics[width=0.45\textwidth]{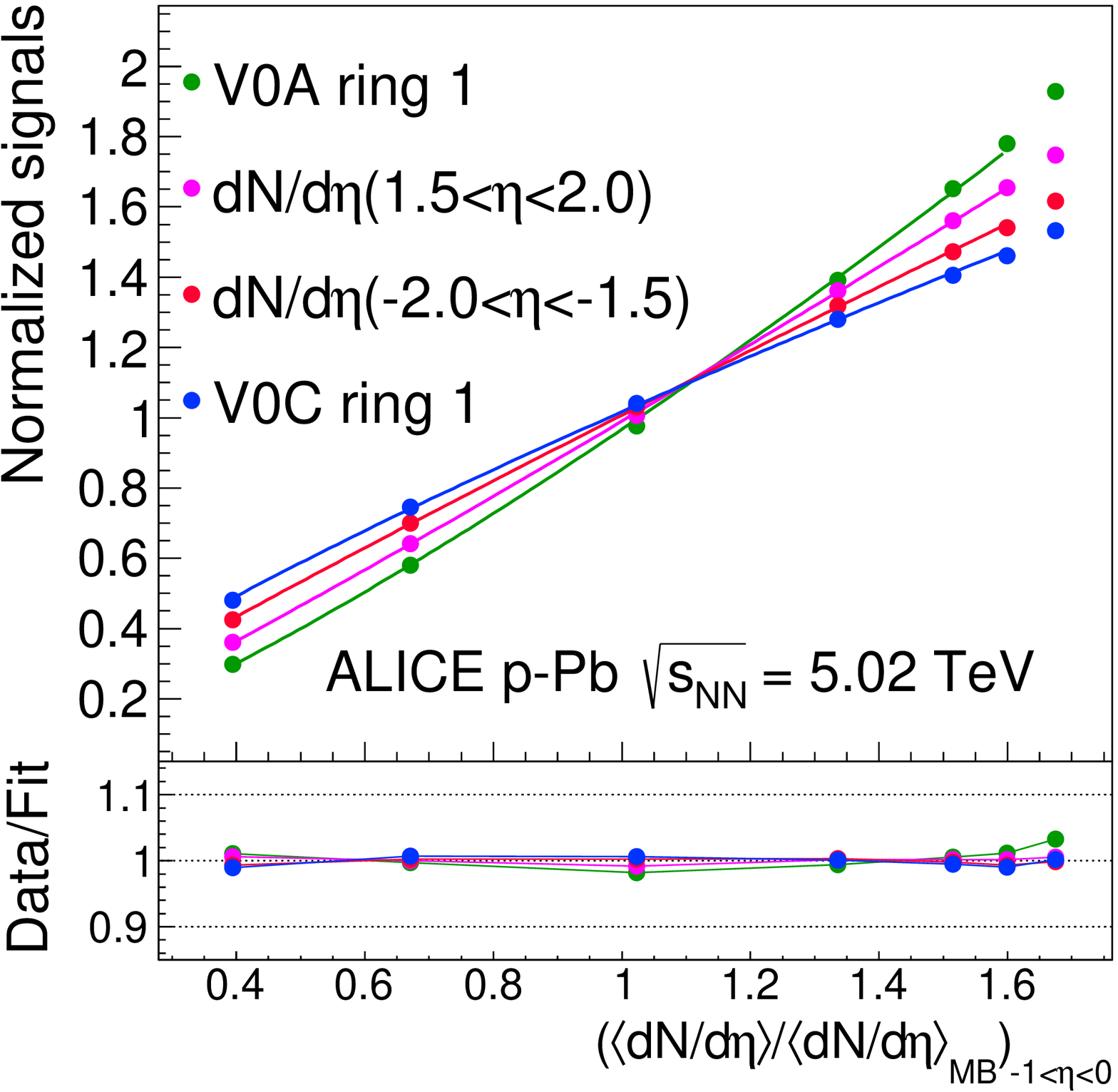} 
\caption{(color online) Left panel: Normalized signal from various observables (the innermost
  ring of \VZEROA\ and \VZEROC\ and two rapidity intervals of \dNdeta)
  versus the normalized charged-particle density averaged over $-1 <
  \eta < 0$. Left panel: fit with the linear function from
  Eq.~\ref{eq:Npartfit}. Right panel: fit with the power-law function
  from Eq.~\ref{eq:Plawfit}. Only data from \pPb\ collisions are shown.
\label{fig:scaling}}
\end{figure}

Figure~\ref{fig:scaling} shows, for bins in ZN centrality, the
correlation between a few selected normalized signals and the
normalized charged-particle density averaged over $-1 < \eta < 0$. The
statistical uncertainty is negligible, while the systematic
uncertainties largely cancel in the ratio to the MB signals.
One can note that the correlation exhibits a clear dependence on the
pseudorapidity of the normalized signal. The slope of the normalized
signals with \dNdeta\ diminishes towards the proton direction (C-side
in \pPb\ collisions). For example, in the innermost ring of the
\VZEROC\ detector the signal amplitude range is about a factor three,
while for the innermost ring of the \VZEROA\ detectors it is about
twice as large.

In the Wounded Nucleon Model~\cite{Bialas:2007eg}, the total
number of participants \Npart\ is expressed in terms of target and
projectile participants.  The charged particle density at mid-rapidity
is thus proportional to \Npart, whereas at higher rapidities the model
predicts a dependence on a linear combination of the number of target
and projectile participants with coefficients which depend on the
rapidity. Close to Pb-rapidity a linear wounded target nucleon scaling
(\Nparttar\ = \Npart\ - 1) is expected.

In order to further understand the relative trends of
the observables in Fig.~\ref{fig:scaling} and to relate them with
geometrical quantities, such as \Npart, one can adopt the Wounded Nucleon Model 
and make the assumption that \dNdeta\ in $-1 < \eta < 0$ is
proportional to \Npart.  In this case, the other observables can be
related to \Npart, assuming linear or power-law dependence. The linear
dependence can be parameterized with $\Npart - \alpha$, where $\alpha$
is a free parameter.  Then the normalized signals can be expressed
with $(\Npart - \alpha)/ \langle \Npart - \alpha \rangle $ and one
obtains the following linear relation:
\begin{equation} \label{eq:Npartfit}
\frac{\langle S \rangle _{\rm i}}{\langle S \rangle _{MB}}  = 
\frac{\langle \Npart \rangle _{\rm MB}}{(\langle \Npart \rangle _{\rm MB} - \alpha )} \cdot \left(\frac{\langle {\rm d}N/{\rm d}\eta \rangle _{\rm i}}{\langle {\rm d}N/{\rm d}\eta \rangle _{\rm MB}}\right)_{-1<\eta<0} - \frac{\alpha}{(\langle \Npart \rangle _{\rm MB} - \alpha )}
\end{equation}
where $\avNpart_{\rm MB}$ =~7.9 is the average number of participating nucleons in
minimum bias collisions. The relation is used to find $\alpha$ for
each observable by a fit to the data.  Analogously, we can also fit a
power-law function as:
\begin{eqnarray} \label{eq:Plawfit}
\frac{\langle S \rangle _{\rm i}}{\langle S \rangle _{\rm MB}} &=& 
\left( \frac{\langle {\rm d}N/{\rm d}\eta \rangle _{\rm i}^\beta}{\sum_{i} w_i \langle {\rm d}N/{\rm d}\eta \rangle _{\rm i}^\beta} \right)_{-1<\eta<0} \nonumber \\
&=&  \left( \frac{\langle {\rm d}N/{\rm d}\eta \rangle _{\rm i}^\beta}{\langle \langle {\rm d}N/{\rm d}\eta \rangle ^\beta \rangle _{\rm MB}}\right)_{-1<\eta<0}
\end{eqnarray}
where the $w_i$ are the width of the centrality classes and $\beta$ is a
fit parameter.  Since we made the assumption that \dNdeta\ in $-1 <
\eta < 0$ is proportional to \Npart, $\beta$ obtained from
Eq.~\ref{eq:Plawfit} equivalently quantifies the deviations from a
perfect \Npart\ ($\beta=1$) scaling.  As can be seen from the lower panels
of Fig.~\ref{fig:scaling} the power-law fit describes the data better,
especially for the observables located further away from
mid-rapidity. This also means that the linear dependence assumed in
Eq.\ref{eq:Npartfit} can only be valid approximately.

\begin{figure}[t!f]
 \centering
 \includegraphics[width=0.45\textwidth]{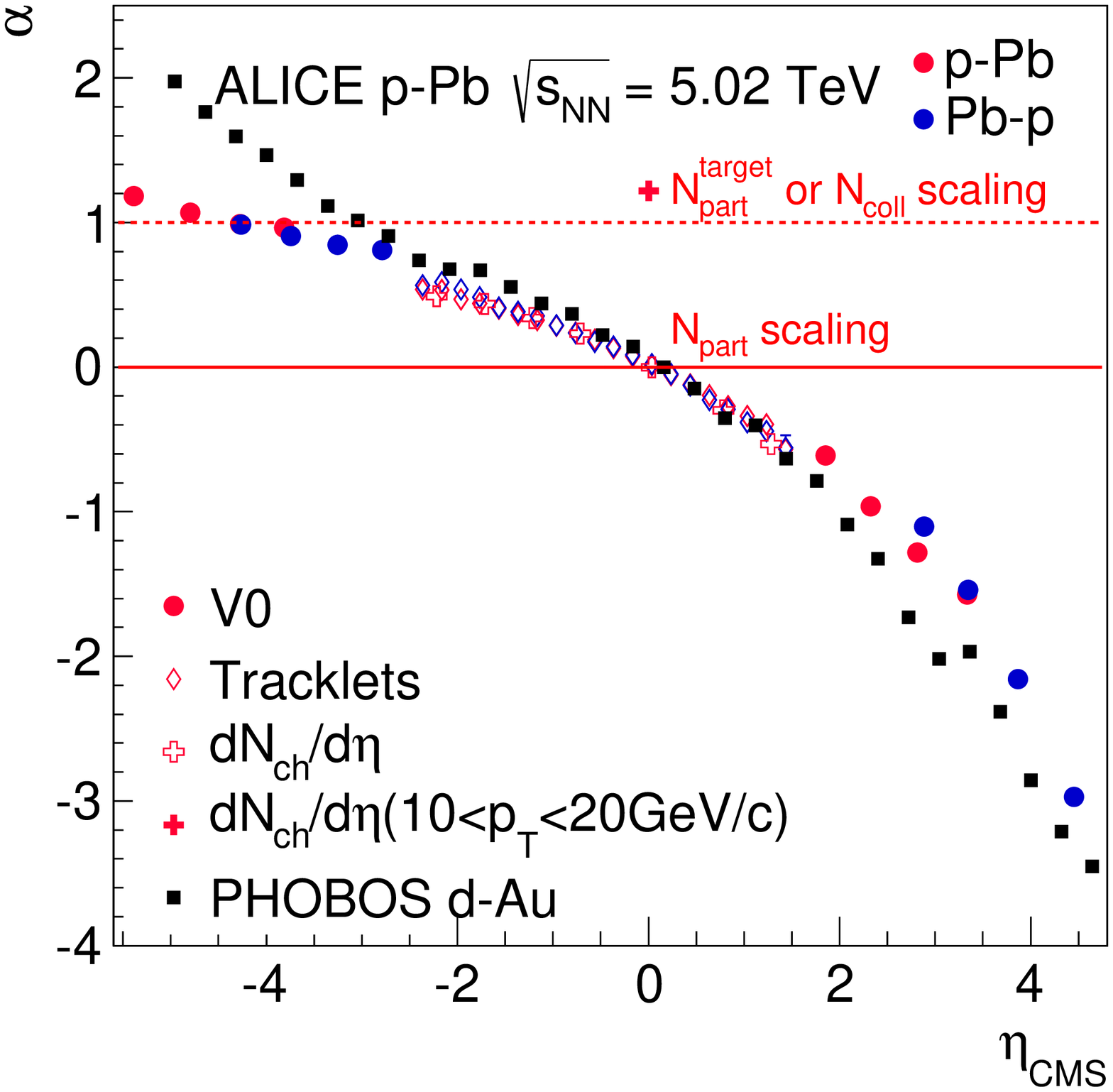}
\hspace{0.5cm}
 \includegraphics[width=0.45\textwidth]{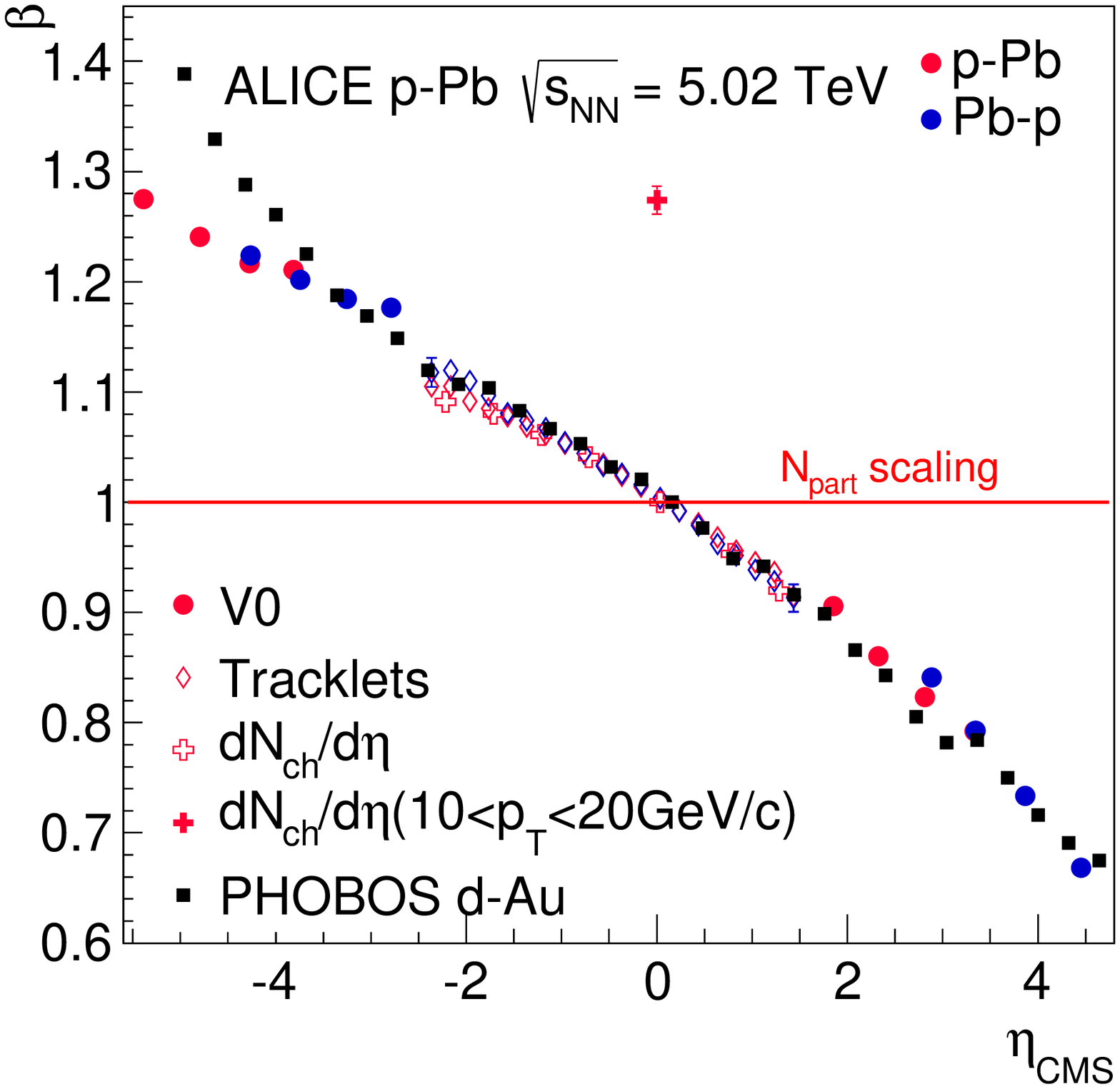}
 \caption{(color online) Results from the fits of Eq.~\ref{eq:Npartfit} (left) and
   Eq.~\ref{eq:Plawfit} (right) of the normalized signals as a
   function of the pseudorapidity covered by the various
   observables. The red horizontal lines (left) indicate the ideal
   \Npart\ and \Ncoll\ geometrical scalings.  For the PHOBOS data,
   $\eta_{\rm cms}$ has been scaled by the ratio of the beam
   rapidities in \pPb\ at $\sqrt{s_{\rm{NN}}}$ = 5.02 TeV and d--Au
   collisions at $\sqrt{s_{\rm{NN}}}$ = 200 GeV.
  \label{fig:fitdNdeta}}
\end{figure}

Fig.~\ref{fig:fitdNdeta} shows the results of the fits in
Eq.~\ref{eq:Npartfit} and \ref{eq:Plawfit} as a function of $\eta_{\rm
  cms}$ of the measured observables. The figure displays data
collected in both \pPb\ and \Pbp\ beam configurations. Since the
\Pbp\ data were taken at high-luminosity (reaching 200 kHz, roughly
corresponding to a luminosity of 10$^{29} \rm s^{−1} \rm cm^{−2}$),
the results are affected by interaction pile-up (probability per bunch
crossing between 3.8-4.3\%). In order to reduce the effect of the
pile-up and to treat \pPb\ and \Pbp\ data consistently, we excluded the
0-5\% centrality class from the fits. Furthermore, in order to take
into account the remaining distortions in the 5-100\% classes, the
\Pbp\ data were corrected using the results for the tracklets (also
shown in Fig.~\ref{fig:fitdNdeta}) in a small $\eta$-region,
($|\eta_{\rm lab}| < 0.2$), where $|\eta_{\rm cms}|$ is nearly
identical for \pPb\ and \Pbp\ configurations. Typically the
absolute correction is 0.05 and 0.01 for the $\alpha$ and $\beta$
parameters, respectively.

The results presented in Fig.~\ref{fig:fitdNdeta} indicate a smooth and
continuous change of the scaling behaviour for charged particle
production with pseudorapidity.  It is worth noting that at large
negative pseudorapidity (Pb-going direction) the values of the
parameters $\alpha$ and $\beta$ reach those obtained for
charged-particle production at high-\pt\ . In contrast, the parameter
values are much lower in the proton-going direction.  Our
data are overlaid with the corresponding fit parameters derived from
PHOBOS charged-particle multiplicity measurements in d--Au collisions
at $\sqrt{s_{NN}}$ = 200 GeV~\cite{Back:2004mr}.  The normalized
charged-particle multiplicity in each pseudorapidity bin is fit
against $\left(\frac{\langle {\rm d}N/{\rm d}\eta \rangle
  _{\rm i}}{\langle {\rm d}N/{\rm d}\eta \rangle
  _{\rm MB}}\right)_{|\eta|<0.1}$ using Eq.~\ref{eq:Npartfit} and
~\ref{eq:Plawfit}. The results obtained in this way are then adjusted
by scaling the x-axis ($\eta_{\rm cms}$) by the ratio of the beam
rapidities in \pPb\ at $\sqrt{s_{NN}}$ = 5.02 TeV and d--Au collisions
at $\sqrt{s_{NN}}$ = 200 GeV. The comparison between PHOBOS and our
data shows a good agreement over a wide $\eta$ range, with some
deviations at large negative pseudorapidity. In particular, the
$\eta$ region covered by the innermost ring of the \VZEROA\ detector
corresponds to the target fragmentation region where extended
longitudinal scaling was observed at RHIC~\cite{Back:2004mr}.  The minimum bias
\Npart\ and \Ncoll\ are obtained by PHOBOS relying on
a tuned HIJING-based Monte Carlo simulation~\cite{Back:2004mr}.

\subsection{Calculation of \avNcoll} \label{subsec:HNcoll}
As discussed in the previous section, selecting the events using the
ZN signal is expected to be free from bias on the bulk multiplicity
or high-\pt\ particle yields.  In order to establish a relationship to the
collision geometry, we exploit the findings from the correlation
analysis described above and make use of observables that are expected
to scale as a linear function of \Ncoll\ or \Npart.

Three sets of \avNcoll\ values are calculated, based on the following assumptions:
\begin{enumerate} 
\item $\Ncoll^{\rm mult}$: the charged-particle multiplicity at
  mid-rapidity is proportional to the number of participants (\Npart);
\item $\Ncoll^{\rm high-\pt}$: the yield of charged
  high-\pt\ particles at mid-rapidity is proportional to the number of
  binary NN collisions (\Ncoll);
\item $\Ncoll^{\rm Pb-side}$: the target-going charged-particle
  multiplicity is proportional to the number of wounded target
  nucleons (\Nparttar\ = \Npart - 1 = \Ncoll).
\end{enumerate}
For the charged-particle multiplicity in the Pb-going side we use the
signal from the innermost ring of the \VZEROA\ detector.  We note that
assumptions 1) and 2) are satisfied for minimum bias collisions,
where we measured a value of $(\dNdetac)/\avNpart$ consistent with that in inelastic
\pp\ collisions (0.97~$\pm$~0.08)~\cite{alice_pA_dndeta}
and an integrated $R_{\rm pA}$(10~<\pt\ <~20~GeV/$c$)~=~0.995 $\pm$ 0.010~(stat.) $\pm$ 0.090~(syst.) (see Sec.~\ref{sec:results}).

\begin{figure}[tbh!f]
 \centering
 \includegraphics[width=0.75\textwidth]{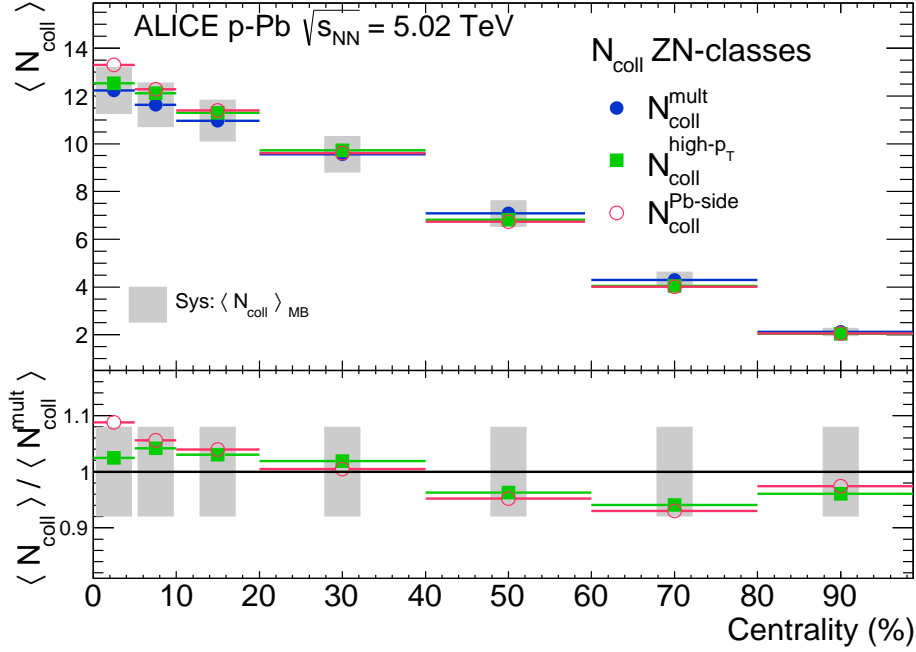}
 \caption{(color online) \avNcoll\ values obtained with the hybrid method under the
   three assumptions discussed in the text. The systematic
   uncertainty, shown as a grey band around the $\Ncoll^{\rm mult}$ points, represents the
   8\% uncertainty on the $\avNcoll_{\rm MB}$.
  \label{fig:NcollH}}
\end{figure}

Therefore, in order to obtain the average number of binary NN
collisions in each centrality interval, the minimum bias value of 
$\avNpart_{\rm MB}$= 7.9, is scaled using the ratio of the multiplicity at
mid-rapidity:

\begin{eqnarray}
\langle \Npart \rangle _{\rm i}^{\rm mult} &=& \langle \Npart \rangle _{\rm MB} \cdot 
\left(\frac{\langle {\rm d}N/{\rm d}\eta \rangle _{\rm i}}{\langle {\rm d}N/{\rm d}\eta \rangle _{\rm MB}}\right)_{-1<\eta<0} \\ 
\langle N_\mathrm{coll} \rangle _{\rm i}^{\rm mult} &=& \langle \Npart \rangle _{\rm i}^{\rm mult} -1 
\end{eqnarray}
In a similar way the minimum bias value of $\avNcoll_{\rm MB}$ = 6.9 is
scaled using the ratio of the yield of high-\pt\ particles at mid-rapidity to obtain $\Ncoll^{\rm high-\pt}$:
\begin{eqnarray} 
\langle N_\mathrm{coll} \rangle _{\rm i}^{\rm high-\pt} &=& \langle N_\mathrm{coll} \rangle _{\rm MB} \cdot 
\frac{\langle S \rangle _{\rm i}}{\langle S \rangle _{\rm MB}} 
\end{eqnarray}
where $S$ stands for the charged-particle yields with $10 < \pt < 20$~GeV/$c$.
Alternatively, one can use the Pb-side multiplicity to obtain $\Ncoll^{\rm Pb-side}$
\begin{eqnarray} 
\langle N_\mathrm{coll} \rangle _{\rm i}^{\rm Pb-side} &=& \langle N_\mathrm{coll} \rangle _{MB} \cdot 
\frac{\langle S \rangle _{\rm i}}{\langle S \rangle _{MB}} 
\end{eqnarray}
where $S$ stands for the raw signal of the innermost ring of \VZEROA.
The obtained values of \avNcoll\ in ZN-centrality
classes are listed in Table~\ref{tab:NcollCompareH} and shown in
Fig.~\ref{fig:NcollH}.
\begin{table*}[thb!f]
 \centering \footnotesize  
\hspace{-1cm}
\begin{tabular}{c|ccc}
Centrality & $\Ncoll^{\rm mult}$ & $\Ncoll^{\rm high-\pt}$ & $\Ncoll^{\rm Pb-side}$\\
 \hline 
 \phantom{0}0 - \phantom{0}\phantom{0}5   & 12.2 & 12.5 & 13.3 \\ 
 \phantom{0}5 - \phantom{0}10             & 11.6 & 12.1 & 12.3 \\ 
 10 - \phantom{0}20                       & 11.0 & 11.3 & 11.4 \\ 
 20 - \phantom{0}40                       & 9.56 & 9.73 & 9.60 \\ 
 40 - \phantom{0}60                       & 7.08 & 6.81 & 6.74 \\ 
 60 - \phantom{0}80                       & 4.30 & 4.05 & 4.00 \\ 
 80 - 100                                 & 2.11 & 2.03 & 2.06 \\
\end{tabular}
 \caption{\avNcoll\ values obtained under the three assumptions
   discussed in the text.
\label{tab:NcollCompareH}} 
\end{table*}

\begin{figure}[tbh!f]
 \centering
 \includegraphics[trim = 0cm 0cm 0cm 0cm, clip  = true, width=0.6\textwidth]{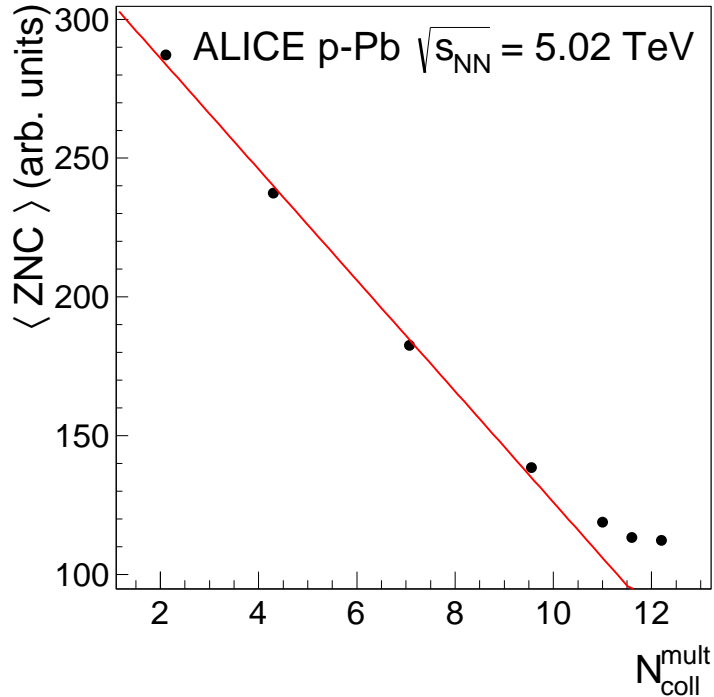}
 \caption{(color online) Signal in the proton-going direction ZNC as a
   function of $\Ncoll^{\rm mult}$. The red line shows a linear fit to
   the first four data points.
 \label{fig:znc_ncoll}}
\end{figure}

The systematic uncertainty is given by the 8\% uncertainty on the
$\avNcoll_{\rm MB}$ (or the 3.4\% uncertainty on the $\avTAB_{\rm
  MB}$) listed in Table~\ref{tab:NcollCompare}. We assign no
uncertainty to the assumptions made for particle scaling. The
differences between the three sets of values do not exceed 9\% in all
centrality classes. This confirms the consistency of the assumptions
used, but it does not prove that any (or all) of the assumptions are
valid.  We note that these values, in particular $\Ncoll^{\rm
  Pb-side}$, agree within 18\% with those calculated with the SNM (see
Fig.~\ref{fig:Ncoll} and Table~\ref{tab:NcollCompare}), except for the
most peripheral reactions, where the SNM is inaccurate.

In addition, we plot in Fig.~\ref{fig:znc_ncoll} the zero degree
signal from neutral particles in the proton-going direction ZNC vs
$\Ncoll ^ {\rm mult}$.  We have excluded events without a signal in
the ZNC, however the qualitative trend does not change when including
those events.  Over a wide range of centralities (10-100\%) a linear
anti-correlation is observed. This is consistent with a longitudinal
energy transfer of the proton proportional to the number of binary
collisions.

\section{Results and implications for particle production} 
\label{sec:results}
\subsection{Charged Particle Density}
The measurement of the centrality dependence of the particle
multiplicity density allows a discrimination between models that
describe the initial state of heavy ion collisions.  In
\cite{alice_pA_dndeta} we described the charged particle pseudorapidity
density in minimum bias collisions. The same analysis was repeated,
dividing the visible cross-section (see Sec.\ref{sec:multiplicity}) into
event classes defined by the centrality estimators described above,
and the \avNpart\ values associated to each centrality interval were
calculated using the methods discussed in Sec.~\ref{sec:GlauberFit},
\ref{sec:geometry}, \ref{sec:hybrid}.

\begin{figure}[t!f]
 \centering 
 \includegraphics[width=0.45\textwidth]{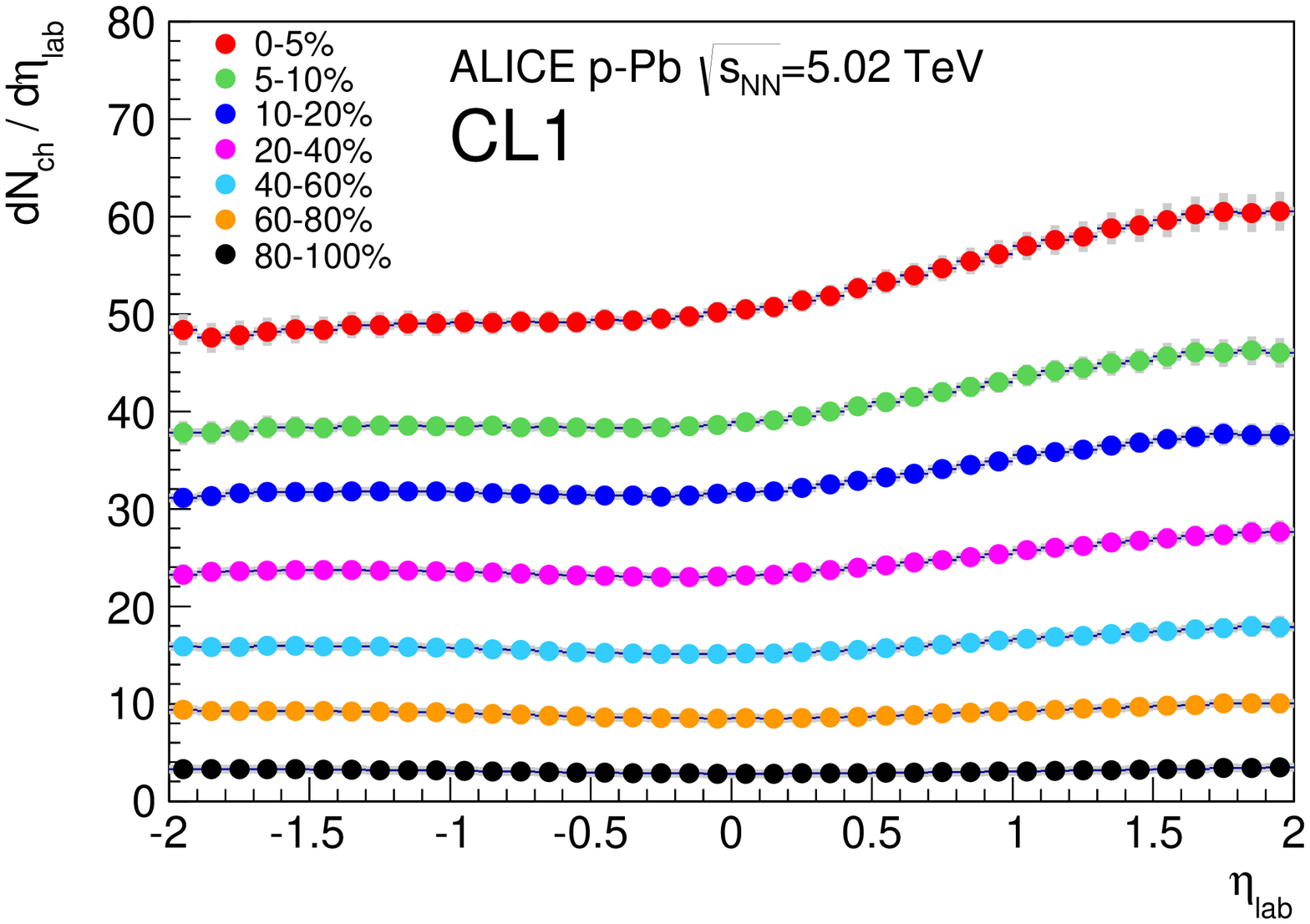}
 \includegraphics[width=0.45\textwidth]{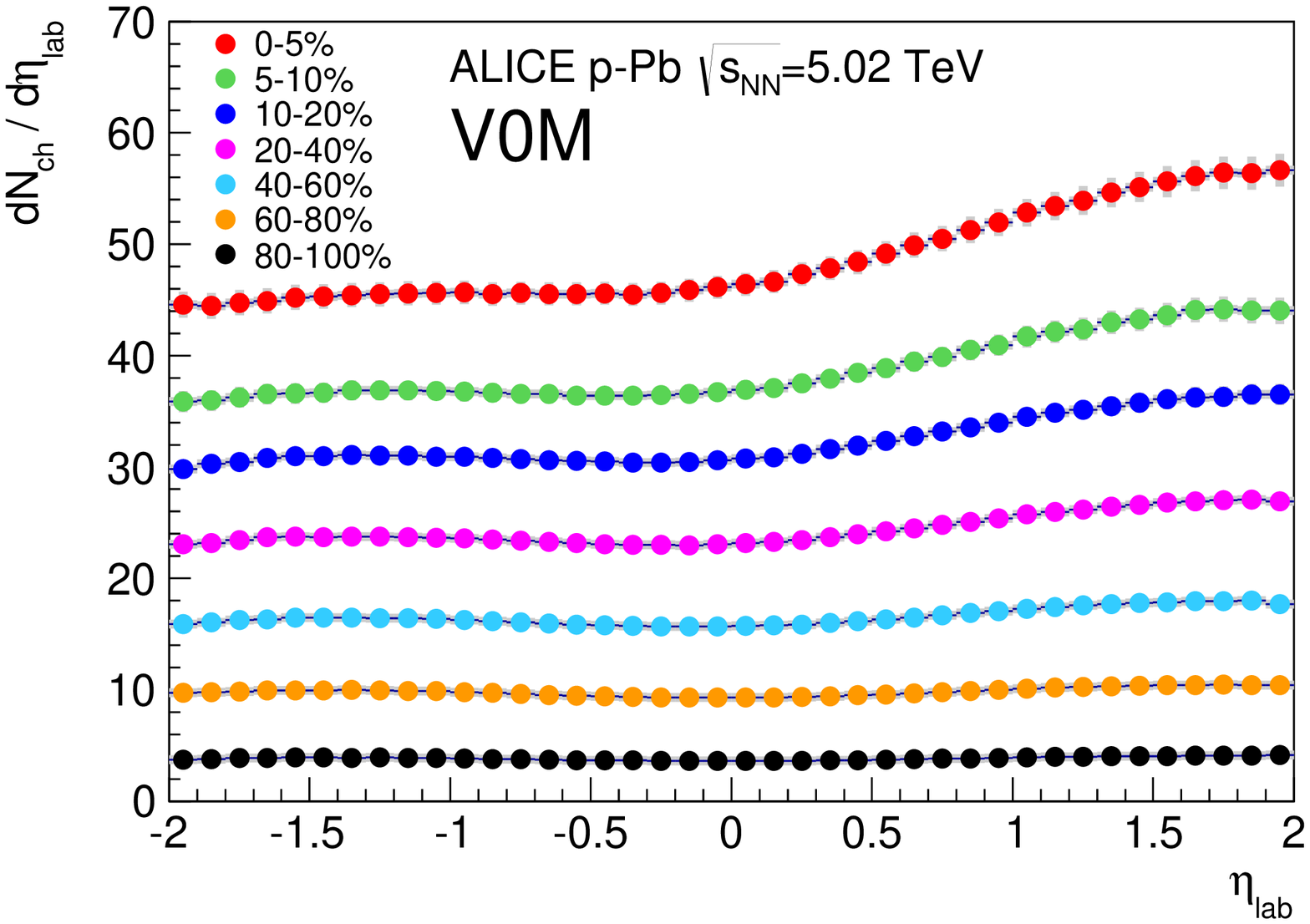}
 \includegraphics[width=0.45\textwidth]{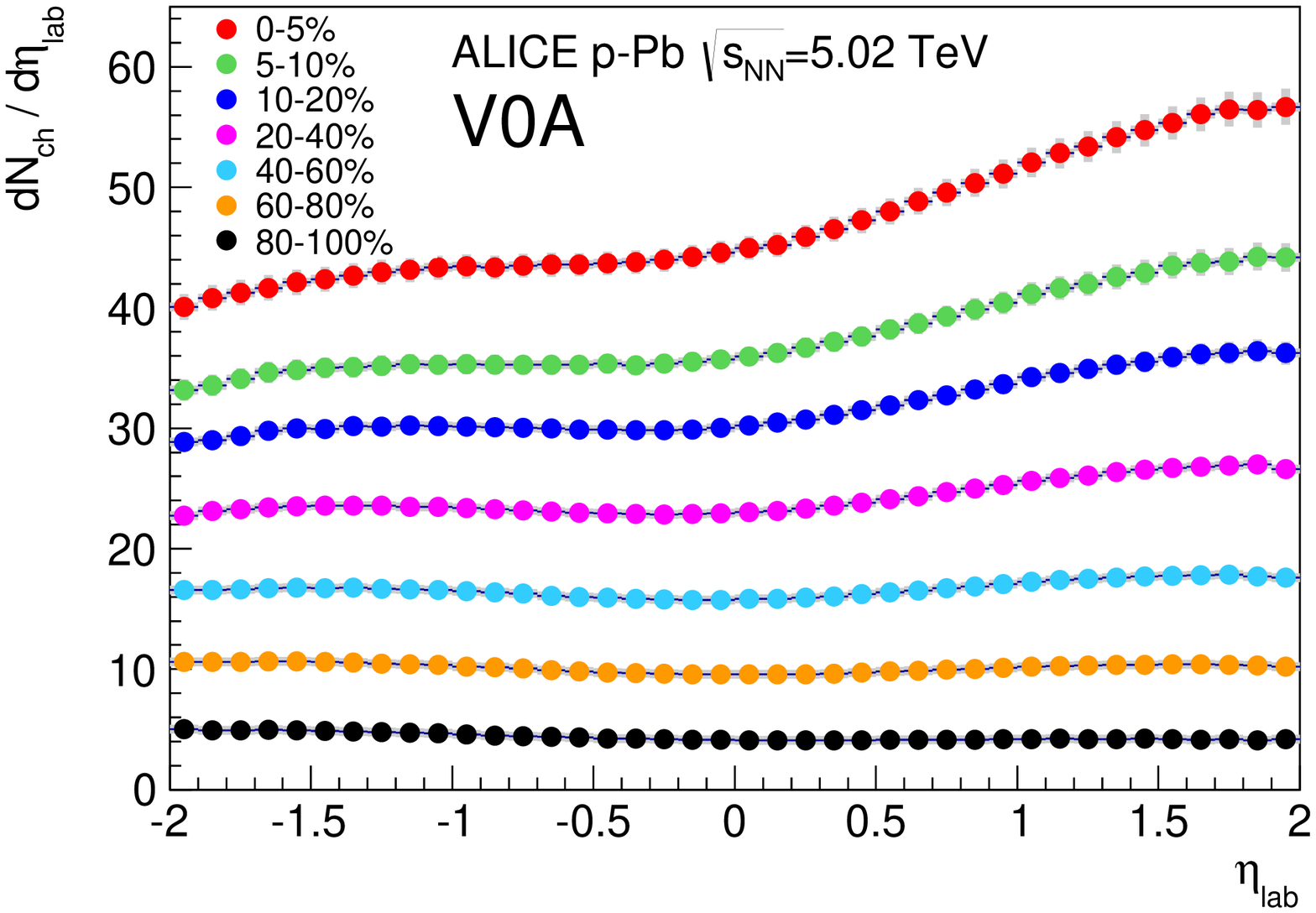}
 \includegraphics[width=0.45\textwidth]{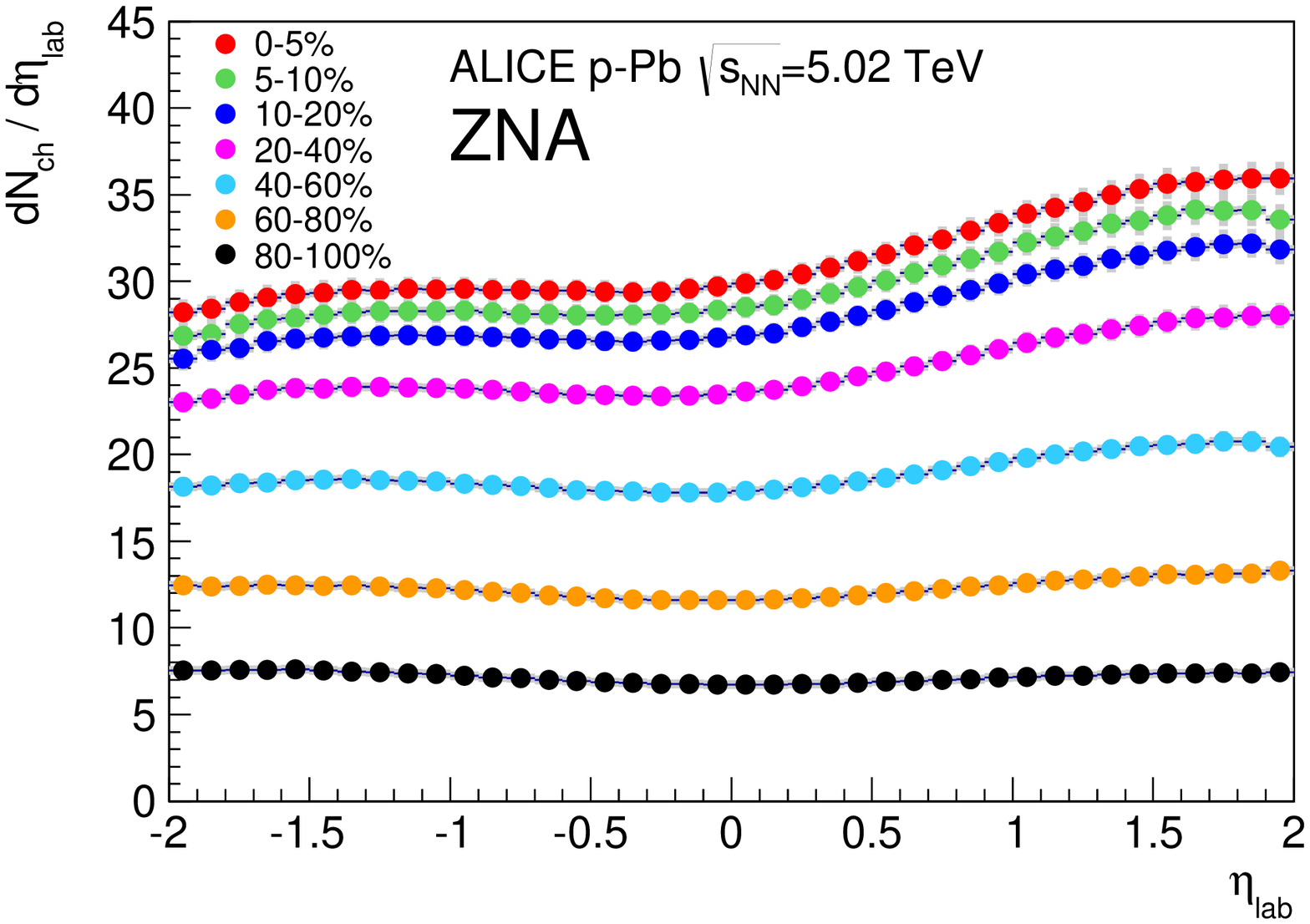}
 \caption{(color online) Pseudorapidity density of charged particles measured in
   \pPb\ collisions at $\snn=5.02$~TeV for various centrality classes
   and estimators. The different panels correspond to different
   centrality estimators: CL1 (top left), V0M (top right), V0A (bottom
   left), ZNA (bottom right).
\label{fig:dndeta}}
\end{figure}

The results of the charged particle multiplicity density as a function
of the pseudorapidity are presented in Fig.~\ref{fig:dndeta} for
different centrality intervals and different centrality estimators.
The fully correlated systematic uncertainty, is given by the
quadrature sum of the 2.2\% minimum bias error detailed in
\cite{alice_pA_dndeta}, and an $\eta$-dependent uncertainty from the
vertex efficiency and the centrality selection.

\begin{figure}[t!f]
 \centering 
 \includegraphics[width=0.75\textwidth]{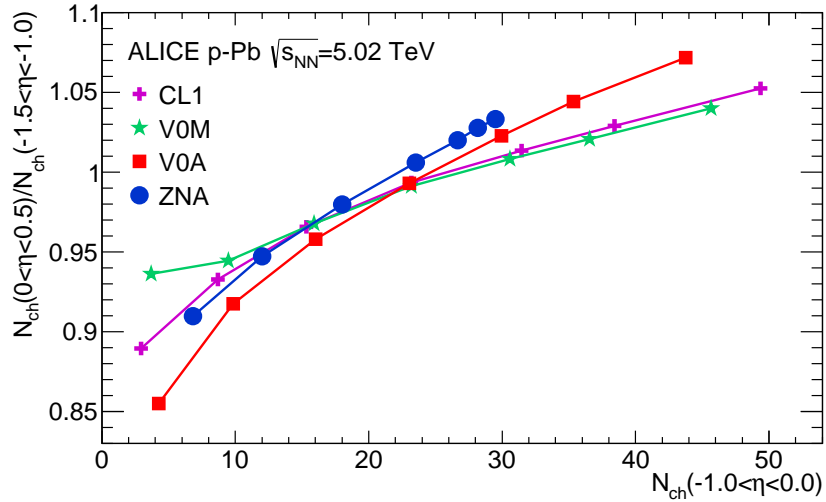}
 \caption{(color online) Asymmetry of particle yield, calculated as ratio of the
   pseudorapidity density integrated in $0<\eta<0.5$ to the one in $ -1.5 < \eta
   < -1$ as a function of the pseudorapidity density integrated at
   mid-rapidity for various centrality classes and estimators.
\label{fig:NchAsym}}
\end{figure}

In peripheral collisions (60-80\% and 80-100\%)
the shape of the distribution is almost fully symmetric and resembles
what is seen in proton-proton collisions. In more central collisions,
the shape of \dNdeta\ becomes progressively more asymmetric, with an
increasing excess of particles produced in the direction of the Pb
beam compared to the proton-going direction.
The shape of the pseudorapidity density function is sensitive to
details of particle production models. For example, it was found
in~\cite{alice_pA_dndeta} that in minimum bias reactions the
$\eta_{\rm lab}$ dependence is described relatively well by
HIJING~\cite{Xu:2012au} or DPMJET\cite{Roesler:2000he}, with a gluon
shadowing parameter tuned to describe experimental data at lower
energy, whereas the saturation
models~\cite{Dumitru:2011wq,Tribedy:2011aa,Albacete:2012xq} exhibit a
steeper $\eta_{\rm lab}$ dependence than the data.  We have quantified
the centrality evolution of the pseudorapidity shape for the different
centrality estimators by analyzing the density at mid-rapidity, and the
asymmetry of particle yield between the proton and the Pb peak
regions, as the ratio of \dNdeta\ at $0<\eta<0.5$ and
$-1.5<\eta<-1.0$, symmetrically around the centre of mass. This is
  shown in Fig.~\ref{fig:NchAsym}.

\begin{figure}[t!f]
 \centering 
 \includegraphics[width=0.495\textwidth]{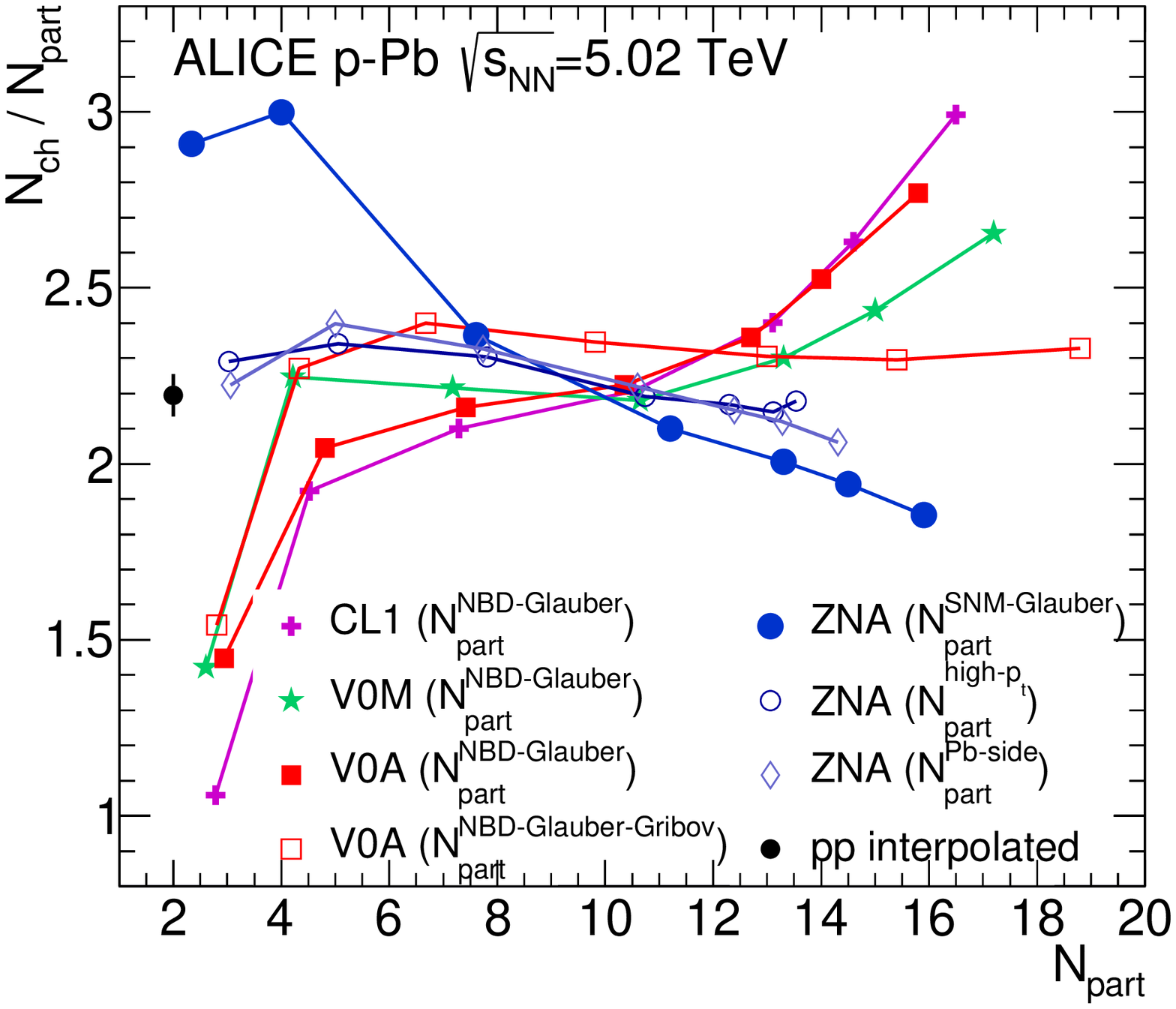}
 \includegraphics[width=0.495\textwidth]{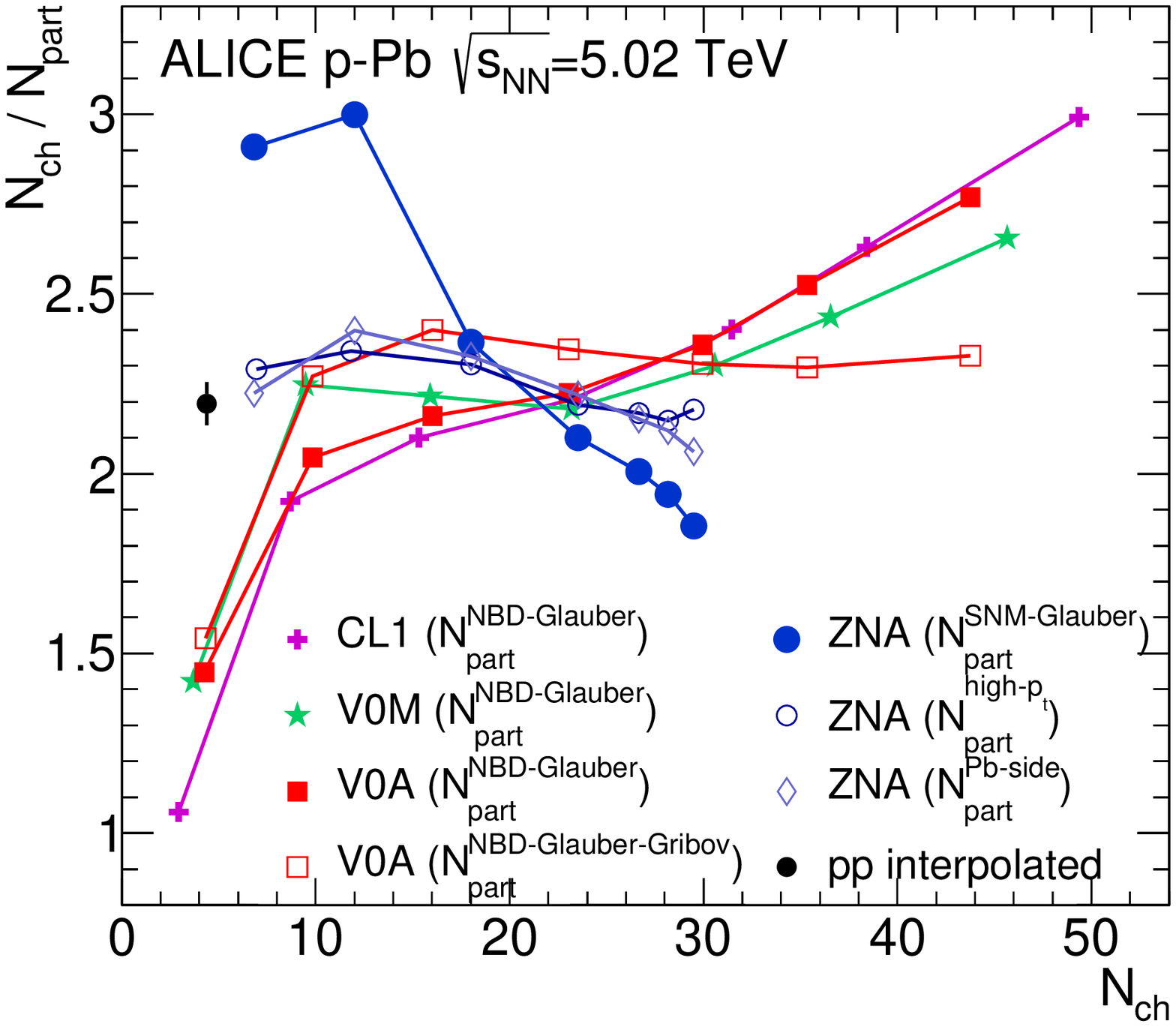}
 \caption{(color online) Pseudorapidity density of charged particles measured in
   p--Pb collisions at mid-rapidity per participant as a function of
   \Npart\ (left), or as a function of the mid-rapidity density (right),
   for various centrality estimators.
\label{fig:dndetaNpart2}}
\end{figure}

Figure ~\ref{fig:dndetaNpart2} shows the \dNdeta\ integrated at
mid-rapidity divided by the number of participants as a function of
\avNpart\ (left) or as a function of \dNdeta (right) for various
centrality estimators. The systematic uncertainty is smaller than the
marker size.  For the V0A centrality estimator, in addition to the
\avNpart\ from the standard Glauber calculation, the results obtained
with the implementation of Glauber-Gribov model (with $\Omega=0.55$)
are also shown.  For CL1, V0M, and V0A, the charged particle density
at mid-rapidity has as steeper than linear increase, as a consequence
of the strong multiplicity bias discussed in Sec.~\ref{sec:bias},
which is strongest in CL1, where the overlap with the tracking region
is maximum.  This trend is not seen in the case of the Glauber-Gribov
model, which shows a relatively constant behaviour for the integrated
yield divided by the number of participant pairs, with the exception
of the most peripheral point.

For ZNA, there is a clear sign of saturation above $\Npart \sim 10$,
as the \avNpart\ values are closer to each other. Most probably, this is
due to the saturation of forward neutron emission.  We note that none
of these curves point towards the \pp\ data point. This suggests that
the geometry bias, present in peripheral collisions, together with the
multiplicity bias for CL1, V0M and V0A, has a large effect on this
centrality class.

In contrast, the results obtained with the hybrid method, where the
$\Npart^{\rm Pb-side}$ and the $\Npart^{\rm high-\pt}$ give very
similar trends, show, within $\pm 10\%$, scaling with \Npart, which
naturally reaches the \pp\ point, well within the quoted uncertainty
of 8\% on the \Npart\ values. In addition, they show that the range in
\Npart\ covered with an unbiased centrality selection is more limited
than what is obtained using estimators based on particle
multiplicity. The latter do not select on the collision geometry but
rather on the final products of the collision. This effect is
emphasized in the right plot, which shows the same quantity
\Nch\ divided by \Npart\ as a function of \Nch. Here the limited range
in \Nch\ reached with the \ZNA\ selection is clearly visible.  This
indicates the sensitivity of the \Npart-scaling behaviour to the
Glauber modelling, as well as the importance of the multiplicity
fluctuations.

\subsection{Nuclear modification factors}
As discussed in section \ref{sec:bias}, the various centrality
estimators induce a bias on the nuclear modification factor depending
on the rapidity range they cover. In contrast to minimum bias
collisions, where $\avNcoll = 6.9$ is fixed by the ratio of the pN and
\pPb\ cross-sections, in general, \Ncoll\ for a given centrality class
cannot be used to scale the \pp\ cross-section, or to calculate
centrality-dependent nuclear modification factors.  For a centrality
selected event sample, we therefore define $\qpa$ as
\begin{equation} \label{eq:Qpa}
\qpa (\pt ; {\rm cent}) =
\frac{{\rm d}N^{\rm pPb}_{\rm cent}/{\rm d}\pt} {\langle {N_{\rm coll}^{\rm Glauber}\rangle {\rm d}N^{\rm pp}/ {\rm d}\pt}}=
\frac{{\rm d}N^{\rm pPb}_{\rm cent}/{\rm d}\pt} {\langle {T_{\rm pPb}^{\rm Glauber} \rangle {\rm d}\sigma^{\rm pp}/{\rm d}\pt}}
\end{equation}
for a given centrality percentile according to a particular centrality
estimator.  In our notation we distinguish $\qpa$ from $\rpa$ because
the former is influenced by potential biases from the centrality
estimator which are not related to nuclear effects. Hence, $\qpa$ can
be different from unity even in the absence of nuclear effects.


The \pt\ distribution of primary charged particles in minimum bias
collisions is given in~\cite{alice_RpA_new}.  The charged particle
spectra are reconstructed with the two main ALICE tracking detectors,
the Inner Tracking System and the Time Projection Chamber, and are
corrected for the detector and reconstruction efficiency using a Monte
Carlo simulation based on the DPMJET event
generator~\cite{Roesler:2000he}. The systematic uncertainties on
corrections are estimated via a comparison to a Monte Carlo simulation
using the HIJING event generator~\cite{hijing}, while the
\pt\ resolution is estimated from the space-point residuals to the
track fit and verified with data.  The total systematic uncertainty
ranges between 3.4\% and 6.7\% in the measured \pt\ range, 0.15-50
GeV/$c$, with a negligible $\eta_{\rm cms}$ dependence.  The nuclear
modification factor is calculated by dividing the data by the
reference \pp\ spectrum scaled by $\avNcoll_{\rm MB}$.  The reference
\pp\ spectrum is obtained at low \pt\ (\pt $<$ 5 GeV/$c$) by
interpolating the data measured at \s\ = 2.76 and 7 TeV, and at high
\pt\ (\pt $>$ 5 GeV/$c$) by scaling the measurements at \s\ = 7 TeV
with the ratio of spectra calculated with NLO pQCD at \s\ = 5.02 and 7
TeV~\cite{Abelev:2013ala}.  The systematic uncertainty, given by the
largest of the relative systematic uncertainties of the spectrum at
2.76 or 7 TeV at low-\pt, and assigned from the relative difference
between the NLO-scaled spectrum for different scales and the
difference between the interpolated and the NLO-scaled data at
high-\pt, ranges from 6.8\% to 8.2\%.  For MB collisions the nuclear
modification factor $R_{\rm pPb}$ is consistent with unity for
\pt\ above 6 GeV/c.

\begin{figure}[t!f]
 \centering
 \includegraphics[width=0.99\textwidth]{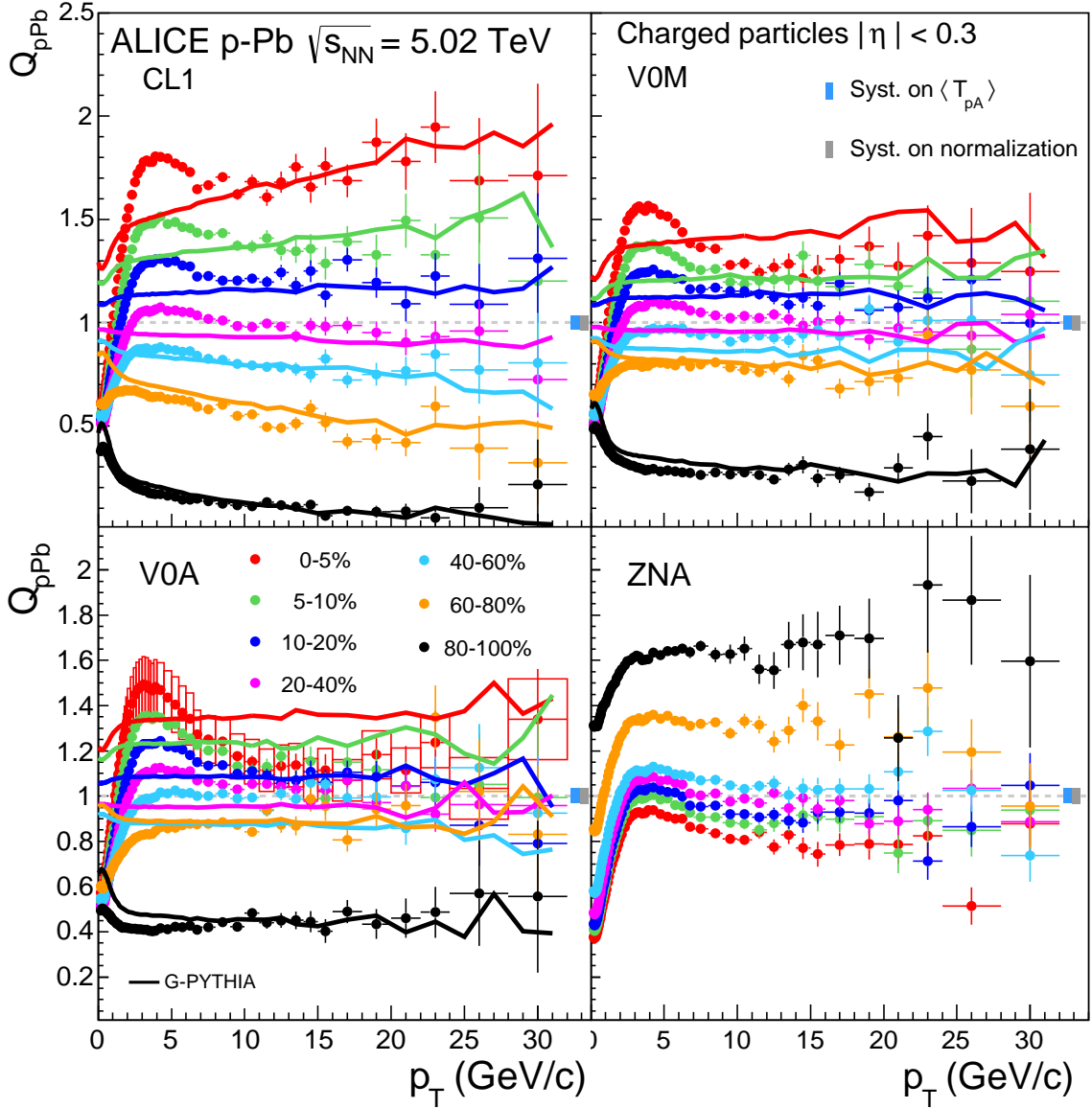}
 \caption{(color online) \qpa\ spectra (points) of all primary charged particles for various 
   centrality classes obtained with the different centrality
   estimators explained in the text.  The lines are from G-PYTHIA calculations. 
   The systematic error on the spectra is
   only shown for the V0A 0-5\% centrality bin and is the same for all
   others. The systematic uncertainty on \pp\ and \pPb\ normalization
   is shown as a gray box around unity at \pt\ $=0$. The systematic
   uncertainty on $\avTAB_{\rm MB}$ is shown as a light blue box
   around unity at high \pt.}
\label{fig:QpA}
\end{figure}

The same analysis was repeated dividing the visible cross-section (see Sec.\ref{sec:multiplicity}) 
in event classes defined by the
centrality estimators described above, and the $\qpa $ were calculated
using the values of \avNcoll\ listed in
Tables~\ref{tab:NcollCompare} and \ref{tab:NcollCompareH}, for each given
estimator.  Figure~\ref{fig:QpA} shows the $\qpa $ for different
centrality estimators and different centrality classes. The
uncertainties of the \pPb\ and \pp\ spectra are added in quadrature,
separately, for the statistical and systematic uncertainties. The
systematic uncertainty on the spectra is only shown for the V0A 0-5\%
centrality bin and is the same for all others, since all the
  corrections are independent of centrality. The total systematic
uncertainty on the normalization, given by the quadratic sum of the
uncertainty on the normalization of the \pp\ data and the
normalization of the \pPb\ data, amounts to 6.0\% and is shown as a
gray box around unity. The systematic uncertainty on \TAB\ is shown as
a light blue box around unity. For simplicity, we draw only the
uncertainty for the minimum bias value $\avTAB_{\rm MB}$.

\begin{figure}[t!f]
 \centering
 \includegraphics[width=0.99\textwidth]{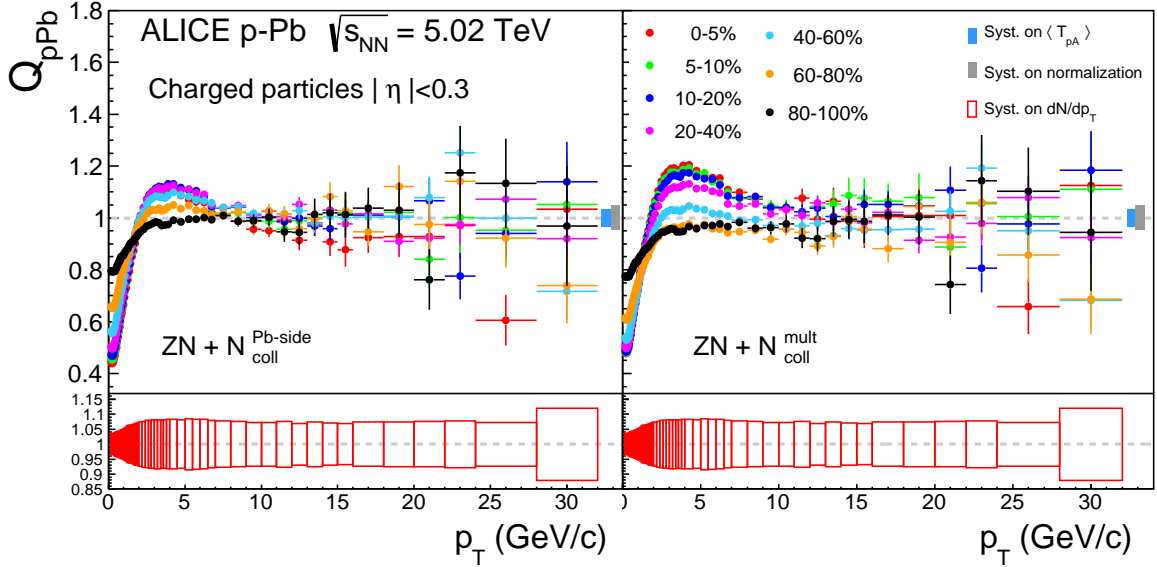}
 \caption{(color online) \qpa\ spectra with the hybrid method. Spectra are calculated
   in ZNA classes with \avNcoll as given in Table \ref{tab:NcollCompareH}, and are obtained with assumptions on
   particle production described in Sec.~\ref{sec:hybrid}.  
  \label{fig:QpAhybrid}}
\end{figure}

As expected, for CL1, V0M and V0A, \qpa\ strongly deviates from unity
at high \pt\ in all centrality classes, with values well above unity
for central collisions and below unity for peripheral collisions.
However the spread between centrality classes reduces with increasing
rapidity gap between the range used for the centrality estimator and
that used for the \pt\ measurement.  There is a clear indication of
the jet-veto bias in the most peripheral CL1 class, where \qpa has a
significant negative slope (\pt $>$ 5 GeV/$c$) since the jet
contribution to the total multiplicity increases with \pt.  This
jet-veto bias diminishes for V0M, and is absent for V0A, where $\qpa <
1 $ for peripheral collisions, indicating that the multiplicity bias
is still present.

In order to study the centrality determination biases further, the
\qpa\ spectra are compared to the G-PYTHIA spectra.  The event
centrality is obtained from the charged particle multiplicity in the
rapidity region covered by each estimator in the same way as in data,
and \avNcoll\ is directly obtained from the Monte Carlo. The
calculation is shown as lines in Fig.~\ref{fig:QpA}.  With this
approach, the general trend at high \pt\ is reasonably well described
for all centrality classes, particularly for CL1. This suggests that
particle production at high \pt\ in \pPb\ collisions indeed can be
approximated by an an incoherent superposition of \pp\ collisions.
The agreement, however, is not as good for the V0A and V0M estimators,
since the model is not adequate for forward particle production,
particularly in the target fragmentation region.  G-PYTHIA also
reproduces the jet-veto bias, as indicated by the good agreement of
the \pt\ dependence in the low and intermediate \pt\ region in the
most peripheral CL1 collisions.

However, for central collisions, the \qpa\ values show a significant
enhancement at intermediate \pt\ $\approx$~3~GeV/$c$ (called the
Cronin effect, a nuclear modification factor above unity at
intermediate \pt, observed at lower energies in
\pA\ collisions~\cite{PhysRevD.11.3105,PhysRevLett.91.072303,Adams20058,Accardi:2002ik}),
which increases with centrality independently of the estimator used.
The enhancement in the intermediate \pt\ region is about 15\%, and the
differences in the height of the peak among centrality estimators are
small with respect to the absolute increases of the \pPb\ yields.  The
enhancement is not reproduced by our model of incoherent superposition
of \pp\ collisions.  In contrast, in the low \pt\ region, below the
Cronin peak, the yield is overestimated by the model. This
overestimate at low \pt\ is expected because this \pt\ region is
dominated by soft processes and therefore is not expected to scale
with \Ncoll.  On the other hand, the intermediate \pt\ region is
expected to be dominated by hard scatterings and should scale with
\Ncoll\ in the absence of nuclear effects. From this we can conclude
that the Cronin enhancement observed is due to nuclear modification
effects, as observed in other
measurements~\cite{Abelev:2012ola,Abelev:2013bla,Abelev:2013haa,ABELEV:2013wsa},
as well as in the minimum bias \rpa~\cite{alice_RpA}.

The bottom right plot of Fig.~\ref{fig:QpA} shows the \qpa\ for the ZNA
centrality selection.  The classes selected by the ZNA present spectra
much more similar to each other than the other estimators, as expected
in the absence of a multiplicity bias.  The height of the Cronin peak
relative to the yield at high-\pt\ is larger with the V0A selection,
which may be seen as a sign of a remaining small bias in V0A, expected
from the G-PYTHIA calculations. However, for peripheral collisions
(60-80\% and 80-100\%), the absolute values of the spectra at high
\pt\ indicate the presence of a bias on \Ncoll\ in the ZNA measurement. 
This is not due to the event selection, but is
due to the inaccurate estimate of \avNcoll\ values for peripheral
events, where a small, absolute uncertainty results in a large
relative deviation in the \qpa calculation.
\begin{figure}[t!]
 \centering
 \includegraphics[width=0.495\textwidth]{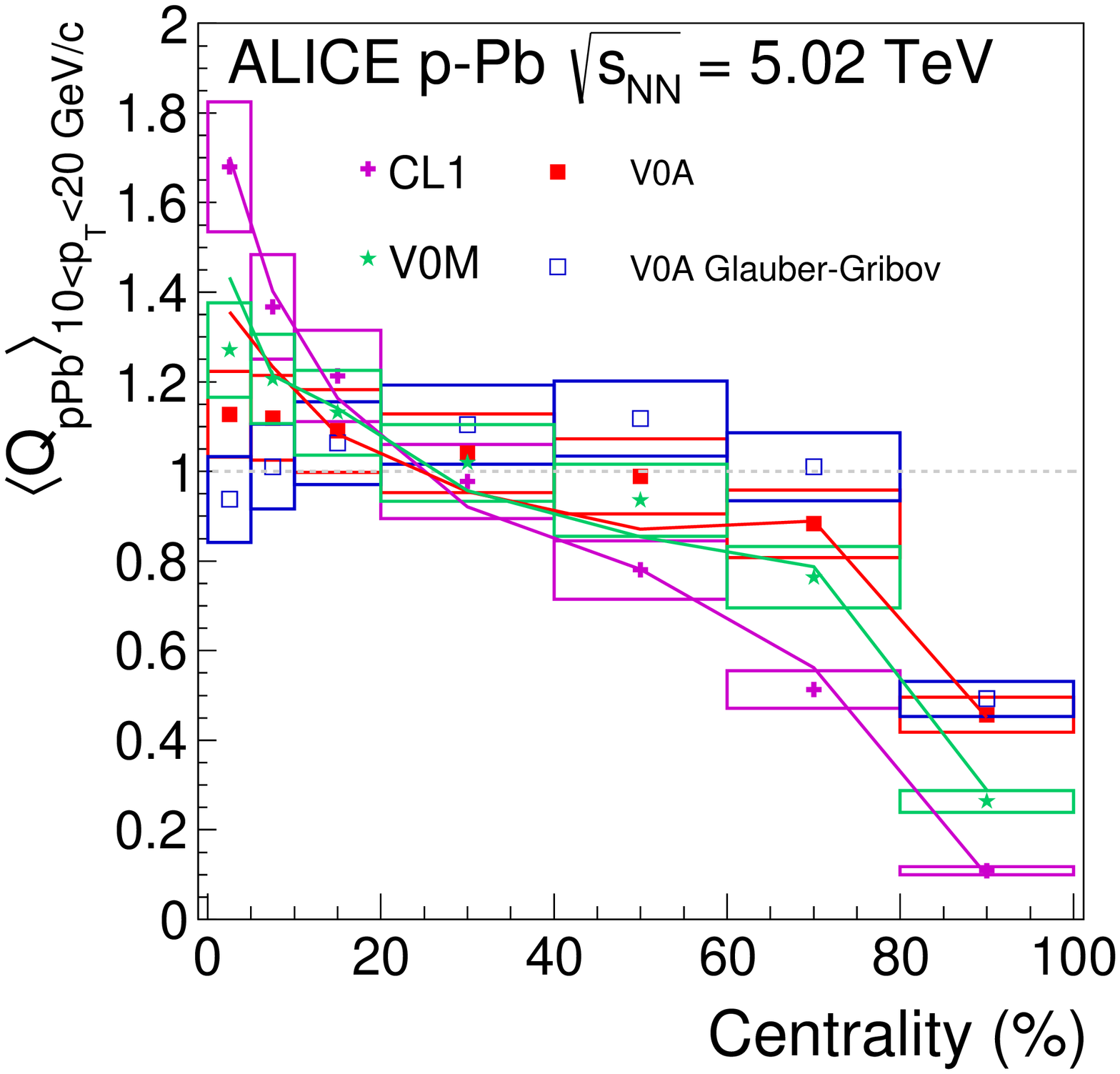}
 \includegraphics[width=0.495\textwidth]{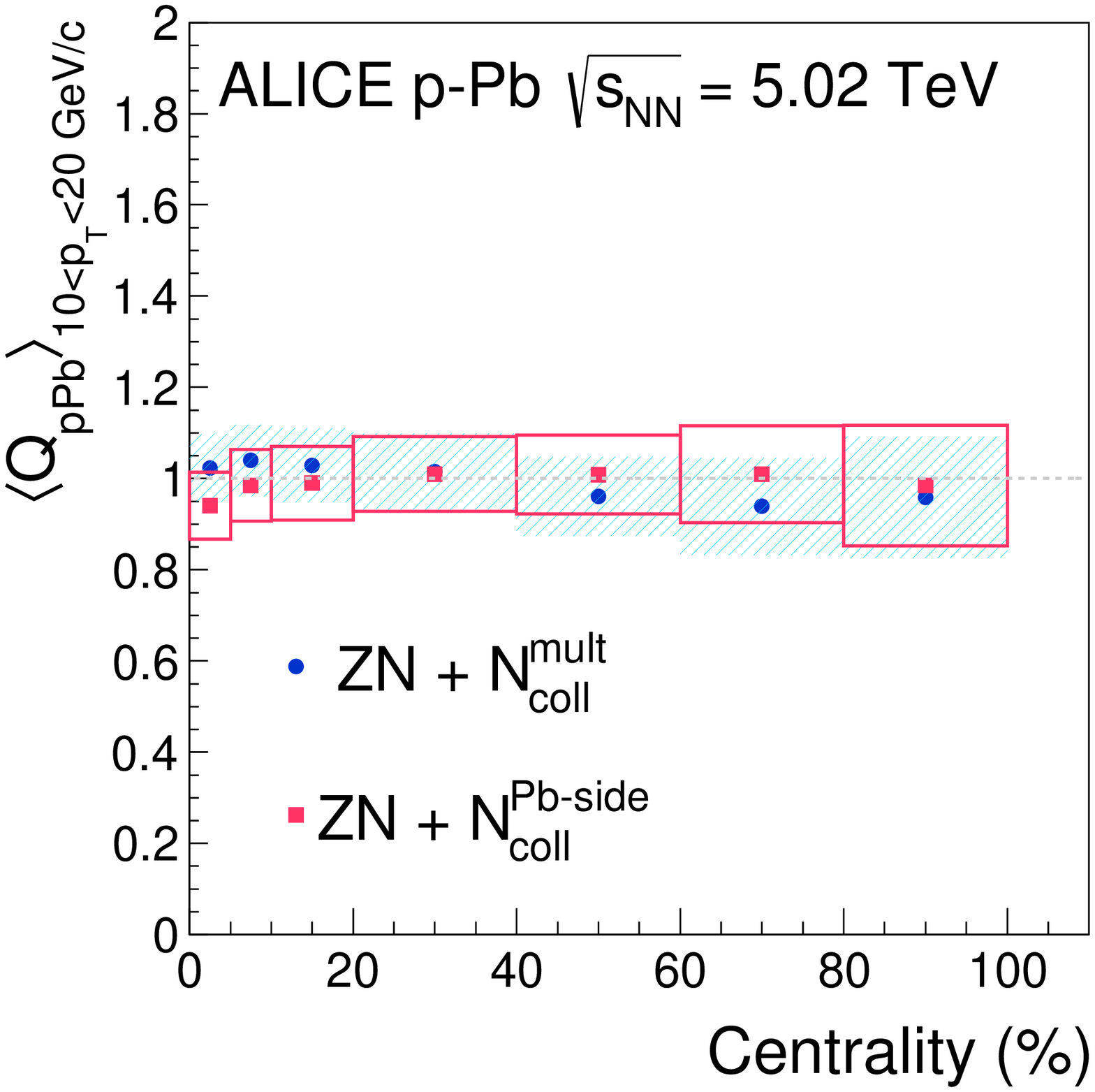}
 \caption{(color online) Average \qpa\ calculated ($10 < \pt < 20 \, {\rm GeV}/c$) as a
   function of centrality, with various centrality estimators. The
   left panel shows results from the data (points) and from the
   G-PYTHIA calculation (lines). The right panel shows the results for
   the hybrid method, where centrality classes are selected with ZNA,
   and \avNcoll\ are calculated with the assumptions on particle
   production described in Sec.~\ref{sec:hybrid}.
  \label{fig:QvsCent}}
\end{figure}

As discussed in Sec.~\ref{sec:hybrid}, the hybrid method uses
centrality classes selected with ZNA and \avNcoll\ values determined
with assumptions on particle production. Fig.~\ref{fig:QpAhybrid}
shows the resulting \qpa\ values, $\qpa^{\rm mult}$ on the left and
$\qpa^{\rm Pb-side}$ on the right panel.  Here it is important to note
that the ratios in the lower right panel in Fig.~\ref{fig:QpA}, and
both panels in Fig.~\ref{fig:QpAhybrid} have the same shape by
construction, and only differ due to the scaling (\Ncoll) of the
reference.  The small differences among the \avNcoll\ values
(Table~\ref{tab:NcollCompareH}) are reflected in consistent \qpa, which also
remain consistent with unity at high-\pt\ for all centrality classes. This
confirms the absence of initial state effects, already observed for
minimum bias collisions.  The Cronin enhancement, which has already
been noted in minimum bias collisions, is observed to be stronger
in central collisions and nearly absent in peripheral collisions.  The
enhancement is also weaker at 5.02 TeV compared to 200 GeV
~\cite{PhysRevD.11.3105}.  The geometry bias, described in
Sec.\ref{subsec:geobias}, is still present and uncorrected, even with
this method. Its effect is limited to only peripheral classes,
resulting in $\qpa <1$ for 80-100\%.

Fig.~\ref{fig:QvsCent} shows the mean \qpa\ at high momentum as a
function of centrality for the various centrality estimators.  The
centrality dependence of $\qpa^{\rm Glauber}$ extracted from
multiplicity distributions is shown on the left. It is reminiscent of
the multiplicity bias, and reproduced by the G-PYTHIA calculation
(lines in the figure).  The mean \qpa\ changes less with increasing
rapidity gap between the centrality estimator and the region where the
\pt\ measurement is performed, as expected from the multiplicity bias.
Instead, the \qpa\ extracted with the hybrid model
(Fig.~\ref{fig:QvsCent} right) is consistent with unity and the
results from the two assumptions used for the \avNcoll\ calculation
are in agreement.

\begin{figure}[t!]
 \centering
 \includegraphics[width=0.995\textwidth]{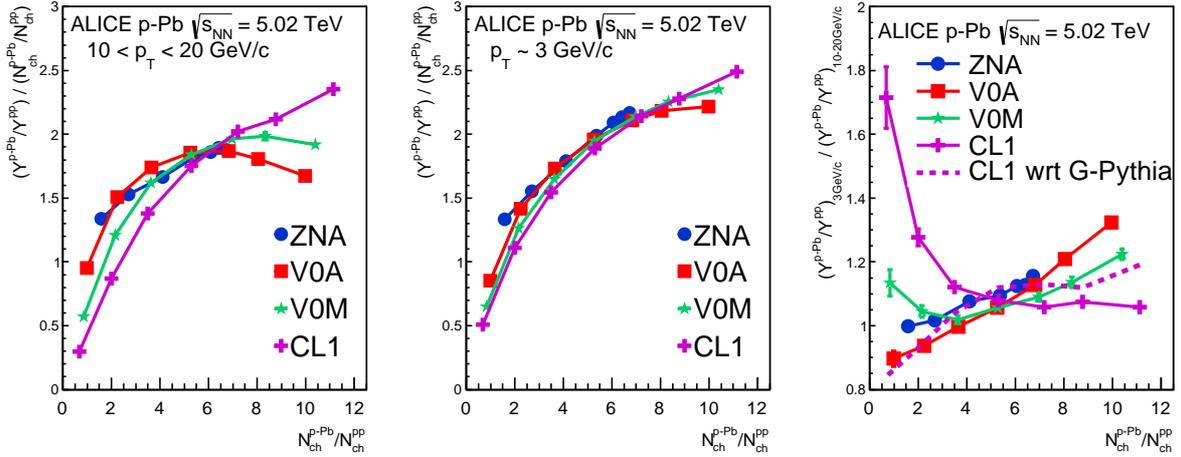}
 \caption{(color online) Left and Middle: ratio of yields in \pp\ and \pPb\ collisions for two momentum ranges, divided by the 
 ratio of \Nch\ in \pp\ and \pPb\ collisions. Right: ratio of the yield around 3 GeV/$c$ to the yield at high-\pt. 
 Values are calculated for different centrality estimators.
 For CL1 we also show the ratio to the value calculated with Q-PYTHIA at 3 GeV/$c$ (right panel only). 
  \label{fig:ratios}}
\end{figure}

To compare the impact of the multiplicity bias from the different
estimators on the nuclear modification factors, the ratio of the
spectra in \pp\ and \pPb\ in different momentum ranges ($Y^{\rm
  pPb}/Y^{\rm pp}$) is divided by the ratio of charged particle
density at mid-rapidity in \pp\ and \pPb\ ($\Nch^{\rm pPb}/\Nch^{\rm
  pp}$) and it is plotted as a function of ($\Nch^{\rm pPb}/\Nch^{\rm
  pp}$) in Fig.~\ref{fig:ratios}. Left and middle panels show the
yield 
at high-\pt\ (10-20~GeV/$c$) 
and 
around the Cronin peak (3 GeV/$c$), respectively.  
Figure~\ref{fig:ratios} clearly shows the shape
bias on particle spectra.  Even for the same average event activity at
mid-rapidity (corresponding to the same point on the x-axis $\Nch^{\rm
  pPb}/\Nch^{\rm pp}$), the \pt\ spectra show a small but significant
dependence on the centrality estimator. This is visible as a different
relative number of particles ($Y^{\rm pPb}/Y^{\rm pp}$) in the
intermediate (3 GeV/$c$) or in the high-\pt\ (10-20 GeV/$c$) region.
Also the height of the Cronin peak relative  to the high-\pt\ yield
depends on the centrality estimator.  
This is shown in the right panel of Fig.~\ref{fig:ratios},
which plots the double ratio of the \pPb\ to \pp\ yields at 3 GeV/$c$
and in 10-20 GeV/$c$ ($(Y^{\rm pPb}/Y^{\rm pp})_{3
  \mathrm{GeV}/c}/(Y^{\rm pPb}/Y^{\rm pp})_{10-20 \mathrm{GeV}/c}$).
Since, for CL1,  $Q_{\rm pPb}$ is not constant at high \pt\ we plot also the 
ratio $(Y^{\rm pPb}/Y^{\rm pp})_{3                                                                                          
  \mathrm{GeV}/c}$ to the value calculated with G-PYTHIA at 3 GeV/$c$.
The Cronin peak is clearly visible
for the V0M and CL1 (with respect to G-PYTHIA) selection, 
and very pronounced for the V0A selection. As
previously noted, the ZNA selection shows a similar trend and similar
value as V0A, when restricted to the \dNdeta\ range common to both
estimators. However the differences are still significant, and the
common range is still rather small. In particular, the height of the
Cronin peak is larger with ZNA than with V0A in the common
\dNdeta\ range, which may be seen as a sign of a remaining small bias
in V0A, confirming what is observed by G-PYTHIA calculations.

The study of the correlation between observables measured in such
different parts of phase space has shown that it is possible to select
similar event classes using estimators that are causally disconnected
after the interaction. This is very important because this suggests that
any such correlation can only arise from the initial geometry of the
collision.
 
\section{Summary}
\label{sec:conclusions}
In summary, we have studied the centrality dependence of charged
particle production, with measurements that comprise the
charged particle pseudorapidity density and the nuclear modification
factor.  The methods to determine centrality in p--A collisions using
multiplicity measurements or zero-degree energy have been presented in
detail.  The former induce a bias on the hardness of the pN collisions
that can be quantified by the number of hard scatterings per pN
collision. Low (high) multiplicity p--Pb corresponds to lower (higher)
than average number of hard scatterings.  For observables based on
centrality estimates from multiplicity, nuclear effects should be
calculated, including this bias when comparing to an incoherent
superposition of pN collisions.

In contrast, the energy deposited at zero degrees by slow nucleons
 in the ZDC is expected to be insensitive to a multiplicity bias. Under
this assumption, but in the absence of a model which properly relates the
ZDC energy to the number of collisions, these are calculated assuming
multiplicity scaling laws in the given kinematic ranges.  In particular,
we assume that the multiplicity at mid-rapidity is proportional to \Npart,
that multiplicity in the target-going direction is proportional to the
number of wounded target nucleons, or that the yield of high-\pt\
particles is proportional to \Ncoll. The equivalence of these
assumptions has been shown and discussed.  Therefore, under these
assumptions, we find i) that nuclear modification factors are
consistent with unity above $\sim$ 8~GeV/c, with no centrality
dependence; ii) that the multiplicity of charged particles at
mid-rapidity scales linearly with the total number of participants;
iii) and that the longitudinal features of \pPb\ collisions at $\snn =
5.02$~TeV, as reflected by the centrality dependence of the
pseudorapidity distributions of charged particles, are very similar to
those seen in d--Au collisions at RHIC energies.  The latter were
interpreted in support of extended longitudinal scaling in the
fragmentation regions.  
These results represent valuable input for the study of the event activity 
dependence of hard probes in p-Pb collision and, hence,
help to establish baselines for the interpretation of the Pb-Pb data.

\ifpreprint
\iffull
\newenvironment{acknowledgement}{\relax}{\relax}
\begin{acknowledgement}
\section*{Acknowledgements}
The ALICE Collaboration would like to thank all its engineers and technicians for their invaluable contributions to the construction of the experiment and the CERN accelerator teams for the outstanding performance of the LHC complex.
The ALICE Collaboration gratefully acknowledges the resources and support provided by all Grid centres and the Worldwide LHC Computing Grid (WLCG) collaboration.
The ALICE Collaboration acknowledges the following funding agencies for their support in building and
running the ALICE detector:
State Committee of Science,  World Federation of Scientists (WFS)
and Swiss Fonds Kidagan, Armenia,
Conselho Nacional de Desenvolvimento Cient\'{\i}fico e Tecnol\'{o}gico (CNPq), Financiadora de Estudos e Projetos (FINEP),
Funda\c{c}\~{a}o de Amparo \`{a} Pesquisa do Estado de S\~{a}o Paulo (FAPESP);
National Natural Science Foundation of China (NSFC), the Chinese Ministry of Education (CMOE)
and the Ministry of Science and Technology of China (MSTC);
Ministry of Education and Youth of the Czech Republic;
Danish Natural Science Research Council, the Carlsberg Foundation and the Danish National Research Foundation;
The European Research Council under the European Community's Seventh Framework Programme;
Helsinki Institute of Physics and the Academy of Finland;
French CNRS-IN2P3, the `Region Pays de Loire', `Region Alsace', `Region Auvergne' and CEA, France;
German Bundesministerium fur Bildung, Wissenschaft, Forschung und Technologie (BMBF) and the Helmholtz Association;
General Secretariat for Research and Technology, Ministry of
Development, Greece;
Hungarian Orszagos Tudomanyos Kutatasi Alappgrammok (OTKA) and National Office for Research and Technology (NKTH);
Department of Atomic Energy and Department of Science and Technology of the Government of India;
Istituto Nazionale di Fisica Nucleare (INFN) and Centro Fermi -
Museo Storico della Fisica e Centro Studi e Ricerche "Enrico
Fermi", Italy;
MEXT Grant-in-Aid for Specially Promoted Research, Ja\-pan;
Joint Institute for Nuclear Research, Dubna;
National Research Foundation of Korea (NRF);
Consejo Nacional de Cienca y Tecnologia (CONACYT), Direccion General de Asuntos del Personal Academico(DGAPA), M\'{e}xico, :Amerique Latine Formation academique – European Commission(ALFA-EC) and the EPLANET Program
(European Particle Physics Latin American Network)
Stichting voor Fundamenteel Onderzoek der Materie (FOM) and the Nederlandse Organisatie voor Wetenschappelijk Onderzoek (NWO), Netherlands;
Research Council of Norway (NFR);
Polish Ministry of Science and Higher Education;
National Science Centre, Poland;
Ministry of National Education/Institute for Atomic Physics and Consiliul Naţional al Cercetării Ştiinţifice - Executive Agency for Higher Education Research Development and Innovation Funding (CNCS-UEFISCDI) - Romania;
Ministry of Education and Science of Russian Federation, Russian
Academy of Sciences, Russian Federal Agency of Atomic Energy,
Russian Federal Agency for Science and Innovations and The Russian
Foundation for Basic Research;
Ministry of Education of Slovakia;
Department of Science and Technology, South Africa;
Centro de Investigaciones Energeticas, Medioambientales y Tecnologicas (CIEMAT), E-Infrastructure shared between Europe and Latin America (EELA), Ministerio de Econom\'{i}a y Competitividad (MINECO) of Spain, Xunta de Galicia (Conseller\'{\i}a de Educaci\'{o}n),
Centro de Aplicaciones Tecnológicas y Desarrollo Nuclear (CEA\-DEN), Cubaenerg\'{\i}a, Cuba, and IAEA (International Atomic Energy Agency);
Swedish Research Council (VR) and Knut $\&$ Alice Wallenberg
Foundation (KAW);
Ukraine Ministry of Education and Science;
United Kingdom Science and Technology Facilities Council (STFC);
The United States Department of Energy, the United States National
Science Foundation, the State of Texas, and the State of Ohio;
Ministry of Science, Education and Sports of Croatia and  Unity through Knowledge Fund, Croatia.
Council of Scientific and Industrial Research (CSIR), New Delhi, India
\end{acknowledgement}
\ifbibtex
\bibliographystyle{utphys}
 
\bibliography{centralityNote}{}
\else
\input{refpreprint.tex}
\fi
\appendix
\clearpage
\section{ALICE Collaboration}
\label{app:collab}



\begingroup
\small
\begin{flushleft}
J.~Adam\Irefn{org39}\And
D.~Adamov\'{a}\Irefn{org82}\And
M.M.~Aggarwal\Irefn{org86}\And
G.~Aglieri Rinella\Irefn{org36}\And
M.~Agnello\Irefn{org110}\textsuperscript{,}\Irefn{org93}\And
A.~Agostinelli\Irefn{org28}\And
N.~Agrawal\Irefn{org47}\And
Z.~Ahammed\Irefn{org130}\And
I.~Ahmed\Irefn{org16}\And
S.U.~Ahn\Irefn{org67}
% ...
\renewcommand\labelenumi{\textsuperscript{\theenumi}~}

\section*{Affiliation notes}
\renewcommand\theenumi{\roman{enumi}}
\begin{Authlist}
\item \Adef{0}Deceased
\item \Adef{idp3771552}{Also at: M.V. Lomonosov Moscow State University, D.V. Skobeltsyn Institute of Nuclear Physics, Moscow, Russia}
\item \Adef{idp4040672}{Also at: University of Belgrade, Faculty of Physics and "Vin\v{c}a" Institute of Nuclear Sciences, Belgrade, Serbia}
\item \Adef{idp4384032}{Permanent Address: Permanent Address: Konkuk University, Seoul, Korea}
\item \Adef{idp4926704}{Also at: Institute of Theoretical Physics, University of Wroclaw, Wroclaw, Poland}
\item \Adef{idp5893632}{Also at: University of Kansas, Lawrence, KS, United States}
\end{Authlist}

\section*{Collaboration Institutes}
\renewcommand\theenumi{\arabic{enumi}~}
\begin{Authlist}

\item \Idef{org1}A.I. Alikhanyan National Science Laboratory (Yerevan Physics Institute) Foundation, Yerevan, Armenia
\item \Idef{org2}Benem\'{e}rita Universidad Aut\'{o}noma de Puebla, Puebla, Mexico
\item \Idef{org3}Bogolyubov Institute for Theoretical Physics, Kiev, Ukraine
\item \Idef{org4}Bose Institute, Department of Physics and Centre for Astroparticle Physics and Space Science (CAPSS), Kolkata, India
\item \Idef{org5}Budker Institute for Nuclear Physics, Novosibirsk, Russia
\item \Idef{org6}California Polytechnic State University, San Luis Obispo, CA, United States
\item \Idef{org7}Central China Normal University, Wuhan, China
\item \Idef{org8}Centre de Calcul de l'IN2P3, Villeurbanne, France
\item \Idef{org9}Centro de Aplicaciones Tecnol\'{o}gicas y Desarrollo Nuclear (CEADEN), Havana, Cuba
\item \Idef{org10}Centro de Investigaciones Energ\'{e}ticas Medioambientales y Tecnol\'{o}gicas (CIEMAT), Madrid, Spain
\item \Idef{org11}Centro de Investigaci\'{o}n y de Estudios Avanzados (CINVESTAV), Mexico City and M\'{e}rida, Mexico
\item \Idef{org12}Centro Fermi - Museo Storico della Fisica e Centro Studi e Ricerche ``Enrico Fermi'', Rome, Italy
\item \Idef{org13}Chicago State University, Chicago, USA
\item \Idef{org14}China Institute of Atomic Energy, Beijing, China
\item \Idef{org15}Commissariat \`{a} l'Energie Atomique, IRFU, Saclay, France
\item \Idef{org16}COMSATS Institute of Information Technology (CIIT), Islamabad, Pakistan
\item \Idef{org17}Departamento de F\'{\i}sica de Part\'{\i}culas and IGFAE, Universidad de Santiago de Compostela, Santiago de Compostela, Spain
\item \Idef{org18}Department of Physics and Technology, University of Bergen, Bergen, Norway
\item \Idef{org19}Department of Physics, Aligarh Muslim University, Aligarh, India
\item \Idef{org20}Department of Physics, Ohio State University, Columbus, OH, United States
\item \Idef{org21}Department of Physics, Sejong University, Seoul, South Korea
\item \Idef{org22}Department of Physics, University of Oslo, Oslo, Norway
\item \Idef{org23}Dipartimento di Elettrotecnica ed Elettronica del Politecnico, Bari, Italy
\item \Idef{org24}Dipartimento di Fisica dell'Universit\`{a} 'La Sapienza' and Sezione INFN Rome, Italy
\item \Idef{org25}Dipartimento di Fisica dell'Universit\`{a} and Sezione INFN, Cagliari, Italy
\item \Idef{org26}Dipartimento di Fisica dell'Universit\`{a} and Sezione INFN, Trieste, Italy
\item \Idef{org27}Dipartimento di Fisica dell'Universit\`{a} and Sezione INFN, Turin, Italy
\item \Idef{org28}Dipartimento di Fisica e Astronomia dell'Universit\`{a} and Sezione INFN, Bologna, Italy
\item \Idef{org29}Dipartimento di Fisica e Astronomia dell'Universit\`{a} and Sezione INFN, Catania, Italy
\item \Idef{org30}Dipartimento di Fisica e Astronomia dell'Universit\`{a} and Sezione INFN, Padova, Italy
\item \Idef{org31}Dipartimento di Fisica `E.R.~Caianiello' dell'Universit\`{a} and Gruppo Collegato INFN, Salerno, Italy
\item \Idef{org32}Dipartimento di Scienze e Innovazione Tecnologica dell'Universit\`{a} del  Piemonte Orientale and Gruppo Collegato INFN, Alessandria, Italy
\item \Idef{org33}Dipartimento Interateneo di Fisica `M.~Merlin' and Sezione INFN, Bari, Italy
\item \Idef{org34}Division of Experimental High Energy Physics, University of Lund, Lund, Sweden
\item \Idef{org35}Eberhard Karls Universit\"{a}t T\"{u}bingen, T\"{u}bingen, Germany
\item \Idef{org36}European Organization for Nuclear Research (CERN), Geneva, Switzerland
\item \Idef{org37}Faculty of Engineering, Bergen University College, Bergen, Norway
\item \Idef{org38}Faculty of Mathematics, Physics and Informatics, Comenius University, Bratislava, Slovakia
\item \Idef{org39}Faculty of Nuclear Sciences and Physical Engineering, Czech Technical University in Prague, Prague, Czech Republic
\item \Idef{org40}Faculty of Science, P.J.~\v{S}af\'{a}rik University, Ko\v{s}ice, Slovakia
\item \Idef{org41}Faculty of Technology, Buskerud and Vestfold University College, Vestfold, Norway
\item \Idef{org42}Frankfurt Institute for Advanced Studies, Johann Wolfgang Goethe-Universit\"{a}t Frankfurt, Frankfurt, Germany
\item \Idef{org43}Gangneung-Wonju National University, Gangneung, South Korea
\item \Idef{org44}Gauhati University, Department of Physics, Guwahati, India
\item \Idef{org45}Helsinki Institute of Physics (HIP), Helsinki, Finland
\item \Idef{org46}Hiroshima University, Hiroshima, Japan
\item \Idef{org47}Indian Institute of Technology Bombay (IIT), Mumbai, India
\item \Idef{org48}Indian Institute of Technology Indore, Indore (IITI), India
\item \Idef{org49}Inha University, Incheon, South Korea
\item \Idef{org50}Institut de Physique Nucl\'eaire d'Orsay (IPNO), Universit\'e Paris-Sud, CNRS-IN2P3, Orsay, France
\item \Idef{org51}Institut f\"{u}r Informatik, Johann Wolfgang Goethe-Universit\"{a}t Frankfurt, Frankfurt, Germany
\item \Idef{org52}Institut f\"{u}r Kernphysik, Johann Wolfgang Goethe-Universit\"{a}t Frankfurt, Frankfurt, Germany
\item \Idef{org53}Institut f\"{u}r Kernphysik, Westf\"{a}lische Wilhelms-Universit\"{a}t M\"{u}nster, M\"{u}nster, Germany
\item \Idef{org54}Institut Pluridisciplinaire Hubert Curien (IPHC), Universit\'{e} de Strasbourg, CNRS-IN2P3, Strasbourg, France
\item \Idef{org55}Institute for Nuclear Research, Academy of Sciences, Moscow, Russia
\item \Idef{org56}Institute for Subatomic Physics of Utrecht University, Utrecht, Netherlands
\item \Idef{org57}Institute for Theoretical and Experimental Physics, Moscow, Russia
\item \Idef{org58}Institute of Experimental Physics, Slovak Academy of Sciences, Ko\v{s}ice, Slovakia
\item \Idef{org59}Institute of Physics, Academy of Sciences of the Czech Republic, Prague, Czech Republic
\item \Idef{org60}Institute of Physics, Bhubaneswar, India
\item \Idef{org61}Institute of Space Science (ISS), Bucharest, Romania
\item \Idef{org62}Instituto de Ciencias Nucleares, Universidad Nacional Aut\'{o}noma de M\'{e}xico, Mexico City, Mexico
\item \Idef{org63}Instituto de F\'{\i}sica, Universidad Nacional Aut\'{o}noma de M\'{e}xico, Mexico City, Mexico
\item \Idef{org64}iThemba LABS, National Research Foundation, Somerset West, South Africa
\item \Idef{org65}Joint Institute for Nuclear Research (JINR), Dubna, Russia
\item \Idef{org66}Konkuk University, Seoul, South Korea
\item \Idef{org67}Korea Institute of Science and Technology Information, Daejeon, South Korea
\item \Idef{org68}KTO Karatay University, Konya, Turkey
\item \Idef{org69}Laboratoire de Physique Corpusculaire (LPC), Clermont Universit\'{e}, Universit\'{e} Blaise Pascal, CNRS--IN2P3, Clermont-Ferrand, France
\item \Idef{org70}Laboratoire de Physique Subatomique et de Cosmologie, Universit\'{e} Grenoble-Alpes, CNRS-IN2P3, Grenoble, France
\item \Idef{org71}Laboratori Nazionali di Frascati, INFN, Frascati, Italy
\item \Idef{org72}Laboratori Nazionali di Legnaro, INFN, Legnaro, Italy
\item \Idef{org73}Lawrence Berkeley National Laboratory, Berkeley, CA, United States
\item \Idef{org74}Lawrence Livermore National Laboratory, Livermore, CA, United States
\item \Idef{org75}Moscow Engineering Physics Institute, Moscow, Russia
\item \Idef{org76}National Centre for Nuclear Studies, Warsaw, Poland
\item \Idef{org77}National Institute for Physics and Nuclear Engineering, Bucharest, Romania
\item \Idef{org78}National Institute of Science Education and Research, Bhubaneswar, India
\item \Idef{org79}Niels Bohr Institute, University of Copenhagen, Copenhagen, Denmark
\item \Idef{org80}Nikhef, National Institute for Subatomic Physics, Amsterdam, Netherlands
\item \Idef{org81}Nuclear Physics Group, STFC Daresbury Laboratory, Daresbury, United Kingdom
\item \Idef{org82}Nuclear Physics Institute, Academy of Sciences of the Czech Republic, \v{R}e\v{z} u Prahy, Czech Republic
\item \Idef{org83}Oak Ridge National Laboratory, Oak Ridge, TN, United States
\item \Idef{org84}Petersburg Nuclear Physics Institute, Gatchina, Russia
\item \Idef{org85}Physics Department, Creighton University, Omaha, NE, United States
\item \Idef{org86}Physics Department, Panjab University, Chandigarh, India
\item \Idef{org87}Physics Department, University of Athens, Athens, Greece
\item \Idef{org88}Physics Department, University of Cape Town, Cape Town, South Africa
\item \Idef{org89}Physics Department, University of Jammu, Jammu, India
\item \Idef{org90}Physics Department, University of Rajasthan, Jaipur, India
\item \Idef{org91}Physik Department, Technische Universit\"{a}t M\"{u}nchen, Munich, Germany
\item \Idef{org92}Physikalisches Institut, Ruprecht-Karls-Universit\"{a}t Heidelberg, Heidelberg, Germany
\item \Idef{org93}Politecnico di Torino, Turin, Italy
\item \Idef{org94}Purdue University, West Lafayette, IN, United States
\item \Idef{org95}Pusan National University, Pusan, South Korea
\item \Idef{org96}Research Division and ExtreMe Matter Institute EMMI, GSI Helmholtzzentrum f\"ur Schwerionenforschung, Darmstadt, Germany
\item \Idef{org97}Rudjer Bo\v{s}kovi\'{c} Institute, Zagreb, Croatia
\item \Idef{org98}Russian Federal Nuclear Center (VNIIEF), Sarov, Russia
\item \Idef{org99}Russian Research Centre Kurchatov Institute, Moscow, Russia
\item \Idef{org100}Saha Institute of Nuclear Physics, Kolkata, India
\item \Idef{org101}School of Physics and Astronomy, University of Birmingham, Birmingham, United Kingdom
\item \Idef{org102}Secci\'{o}n F\'{\i}sica, Departamento de Ciencias, Pontificia Universidad Cat\'{o}lica del Per\'{u}, Lima, Peru
\item \Idef{org103}Sezione INFN, Bari, Italy
\item \Idef{org104}Sezione INFN, Bologna, Italy
\item \Idef{org105}Sezione INFN, Cagliari, Italy
\item \Idef{org106}Sezione INFN, Catania, Italy
\item \Idef{org107}Sezione INFN, Padova, Italy
\item \Idef{org108}Sezione INFN, Rome, Italy
\item \Idef{org109}Sezione INFN, Trieste, Italy
\item \Idef{org110}Sezione INFN, Turin, Italy
\item \Idef{org111}SSC IHEP of NRC Kurchatov institute, Protvino, Russia
\item \Idef{org112}SUBATECH, Ecole des Mines de Nantes, Universit\'{e} de Nantes, CNRS-IN2P3, Nantes, France
\item \Idef{org113}Suranaree University of Technology, Nakhon Ratchasima, Thailand
\item \Idef{org114}Technical University of Split FESB, Split, Croatia
\item \Idef{org115}The Henryk Niewodniczanski Institute of Nuclear Physics, Polish Academy of Sciences, Cracow, Poland
\item \Idef{org116}The University of Texas at Austin, Physics Department, Austin, TX, USA
\item \Idef{org117}Universidad Aut\'{o}noma de Sinaloa, Culiac\'{a}n, Mexico
\item \Idef{org118}Universidade de S\~{a}o Paulo (USP), S\~{a}o Paulo, Brazil
\item \Idef{org119}Universidade Estadual de Campinas (UNICAMP), Campinas, Brazil
\item \Idef{org120}University of Houston, Houston, TX, United States
\item \Idef{org121}University of Jyv\"{a}skyl\"{a}, Jyv\"{a}skyl\"{a}, Finland
\item \Idef{org122}University of Liverpool, Liverpool, United Kingdom
\item \Idef{org123}University of Tennessee, Knoxville, TN, United States
\item \Idef{org124}University of the Witwatersrand, Johannesburg, South Africa
\item \Idef{org125}University of Tokyo, Tokyo, Japan
\item \Idef{org126}University of Tsukuba, Tsukuba, Japan
\item \Idef{org127}University of Zagreb, Zagreb, Croatia
\item \Idef{org128}Universit\'{e} de Lyon, Universit\'{e} Lyon 1, CNRS/IN2P3, IPN-Lyon, Villeurbanne, France
\item \Idef{org129}V.~Fock Institute for Physics, St. Petersburg State University, St. Petersburg, Russia
\item \Idef{org130}Variable Energy Cyclotron Centre, Kolkata, India
\item \Idef{org131}Warsaw University of Technology, Warsaw, Poland
\item \Idef{org132}Wayne State University, Detroit, MI, United States
\item \Idef{org133}Wigner Research Centre for Physics, Hungarian Academy of Sciences, Budapest, Hungary
\item \Idef{org134}Yale University, New Haven, CT, United States
\item \Idef{org135}Yonsei University, Seoul, South Korea
\item \Idef{org136}Zentrum f\"{u}r Technologietransfer und Telekommunikation (ZTT), Fachhochschule Worms, Worms, Germany
\end{Authlist}
\endgroup

\else
\ifbibtex
\bibliographystyle{\bibstname}
\bibliography{centralityNote}{}
\else
\input{refpreprint.tex} 
\fi
\appendix
\fi
\else
\iffull
\vspace{0.5cm}
\input{refpaper.tex}
\newpage
\appendix
\else
\appendix
\ifbibtex
\bibliographystyle{\bibstname}
\bibliography{centralityNote}{}
\else
\input{refpaper.tex}
\fi
\fi
\fi
\end{document}